\documentclass[twocolumn,tighten]{aastex63}

\usepackage{amsmath}

\newcommand{\beam}{$\theta_{\mbox{\scriptsize maj}}\times\theta_{\mbox{\scriptsize min}}$}
\newcommand{\mjyperbeam}{mJy\,beam$^{-1}$}
\newcommand{\ujyperbeam}{$\mu$Jy\,beam$^{-1}$}
\newcommand{\rah}{$^{\mbox{\scriptsize h}}$}
\newcommand{\ram}{$^{\mbox{\scriptsize m}}$}
\newcommand{\ras}{$^{\mbox{\scriptsize s}}$}
\newcommand{\decd}{$^{\circ}$}
\newcommand{\decm}{$'$}
\newcommand{\decs}{$''$}

\newcommand{\amax}{$a_{\mbox{\scriptsize max}}$ }

\received{August 09, 2020}
\revised{Month Date, Year}
\accepted{Month Date, Year}
\submitjournal{ApJ}

\shorttitle{JVLA 9 mm polarizatio on OMC-3/MMS~6}
\shortauthors{Liu et al.}
\graphicspath{{./}{figures/}}
\begin{document}

\title{Magnetically regulated disk-formation in the inner 100 au region of the Class 0 young stellar object OMC-3/MMS~6 resolved by JVLA and ALMA}

\correspondingauthor{Hauyu Baobab Liu}
\email{hyliu@asiaa.sinica.edu.tw}

%\collaboration{1}{(AAS Journals Data Scientists collaboration)}

\author[0000-0003-2300-2626]{Hauyu Baobab Liu}
\affil{Institute of Astronomy and Astrophysics, Academia Sinica, 11F of Astronomy-Mathematics Building, AS/NTU No.1, Sec. 4,
Roosevelt Rd, Taipei 10617, Taiwan, ROC}

%\nocollaboration{0}

%% Note that the \and command from previous versions of AASTeX is now
%% depreciated in this version as it is no longer necessary. AASTeX 
%% automatically takes care of all commas and "and"s between authors names.

%% AASTeX 6.3 has the new \collaboration and \nocollaboration commands to
%% provide the collaboration status of a group of authors. These commands 
%% can be used either before or after the list of corresponding authors. The
%% argument for \collaboration is the collaboration identifier. Authors are
%% encouraged to surround collaboration identifiers with ()s. The 
%% \nocollaboration command takes no argument and exists to indicate that
%% the nearby authors are not part of surrounding collaborations.

%% Mark off the abstract in the ``abstract'' environment. 
\begin{abstract}
We have carried out polarization calibration for archival JVLA ($\sim$9 mm) full polarization observations towards the Class~0 young stellar object (YSO) OMC-3/MMS~6 (also known as HOPS-87), and then compared with the archival ALMA 1.2 mm observations.
We found that the innermost $\sim$100 au region of OMC-3/MMS~6 is likely very optically thick (e.g., $\tau\gg$1) at $\sim$1 mm wavelength such that the dominant polarization mechanism is dichroic extinction.
It is marginally optically thin (e.g., $\tau\lesssim$1) at $\sim$9 mm wavelength such that the JVLA observations can directly probe the linearly polarized emission from non-spherical dust.
Assuming that the projected long axis of dust grains is aligned perpendicular to magnetic field (B-field) lines, we propose that the overall B-field topology resembles an hourglass shape, while this "hourglass" appears $\sim$40$^{\circ}$ inclined with respect to the previously reported outflow axis.
The geometry of this system is consistent with a magnetically regulated dense (pseudo-)disk.
% Based on the measurements of Stokes I spectral indices we favor a maximum grain size of $a_{\max}\le$100 $\mu$m in the (pseudo-)disk.
% Assuming $a_{\max}\le$100 $\mu$m and quoting the default DSHARP opacity table,  the observed 29.45 GHz flux density corresponds to an overall dust mass of $\sim$14000 $M_{\oplus}$ within a $\sim$43 au radius.
Based on the observed 29.45 GHz flux density and assuming a dust absorption opacity $\kappa ^{\mbox{\tiny abs}} _ {\mbox{\tiny 29.45 GHz}}=$0.0096 cm$^{2}$\,g$^{-1}$, the derived overall dust mass within a $\sim$43 au radius is $\sim$14000 $M_{\oplus}$.
From this case study, it appears to us that some previous 9 mm surveys towards Class~0/I YSOs might have systematically underestimated dust masses by one order of magnitude, owing to that they assumed the too high dust absorption opacity ($\sim$0.1 cm$^{2}$\,g$^{-1}$) for $\sim$9 mm wavelengths but without self-consistently considering the dust scattering opacity.
\end{abstract}

%% Keywords should appear after the \end{abstract} command. 
%% See the online documentation for the full list of available subject
%% keywords and the rules for their use.
\keywords{evolution --- ISM: individual objects (OMC-3/MMS~6) --- stars: formation}
%% From the front matter, we move on to the body of the paper.
%% Sections are demarcated by \section and \subsection, respectively.
%% Observe the use of the LaTeX \label
%% command after the \subsection to give a symbolic KEY to the
%% subsection for cross-referencing in a \ref command.
%% You can use LaTeX's \ref and \label commands to keep track of
%% cross-references to sections, equations, tables, and figures.
%% That way, if you change the order of any elements, LaTeX will
%% automatically renumber them.
%%
%% We recommend that authors also use the natbib \citep
%% and \citet commands to identify citations.  The citations are
%% tied to the reference list via symbolic KEYs. The KEY corresponds
%% to the KEY in the \bibitem in the reference list below. 

\section{Introduction} \label{sec:introduction}

\begin{deluxetable*}{ l | c c }
\tablecaption{Summary for the JVLA Ka band observations\label{tab:jvla}}
  \tablehead{
  \colhead{Array config.}  & \colhead{C array} &  \colhead{A array }
%     \colhead{Flux calib.}  &  \colhead{Passband calib.} &   \colhead{Phase calib.}   & \colhead{Gain calib.flux}
%  & \colhead{\footnotesize Obs. ID} \\  
%  &{\footnotesize UTC (YYYYMMDD)} &  &    {\footnotesize (GHz)}   & {\footnotesize (meters)}  &   &   & {\footnotesize (Jy)} & 
  } 
  \startdata
Observing date (UTC) & 2016 March 13 & 2016 December 01 \\
Projected baseline lengths (meters) & 45-3210 & 580-36480 \\
Flux calib. & 3C147 & 3C48 \\
Passband calib. & 3C84 & 3C84 \\
Gain calib. & J0541-0541 & J0541-0541 \\
Gain calib. flux (Jy) & 0.44$^{1}$ & 0.48$^{2}$ \\
Passband calib. flux (Jy) & 25.9$^{1}$ & 35.9$^{2}$ \\
Synthesized beam$^{3}$  & \beam$=$1\farcs3$\times$0\farcs69 (P.A.$=$42$^{\circ}$) & \beam$=$0\farcs12$\times$0\farcs067, (P.A.$=$6.1$^{\circ}$) \\
rms noise (\ujyperbeam)$^{3}$ & 14 & 9.0 \\
  \enddata
  \tablecomments{(1) Measured at 33 GHz frequency. (2) Measured at 33.3 GHz frequency. (3) Measured from the Briggs Robust$=$2 weighted images which was produced utilizing all available bandwidths.}
\end{deluxetable*}

Observing linear polarization of dust emission at (sub)millimeter bands have been an essential method of resolving magnetic field (B-field) topology in the interstellar medium from parsec scales down to $\sim$1000 au scales (for a review see \citealt{Hildebrand2000}).
This approach has important applications in the studies about magnetic support against the self-gravitational collapse/fragmentation of molecular clouds, and the re-distribution of angular momentum around circumstellar disks due to magnetic torque.
Typically, the observed linear polarization was attributed to optically thin emission of non-spherical dust grains, of which the projected long axes are aligned perpendicular to B-field lines.
In such cases, the projected B-field orientation can be inferred by the 90$^{\circ}$ rotated electric field (E-field) polarization position angle.
Based on this approach, the recent interferometric surveys at 0.85-1.3 mm wavelengths (e.g. \citealt{Hull2014,Zhang2014,Cox2018,Galametz2018}) have suggested that B-field in the low or high-mass star-forming cores or envelopes may align either perpendicular or parallel to outflow axes.

On the other hand, based on the resolved continuum brightness temperature towards the low-mass Class~0 young stellar object (YSO) IRAS~16293-2422B, \citet{Liu2018} proposed that the resolved dust polarization from this specific source may be attributed to the dichroic extinction of aligned, non-spherical dust against the unpolarized background dust thermal emission (e.g., eminated from the embedded protoplanetary disk).
\citet{Lee2018} also interpreted the $\sim$350 GHz linear polarization in the inner $\sim$100 au region of the Class~0 YSO HH212 by dichroic extinction.
If this is indeed the case, then the projected B-field orientations inferred from the typical way are exactly 90$^{\circ}$ offset from the actual orientations.
The conclusions of those previous surveys thus may have been confused due to mis-identifying the polarization mechanisms. 

Previously, dust polarization due to dichroic extinction has only been considered in the interpretation of observations at infrared or shorter wavelengths (e.g., \citealt{Reissl2017}). 
This was because that it is usually considered that the dust emission in circumstellar cores/envelopes is optically thin long-ward of submillimeter wavelengths.
On the contrary, the spectral energy distribution (SED) analyses of \citet{Li2017} towards a sample of Class~0 YSOs have indicated that the dust optical depth in fact can be higher than 10 at 1~mm wavelength (see also \citealt{Galvan2018}).

\begin{figure*}
    \hspace{-1.5cm}
    \begin{tabular}{ p{9cm} p{9cm} }
         \includegraphics[width=9.3cm]{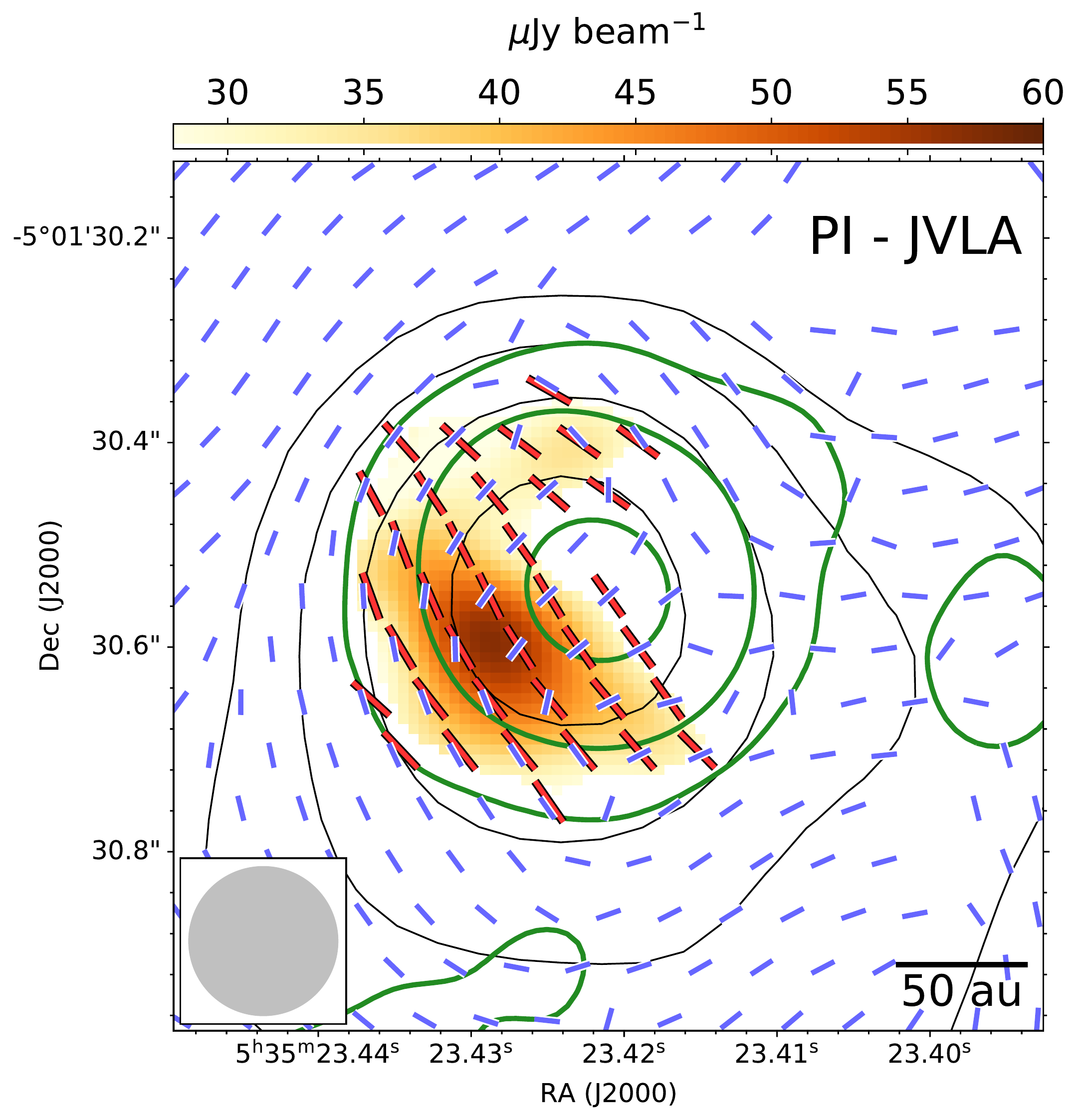} &
         \includegraphics[width=9.3cm]{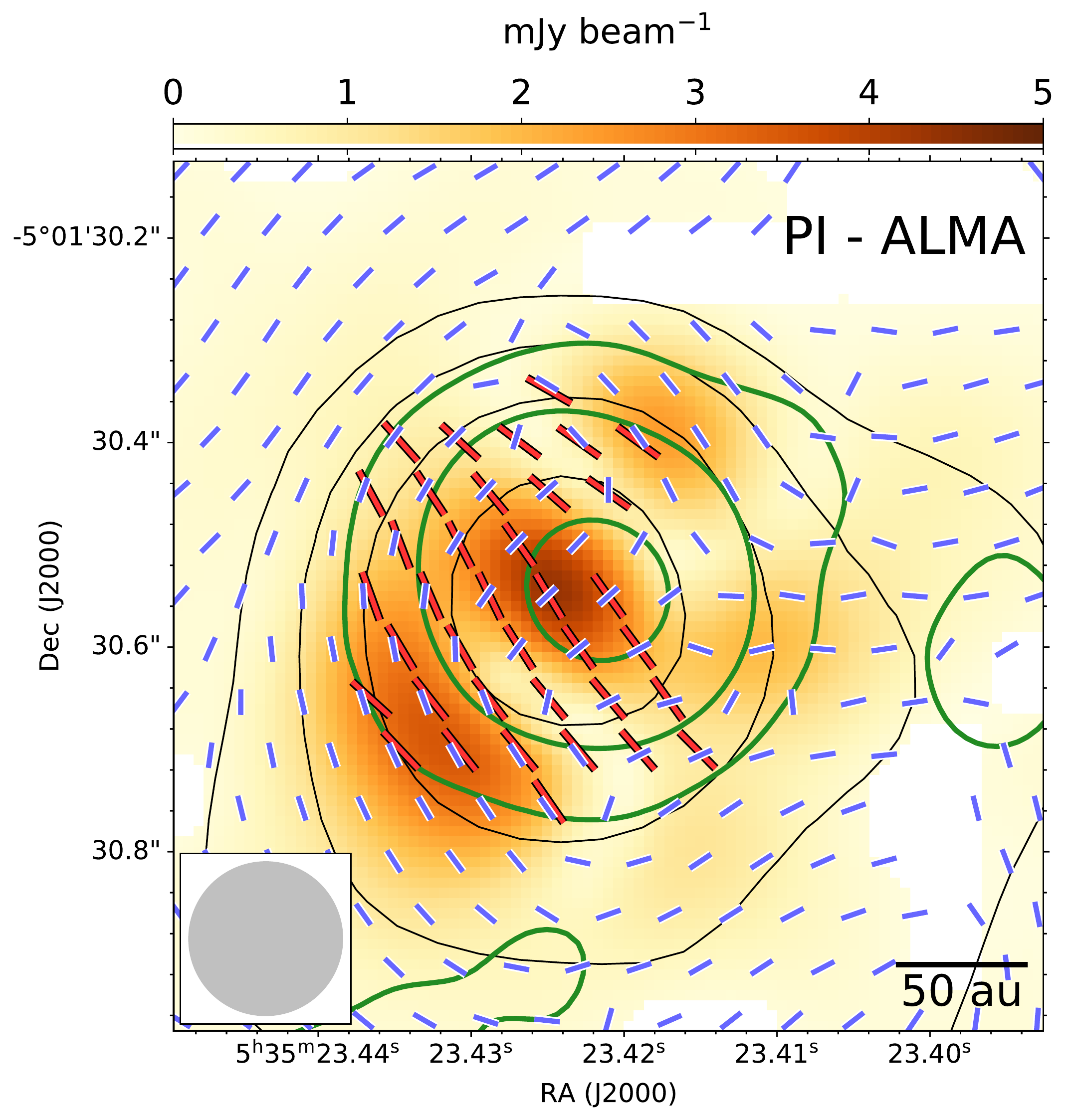} \\ 
    \end{tabular}
    \caption{Polarization images towards OMC-3/MMS~6. Color images in the left and right panels show the polarization intensity (PI) images taken by JVLA and ALMA, respectively. Red and blue line segments show the (E-field) polarization position angle (PA) detected by JVLA and ALMA, respectively. Black contours in both panels show the ALMA Stokes I image ([375-$\sigma$, 750-$\sigma$, 1500-$\sigma$, 3000-$\sigma$]; 1-$\sigma\sim$27 mK)). Green contours in both panels show the JVLA Stokes I image ([5-$\sigma$, 25-$\sigma$, 125-$\sigma$]; 1-$\sigma\sim$0.43 K). All images presented in this figure have been smoothed to a circular 0\farcs15 (58 au) synthesized beam, which are shown in the bottom left. The measured Stokes I, Q, U, PI, PA, and polarization percentages at the centroids of the presented line segments, are given in Table \ref{tab:jvla_segments}. }
    \label{fig:polmap}
\end{figure*}

For the specific Class~0 YSO, NGC1333~IRAS4A1, the hypothesized high optical depth has been unambiguously verified by the spectral line observations at 1-10 mm bands (\citealt{Sahu2019,Su2019,DeSimone2020}).
Short-ward of the $\sim$1 mm wavelength, dust emission is essentially de-polarized in this source owing to the high optical depths, while dust self-scattering (e.g., \citealt{Kataoka2015}) is unlikely to be an efficient polarization mechanism due to that it is hard to produce highly anisotropic radiation fields in environments with very high optical depths.
Therefore, dichroic extinction, to our knowledge, is the only probable efficient linear polarization mechanism in this source at $\sim$1 mm wavelength.
In this case, when comparing the PA measured at $\sim$1 mm wavelengths with those measured at optically thinner, longer wavelength bands, one expect to see an exact 90$^{\circ}$ relative offset.
It has been verified for the first time by the presently lone observational case study of \citet{Ko2020}.

% Besides the correction of B-field position angles, these studies also have implied that the earlier studies in general may have underestimated the dust masses and overestimated the maximum grain sizes.

There has been an argument of whether or not NGC1333~IRAS4A1 (and the not yet confirmed case, IRAS~16293-2422B) are merely the rather massive exceptions.
Whether or not dichroic extinction can be a prominent linear polarization mechanism around any other circumstellar core/envelope at $\lesssim$1 mm wavelengths remains an open question.

In this paper, we report the case study towards an optically much thinner Class~0 YSO, OMC-3/MMS~6 which is located at $\sim$388 pc distance (\citealt{Kounkel2017}; c.f., \citealt{Takahashi2012,Takahashi2019}).
By comparing the archival NRAO Karl G. Jansky Very Large Array (JVLA) $\sim$9 mm observations with the Atacama Large Millimeter Array (ALMA) $\sim$1 mm observations, our aim is to test whether or not we can resolve the 90$^{\circ}$ relative PA offset expected from dichroic extinction.
If true, then this source would be the second case that is verified with linear polarization due to dichroic extinction at $\sim$1 mm wavelength. 
Such a result will infer that this polarization mechanism is not only limited to cases with exceptionally high optical depth but could be rather common instead.
Our data reduction is outlined in Section \ref{sec:obs}.
The results are shown in Section \ref{sec:result}.
We will focus on the innermost region where the polarized intensity is significantly detected by the JVLA observations, given that the ALMA polarization results on the more extended spatial scales have been very well introduced by \citet{Takahashi2019}.
Our toy model for qualitatively interpreting these observations is described in Section \ref{sec:discussion}.
Our conclusion is given in Section \ref{sec:conclusion}.
We provide our polarization measurements in Appendix Section \ref{sec:images}.
The details of how we constructed the toy model and performed radiative transfer are introduced in Appendix Sections \ref{sec:radiativetransfer} and \ref{sec:toy}.

\section{Data reduction} \label{sec:obs}

\subsection{Archival ALMA data} \label{sub:ALMA}
We retrieved the archival, full polarization ALMA observations taken from project 2015.1.00341.S. 
They have been published in \citet{Takahashi2019}. % where the technical details were very well summarized with tables.
There were \beam$=$0\farcs14$\times$0\farcs12 (54 au$\times$47 au; P.A.$=$-50$^{\circ}$) observations (projected baseline ranges: 16-3200 $k\lambda$; hereafter the low-resolution ALMA image) taken in October of 2016; and \beam$=$0\farcs022$\times$0\farcs020 (8.5 au$\times$7.8 au; P.A.$=$-82$^{\circ}$) observations (projected baseline ranges: 76–14700 $k\lambda$; hereafter the high-resolution ALMA image) taken in October of 2015.
We directly utilize the official image products delivered from the quality assurance 2 (QA2).
% This is because that the JVLA observations we are comparing with only had sensitivity to detect polarized intensity from the innermost region where the ALMA observations achieved very high S/N.
% Re-calibrating the ALMA data will mainly affect the fainter, rather spatially extended region, which is not relevant to our present science purpose.

In this paper, we only quantitatively compare the low-resolution ALMA observations with the JVLA observations, given that they have similar angular resolutions.
We smoothed the low-resolution ALMA images to a 0\farcs15 ($\sim$58 au) circular synthesized beam before making quantitative comparison with the JVLA observations (Section \ref{sub:JVLA}).
We re-interpreted the polarization line segments resolved from the high-resolution observations qualitatively, although we caution that the interpretation based on the observations of only one single frequency band would need to be tested by observations at other frequencies (more in Section \ref{sec:discussion}).
At the mean observing frequency 265.968 GHz, the high-resolution ALMA image resolved a peak brightness temperature ($T_{\mbox{\scriptsize b}}$) of $\sim$120 K, which might have been biased low due to missing short-spacing besides the potential absolute flux calibration errors. % although was not serious.
The higher quality image presented in \citet{Takahashi2019} shows a peak $T_{\mbox{\scriptsize b}}$ of $\sim$190 K.

\begin{figure*}
    \hspace{-1cm}
    \begin{tabular}{ p{8.5cm} p{8.5cm} }
         \includegraphics[width=8.7cm]{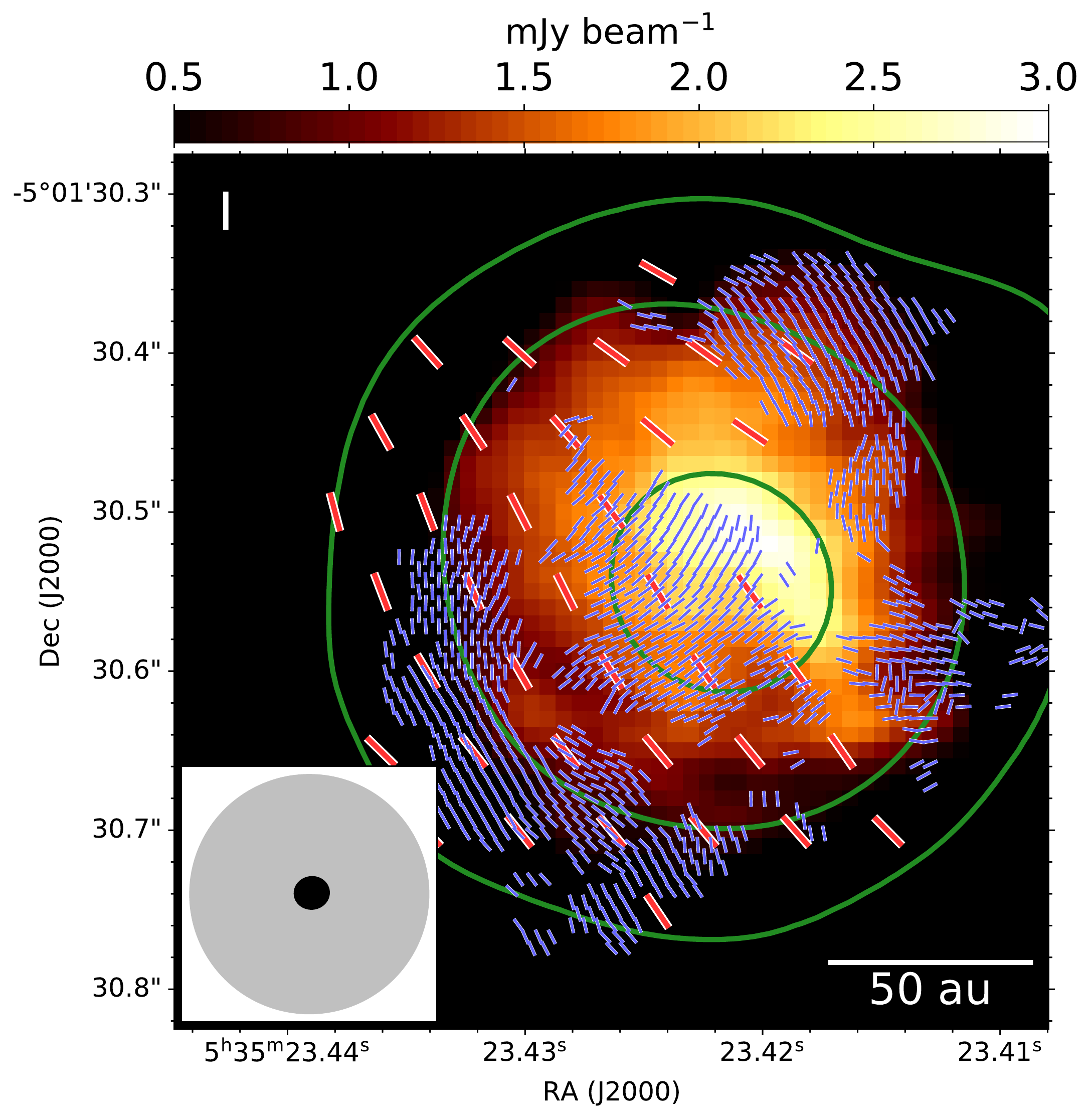}  &
         \includegraphics[width=8.7cm]{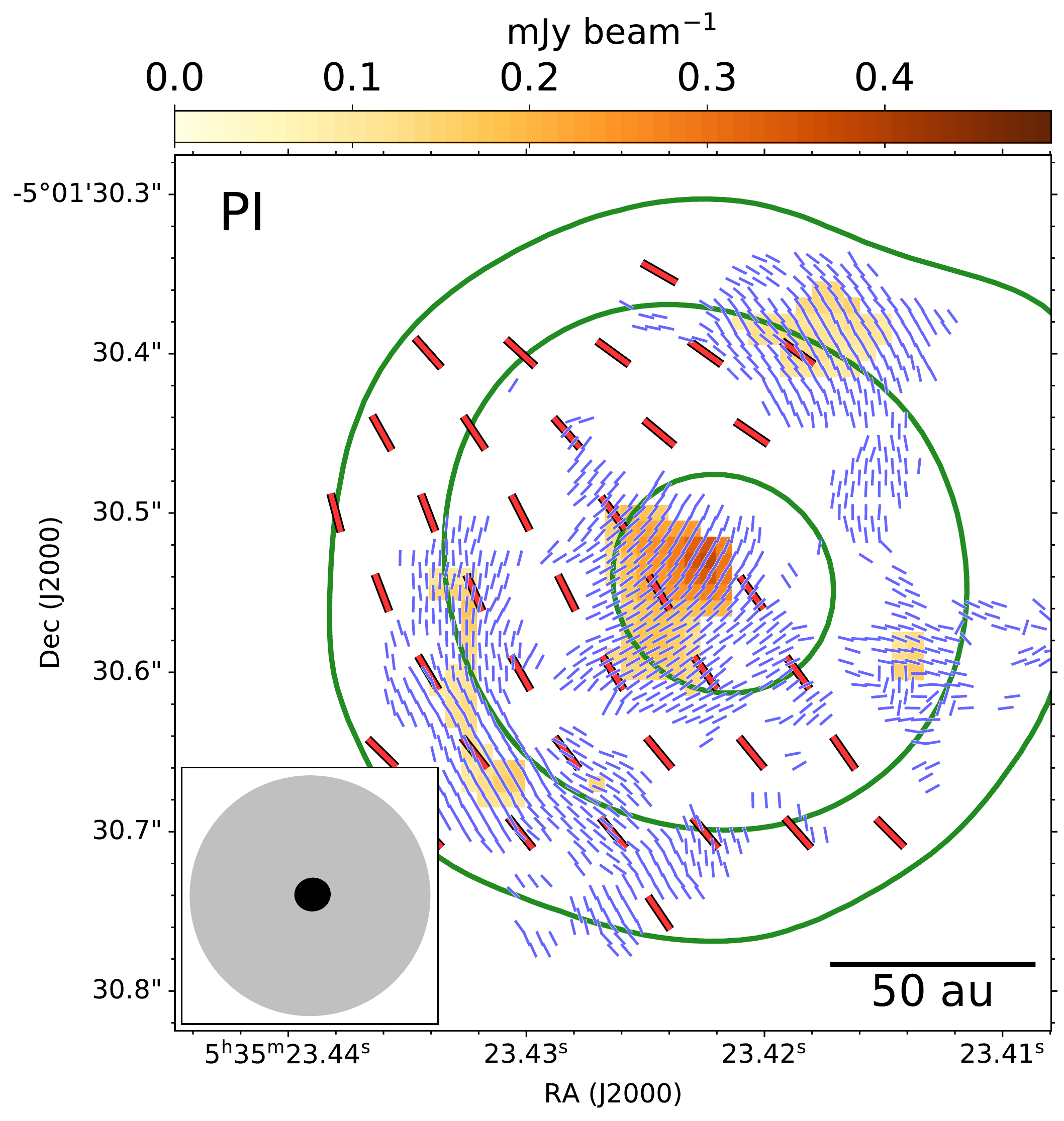}  \\
    \end{tabular}

    \caption{Images showing the inner $\sim$100 au (radius) region of OMC-3/MMS~6.
    {\it Left :--} Stokes I intensity image taken from the high-resolution ALMA observations (color).
    {\it Right :--} PI intensity taken from the high-resolution ALMA observations (color).
    Green contours and red line segments are the same with those presented in Figure \ref{fig:polmap}. Blue line segments show the E-field line segments measured from the high-resolution ALMA observations.  The synthesized beam of the JVLA image and the high-resolution ALMA image are presented in bottom left.
    Gray ellipse in all panels show the 0\farcs15 circular synthesized beam; black ellipse shows the synthesized beam of the high-resolution ALMA image.
    }
    \label{fig:ALMAhires}
\end{figure*}

\begin{figure*}
    \hspace{-1cm}
    \begin{tabular}{ p{8.5cm} p{8.5cm} }
         \includegraphics[width=8.7cm]{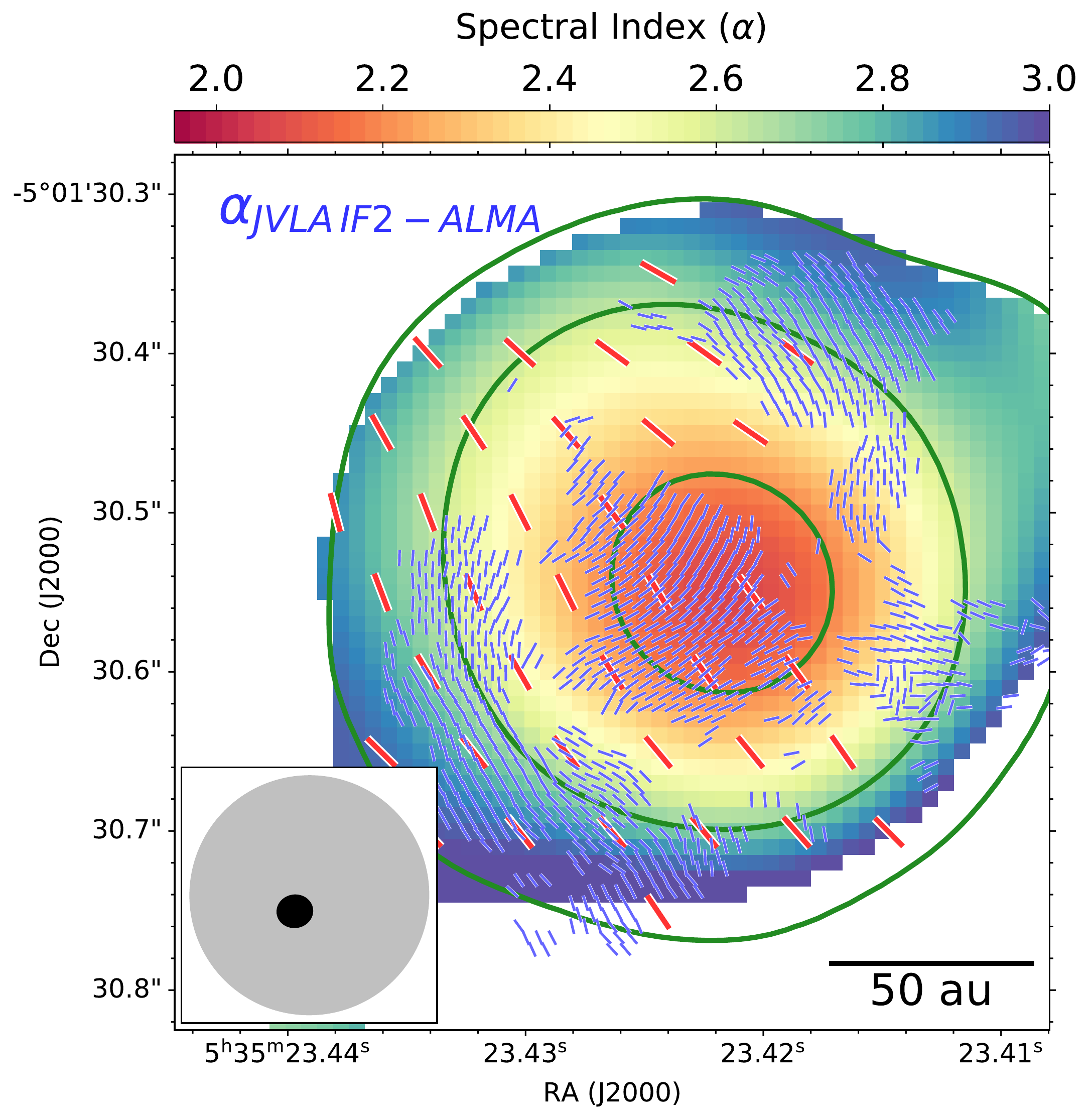}  &
         \includegraphics[width=8.7cm]{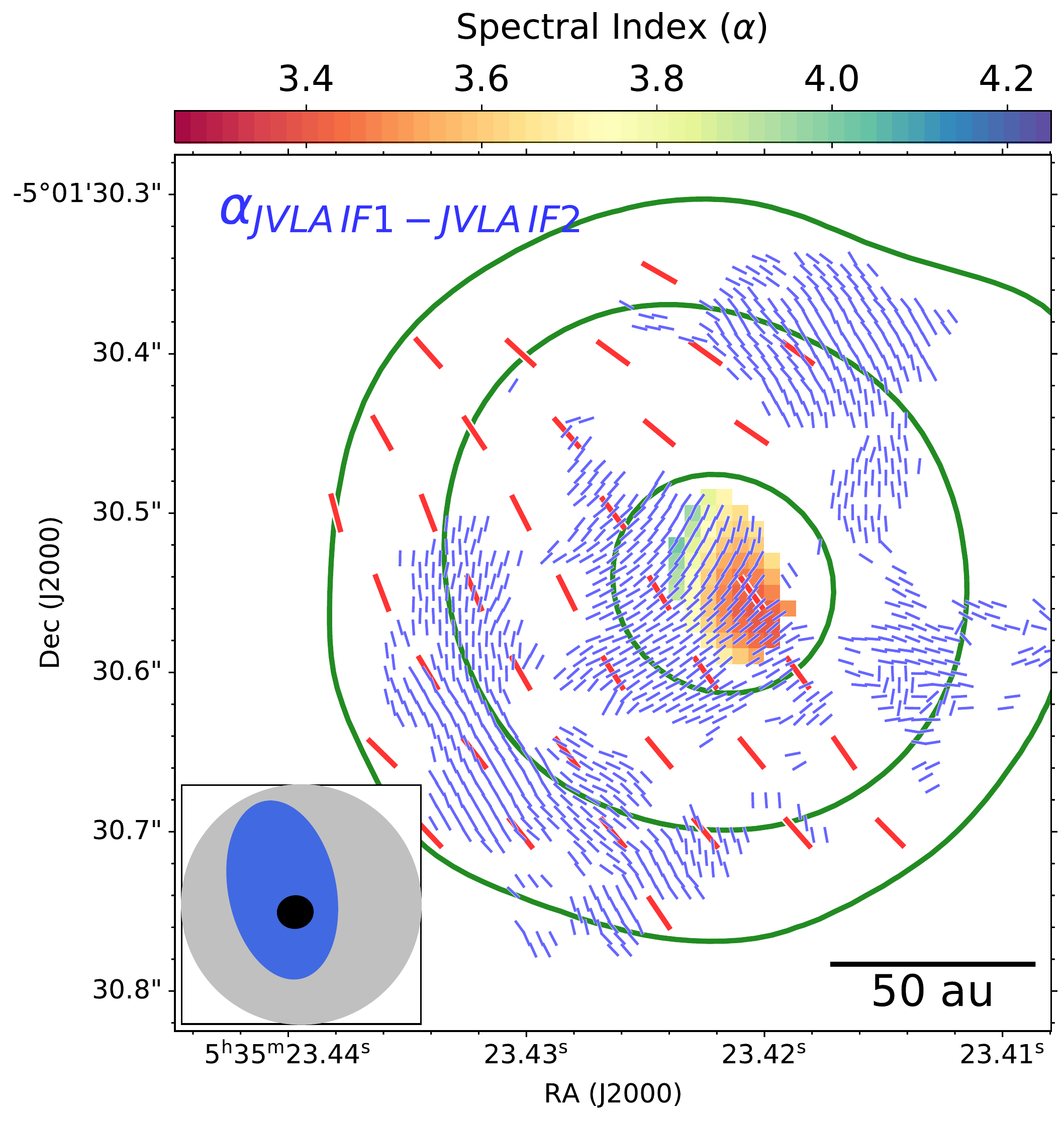}  \\
    \end{tabular}
    \caption{Spectral indices in the inner $\sim$100 au (radius) region of OMC-3/MMS~6.
    {\it Left :--} The spectral index map generated by comparing the low-resolution ALMA image and the JVLA IF2 image (i.e.,  $\alpha_{\mbox{\tiny IF2-ALMA}}$; see Section \ref{subsub:StokesI}), where the spectral indices are presented for the region that the spectral index uncertainties (assuming Gaussian noise and standard error propagation) are lower than 0.1. The detected lowest and highest spectral indices in this map are 2.1 and 3.0, respectively.
    {\it Right :--} The spectral index map generated by comparing the JVLA IF1 and IF2 images (i.e.,  $\alpha_{\mbox{\tiny IF1-IF2}}$; see Section \ref{subsub:StokesI}), where the spectral indices are presented for the region that the spectral index uncertainties are lower than 0.3 (as a compromise due to the limited S/N). The detected lowest and highest spectral indices in this map are 3.4 and 4.1, respectively.  Green contours and red line segments are the same with those presented in Figure \ref{fig:polmap}. Blue line segments show the E-field line segments measured from the high-resolution ALMA observations.  The synthesized beam of the JVLA image and the high-resolution ALMA image are presented in bottom left.
    Gray ellipse in all panels show the 0\farcs15 circular synthesized beam; black ellipse shows the synthesized beam of the high-resolution ALMA image. The blue ellipse in the bottom right panel shows the synthesized beam of the spectral index map presented in that panel.
    }
    \label{fig:ALMAhiresSpid}
\end{figure*}

\subsection{JVLA Ka band data} \label{sub:JVLA}
We have retrieved the archival, full polarization JVLA observations towards OMC-3/MMS~6 which was taken at Ka band in the C and A array configurations (project code: 16A-197).
The Stokes~I images of these JVLA observations have been published in \citet{Tobin2020}.
All these observations utilized the 3 bit sampler, and configured the backend to provide an 8 GHz bandwidth coverage by 64 spectral windows which covered the frequency ranges of 27.0-30.9 GHz (IF1) and 34.9-38.8 GHz (IF2).
The pointing center for the observations on OMC-3/MMS~6 (HOPS-87) was R.A.= 05$^{\mbox{\scriptsize{h}}}$35$^{\mbox{\scriptsize{m}}}$22.891$^{\mbox{\scriptsize{s}}}$ (J2000), decl.=-05$^{\circ}$01$'$24$''$.21 (J2000), which was offset from OMC-3/MM~6 to simultaneously cover another source, HOPS-88.
The C configuration track we retrieved also pointed on R.A.= 05$^{\mbox{\scriptsize{h}}}$35$^{\mbox{\scriptsize{m}}}$18.915$^{\mbox{\scriptsize{s}}}$ (J2000), decl.=-05$^{\circ}$00$'$50$''$.87 (J2000) and R.A.= 05$^{\mbox{\scriptsize{h}}}$35$^{\mbox{\scriptsize{m}}}$21.400$^{\mbox{\scriptsize{s}}}$ (J2000), decl.=-05$^{\circ}$13$'$17$''$.50 (J2000) to observe the other two YSOs, HOPS-91 and HOPS-409.
Further information for these observations have been summarized in Table \ref{tab:jvla}.

We manually followed the standard data calibration strategy using the Common Astronomy Software Applications \citep[CASA;][]{McMullin2007} package release 5.6.2-3.
We adopted the Perley-Butler 2017 flux standards \citep{Perley2017}.
After implementing antenna position corrections, weather information, gain-elevation curve and opacity model, we bootstrapped delay fitting and passband calibrations, and then performed complex gain calibration.
We performed gain phase self-calibration for the C array configuration observations on OMC-3/MMS~6, HOPS-91, and HOPS-409 by combining all available spectral windows and using the solution intervals of 30 seconds, 180 seconds, and 90 seconds, respectively.
We performed gain phase self-calibration for the A array configuration observations by combining all available spectral windows and using a 600 seconds solution interval due to the limited signal-to-noise (S/N) ratios.

These data have been known to be impacted by a delay bug of the JVLA\footnote{https://science.nrao.edu/facilities/vla/data-processing/vla-atmospheric-delay-problem}.
The CASA version we used has been patched to suppress the related delay errors.
The residual delay errors at the mean observing frequency have further been removed during our gain phase self-calibration.
Since we combined spectral windows during the gain phase self-calibration, there might be small, frequency dependent residual delay errors which we cannot remove.
Nevertheless, we do not expect the small frequency dependent residual delay errors to be significant for our present science purpose.

For the C and A array configuration data, we calibrated the cross-hand delay and absolute polarization position angle based on the observations on 3C147 and 3C48.
When deriving these solutions, we limited the projected {\it uv} distance ranges of 3C48 to 0-500 $k\lambda$ since it was seriously spatially resolved at longer baselines.
We calibrated the polarization leakage based on the observations on 3C84. 
The imaging processes and the achieved image quality are outlined in the following sections.

\begin{figure}
    \hspace{-1.5cm}
    \begin{tabular}{ p{9cm} }
         \includegraphics[width=9.5cm]{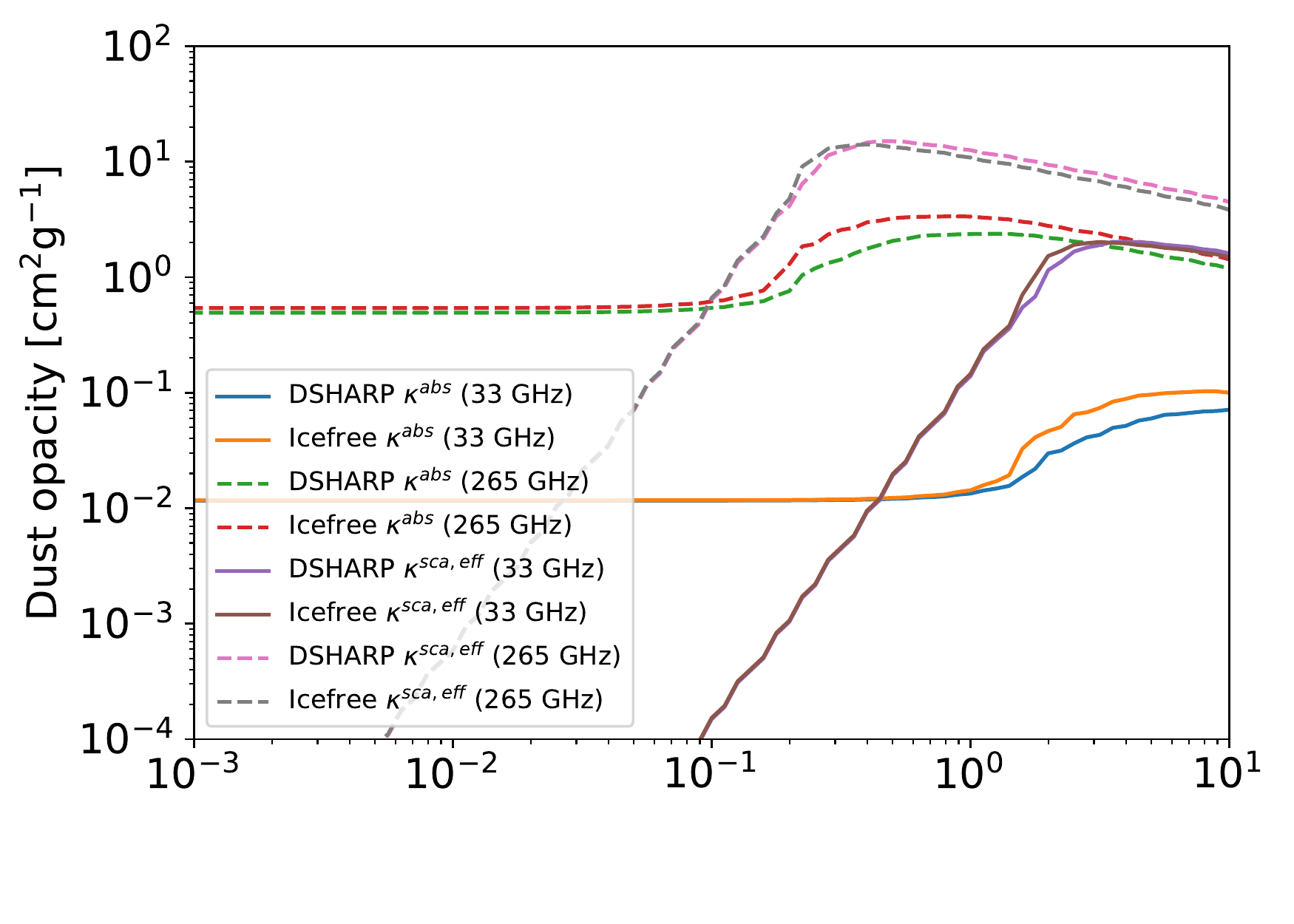} \\
         \vspace{-1.4cm}\includegraphics[width=9.5cm]{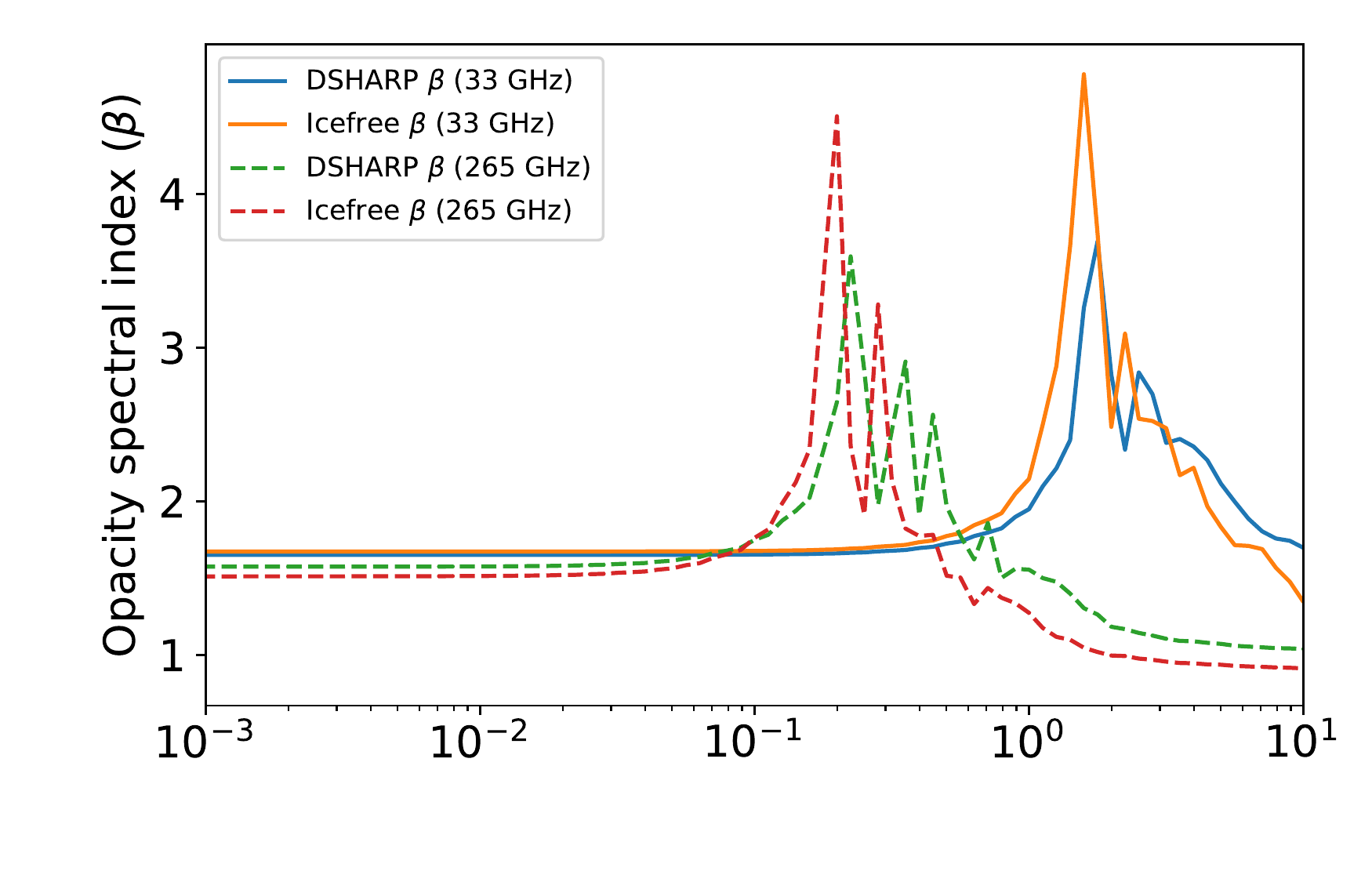} \\
         \vspace{-1.15cm}\includegraphics[width=9.5cm]{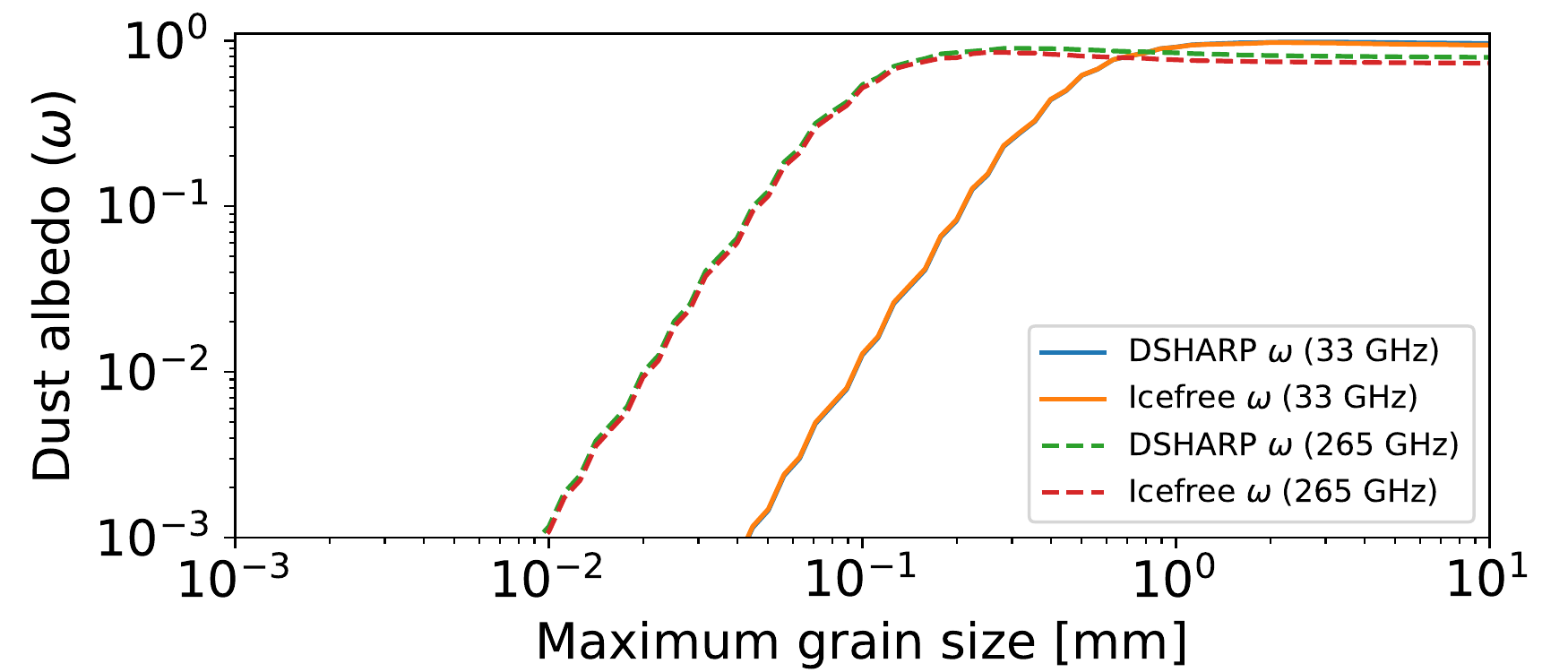} \\
    \end{tabular}
    \caption{The DSHARP-default and DSHARP-icefree dust opacity models at 33 and 265 GHz (quoted from \citealt{Birnstiel2018}). Horizontal axes of all panels show the maximum grain size ($a_{\mbox{\scriptsize max}}$) Top panel shows the absorption ($\kappa^{\mbox{\tiny abs}}$) and the effective scattering opacity which were derived using the  Henyey-Greenstein scattering approximation ($\kappa^{\mbox{\tiny sca, eff}}$). Middle panel shows the opacity spectral indices ($\beta$) derived from $\kappa^{\mbox{\tiny abs}}$. Bottom panel shows the effective albedo ($\kappa^{\mbox{\tiny sca, eff}} / (\kappa^{\mbox{\tiny abs}} + \kappa^{\mbox{\tiny sca, eff}} )$ ).
    }
    \label{fig:opacity}
\end{figure}

\subsubsection{Full Stokes imaging}\label{subsub:IQUimg}
We produced images using the CASA \texttt{tclean task}.
To confirm that using 3C48 and 3C147 as cross-hand delay and polarization position angle calibrators do yield consistent results, we first performed full Stokes (I, Q, U, and V) \texttt{mtmfs clean} \citep{Rau2011} imaging (adopting \textit{nterm}$=$2 and with Briggs Robust$=$2 weighting) for the C and A array configuration data separately, jointly with the two IFs.
We only used point-like {\tt clean} components due to that the convergence behavior of the multi-scale {\tt clean} is nonlinear and is sensitive to how the weighting terms are set.
The achieved synthesized beams and the root-mean-square (RMS) noise are summarized in Table \ref{tab:jvla}.
The observed peak Stokes I intensities at OMC-3/MMS~6, HOPS-91, and HOPS-409 in the C array configuration observations are 2.5, 0.59, and 1.3 \mjyperbeam, respectively.

For all of the covered target sources, we did not detect Stokes V at $>$3-$\sigma$ from both the C and the A array configuration observations.
We did not detect Stokes Q and U at $>$3-$\sigma$ from HOPS-91 and HOPS-409 and will omit further discussion on them given that their Stokes I properties have been very well documented by \citet{Tobin2020}.
We provide the Stokes I, Q, U images and the polarization intensity (PI) images of OMC-3/MMS~6 in Appendix Section \ref{sec:images}.

For OMC-3/MMS~6, we detected Stokes Q ($\sim$2-$\sigma$) and U ($\sim$7-$\sigma$) from the C array configuration observation. 
Given that the achieved S/N in OMC-3/MMS~6 is high, the non-detection of Stokes Q and U from HOPS-409 is sufficient to rule out the hypothesis that the detected polarized intensities were spurious.
We detected Stokes U ($\sim$4-$\sigma$) from the A array configuration observations after tappering the data with a 0\farcs2 arcsecond 2-dimensional Gaussian function (\beam$=$0\farcs27$\times$0\farcs22, P.A.$=$25$^{\circ}$; RMS noise$\sim$14 \ujyperbeam).
In spite of the very different signal-to-noise (S/N) ratios and angular resolutions, the observed Stokes U from these two array configurations appear qualitatively consistent: the detected Stokes U intensities are positive and the Stokes U peak positions are southeast of the Stokes I peak positions.
After confirming this, we proceeded to jointly imaged the C and A array configuration data to yield a better sensitivity.

The final C+A array configuration image achieved a \beam=0\farcs14$\times$0\farcs087 (54 au$\times$34 au, P.A.$=$6.6$^{\circ}$) synthesized beam and a RMS noise of 7.5 \ujyperbeam.
We produced the PI, position angle (PA), and percentage ($P$) maps based on the Stokes I, Q, and U images.
To avoid bias and spurious detections (c.f., \citealt{Vaillancourt2006}), we only utilized the pixels where the Stokes I intensities were above 5-$\sigma$ significance (approximately, within a 2-dimensional sphere of $\sim$0\farcs4 diameter, i.e., $\sim$70 synthesized Gaussian beam area in terms of standard deviation), and where absolute values of the Stokes Q or U intensities were above 3-$\sigma$ significance.
Assuming Gaussian random noise, the expected numbers of independent Stokes Q and U spurious detection are 0.2.
To suppress the positive bias of the polarized intensity PI, we evaluted PI as $\sqrt{Q^{2}+U^{2}-\sigma^{2}}$ where $\sigma$ is the RMS noise.
We smoothed the final A+C array configuration images to a 0\farcs15 circular synthesized beam before making quantitative comparison with the ALMA observations (more in Section \ref{sub:ALMA}).
We performed primary beam corrections to the Stokes I, Q, U, and PI images at the end of these processes.

\subsubsection{Stokes I imaging}\label{subsub:StokesI}
The procedure introduce in Section \ref{subsub:IQUimg} was to maximize sensitivity.
In this section, we introduce the optimized imaging strategy for Stokes I intensity for the purpose of robustly measuring spectral indices ($\alpha$) between the IF1 and IF2 of the JVLA data and between the IF2 of the JVLA data and the low-resolution ALMA data (Section \ref{sub:ALMA}).
All images introduced in this section were produced using the CASA \texttt{tclean task} with the \texttt{mtmfs} parameter \textit{nterm}$=$2.
We adopted a limited {\it uv} distance range of 16-3200 $k\lambda$, which is the same with the {\it uv} distance range of the low-resolution ALMA data.

\paragraph{Comparing IF1 and IF2 :}
We performed the \texttt{mtmfs clean} imaging for the IF1 and IF2 data separately.
The JVLA observations resolved spatial variations of the spectral indice.
To suppress the effect of synthesized beam smearing yet achieve sufficient S/N to constrain $\alpha$, we adopted Briggs Robust$=$0 weighting for both IFs.

We achieved a \beam$=$0\farcs11$\times$0\farcs067 (44 au$\times$26 au; P.A.=12$^{\circ}$) synthesized beam and a 13 \ujyperbeam RMS noise at IF1 (mean frequency $\nu_{\mbox{\tiny IF1}}$: 29.45 GHz); and achieved a \beam$=$0\farcs088$\times$0\farcs056 (34 au$\times$22 au; P.A.=13$^{\circ}$) synthesized beam and a 15 \ujyperbeam RMS noise at IF2 (mean frequency $\nu_{\mbox{\tiny IF2}}$: 36.65 GHz).

We smoothed the IF2 image to the synthesized beam of the IF1 image and then produced the spatially resolved spectral index\footnote{ $\alpha_{\mbox{\tiny IF1-IF2}}= (\log F_{\mbox{\tiny IF1}} - \log F_{\mbox{\tiny IF2}}) / (\log\nu_{\mbox{\tiny IF1}} - \log\nu_{\mbox{\tiny IF2}})   $} $\alpha_{\mbox{\tiny IF1-IF2}}$ distribution map.
Given that the IF1 and IF2 data were taken simultaneously, the derived $\alpha_{\mbox{\tiny IF1-IF2}}$ distribution was not systematically biased by the absolute flux calibration errors (i.e., the uncertainties were dominated by thermal noise).
We will introduce the results of $\alpha_{\mbox{\tiny IF1-IF2}}$ in Section \ref{sub:spidfeature}.

\begin{figure}
    \hspace{-2cm}
    \begin{tabular}{ p{9cm} }
       \includegraphics[width=10cm]{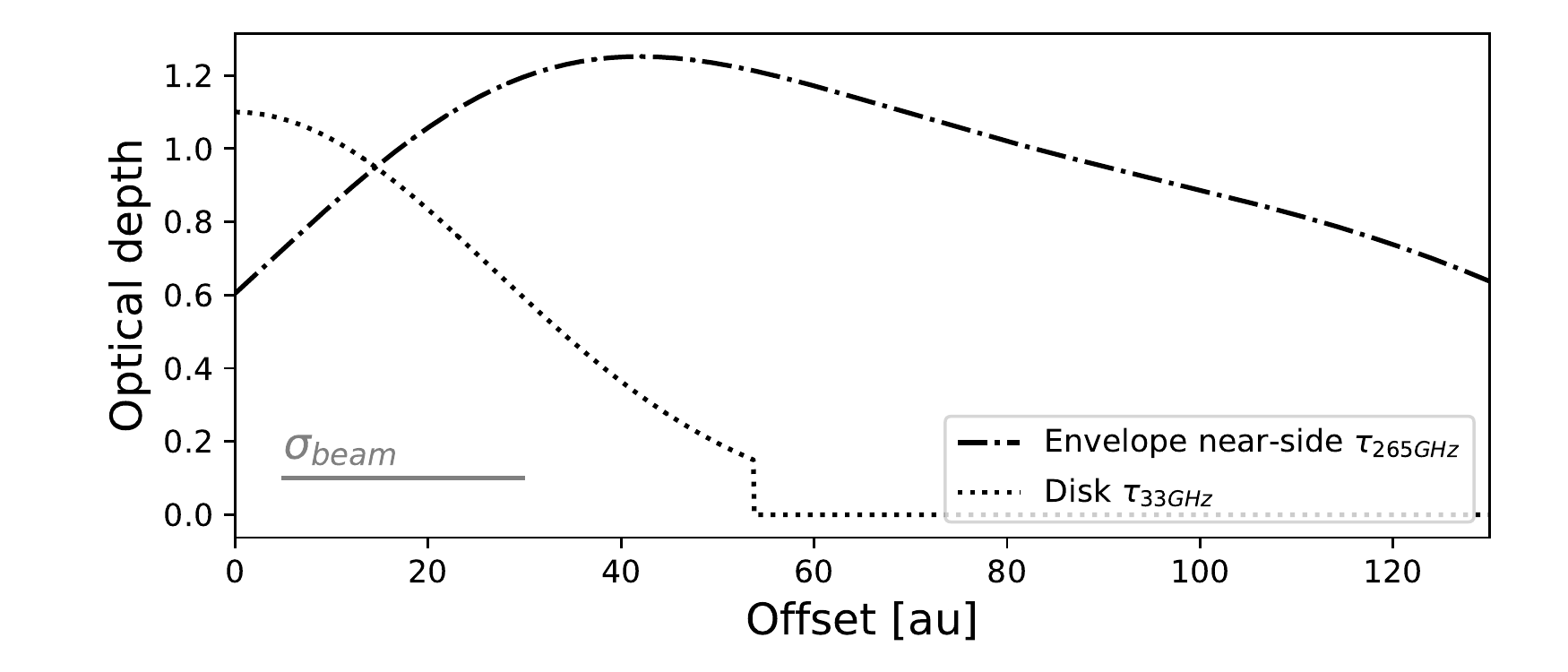} \\
       \vspace{-0.35cm}\includegraphics[width=10cm]{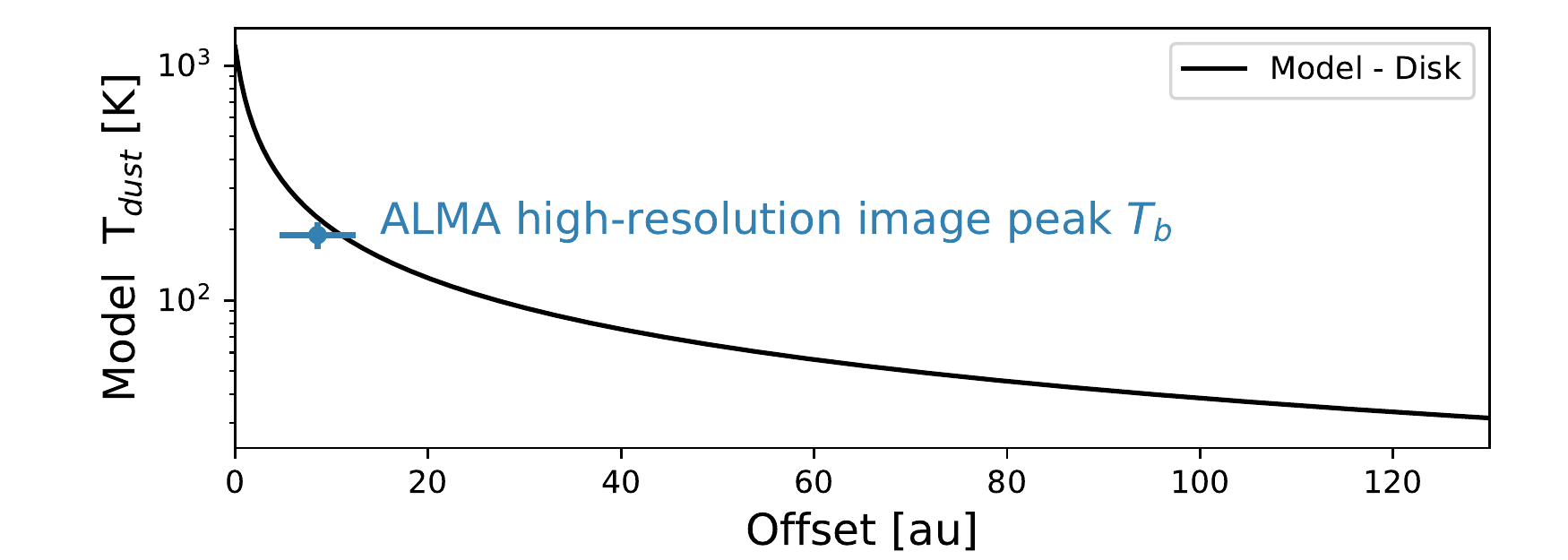} \\
       \vspace{-0.35cm}\includegraphics[width=10cm]{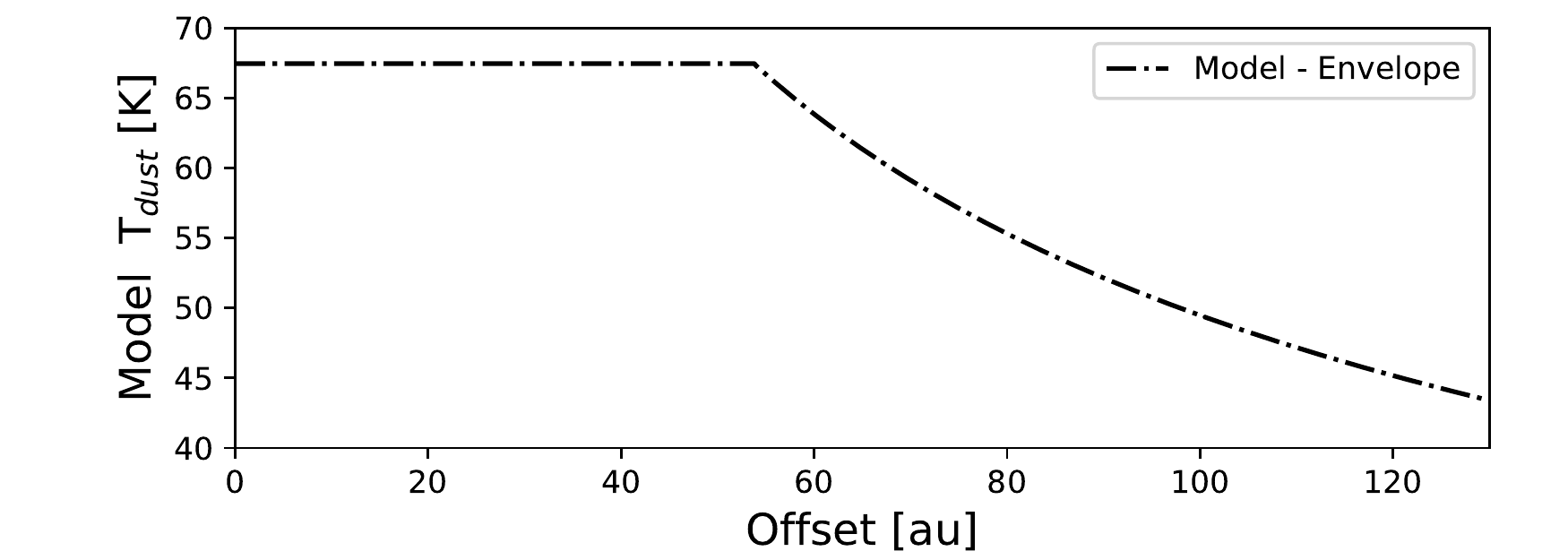} \\
    \end{tabular}
    \caption{Radial dust temperature and optical depth profiles of our toy model for radiative transfer. This model is consistent of a disk which is located background to the inner circumstellar envelope. Top panel shows the 265 GHz optical depth ($\tau_{\mbox{\scriptsize 225 GHz}}$) of the envelope and the 33 GHz optical depth ($\tau_{\mbox{\scriptsize 33 GHz}}$) of the disk. Middle and bottom panels show the dust temperatures in the disk and envelope, respectively. For simplicity, this toy model assumed a constant dust temperature over each line-of-sight. Horizontal axis is the projected separation from the location of the host protostar. }
    \label{fig:model}
\end{figure}

\paragraph{Comparing JVLA IF2 with ALMA:}

The JVLA observations are more limited by brightness temperature sensitivity than the ALMA observations.
To compare them, we imaged the IF2 data using Briggs Robust$=$2 weighting, which yielded a \beam$=$0\farcs13$\times$0\farcs082 (50 au$\times$32 au; P.A.$=$6.8$^{\circ}$) synthesized beam and a 14 \ujyperbeam RMS noise level.
The peak intensity and the flux density (estimated by summing over the region above the 3-$\sigma$ contour) are 1.7 \mjyperbeam (145 K) and 3.9$\pm$0.058 mJy, respectively.
The recovered flux density is reasonably well consistent with that of the Briggs Robust$=$0 weighted image (Section \ref{sub:spidfeature}).
We smoothed this JVLA IF2 image to a 0\farcs15 (58 au) circular synthesized beam and then produce the spatially resolved $\alpha_{\mbox{\tiny IF1-ALMA2}}$ distribution map by comparing with the low-resolution ALMA data introduced in Section \ref{sub:ALMA}.
Assuming the nominal $\sim$10\% absolute flux calibration errors for both the ALMA and the JVLA observations, the derived $\alpha_{\mbox{\tiny IF1-ALMA2}}$ can only be biased by up to $\sim\pm$0.1 thanks to the large frequency leverage arm.

\section{Results}\label{sec:result}
% outflow position angle -6 deg (blue in the north, red in the south, from \citep{Takahashi2012}
% Central position: 5:35:23.4210, -5:01:30.550
% E-field PA at central position: 34 deg (i.e., B-field: -56 deg)
In this section, we objectively report the resolved Stokes I, PI, and PA features in Section \ref{sub:intensityfeature}, and report the resolved spectral indices in Section \ref{sub:spidfeature}.
Our interpretation is provided in Section \ref{sec:discussion}.

\subsection{Stokes I and polarization properties}\label{sub:intensityfeature}
Figure \ref{fig:polmap} shows the JVLA (33 GHz) and the ALMA (265 GHz) low-resolution polarization images introduced in Sections \ref{sub:ALMA} and \ref{subsub:IQUimg}, which have been smoothed to 0\farcs15 angular resolution.
From this figure we see that outward of the $\sim$50 au projected radius around the 33 GHz Stokes I peak, the E-field PA observed at 33 GHz and 265 GHz are consistent with each other; inward of the $\sim$50 au projected radius they present an $\sim$90$^{\circ}$ relative offset.
The E-field PA at the location of the 33 GHz Stokes I peak is 34$^{\circ}$.
Such a discrepancy is also reproduced in the comparison between the JVLA image and the high-resolution ALMA image (Figure \ref{fig:ALMAhires}).
The E-field PA resolved by the high-resolution ALMA image presents bi-modality: the PA observed at inward and outward of $\sim$50 au projected radius show a general, $\sim$90$^{\circ}$ relative offset with respect to each other.

The 33 GHz PI distribution presents a single peak southeast of the Stokes I peak (Figure \ref{fig:polmap}). 
The 265 GHz PI distribution resolved by the low- and high-resolution ALMA images (Figures \ref{fig:polmap}, \ref{fig:ALMAhires}) presents three significant peaks: the central peak which locates close to the 33 GHz Stokes I peak is the strongest; the other two peaks are southeast and northwest of the central peak; their projected separations from the central peak are comparable.

It is intriguing that the 265 GHz Stokes I intensity distribution resolved by the high-resolution ALMA image appears lopsided with respected to the 33 GHz Stokes I peak (Figure \ref{fig:ALMAhires}).
The overall morphology of the 265 GHz Stokes I intensity distribution resolved by the high-resolution ALMA image may resemble what was resolved by the previous high angular resolution ALMA $\sim$350 GHz observations towards the Class~0 YSO, HH212 \citep{Lee2017}:
$\sim$50 au southeast of the 33 GHz Stokes I peak there is a 265 GHz dark lane which aligns in the northeast-southwest direction; this dark lane is sandwiched by brighter 265 GHz emission features.
% The 33 GHz PI peak is approximately co-located with the 265 GHz dark lane.

\subsection{Spectral indices}\label{sub:spidfeature}

The observed peak intensities in the JVLA IF1 and IF2 images are 0.70 \mjyperbeam and 1.2 \mjyperbeam, which correspond to the peak brightness temperatures ($T_{\mbox{\scriptsize b}}$) of 130 K and 220 K, respectively.
The flux densities ($F_{\nu}$; IF1: 1.6$\pm$0.049 mJy; IF2: 3.8$\pm$0.056 mJy) at both IFs were measured by summing over the region above the 3-$\sigma$ contour of the IF1 image.
Based on these measured flux densities, we obtained a mean spectral index $\alpha_{\mbox{\tiny IF1-IF2}}$ of 3.95$\pm$0.20.
We have imaged the C and A array configuration data separately and obtained consistent measurements of $\alpha_{\mbox{\tiny IF1-IF2}}$.
Given that the C and A array configuration observations were based on different absolute flux calibrators (Table \ref{tab:jvla}), this consistency means that the chance that the $\alpha_{\mbox{\tiny IF1-IF2}}$ value is seriously biased by passband calibration errors is small.

Figure \ref{fig:ALMAhiresSpid} shows the spectral index distributions measured between the ALMA and the JVLA observing frequencies ($\alpha_{\mbox{\tiny IF2-ALMA}}$), and the intra-band spectral indices ($\alpha_{\mbox{\tiny IF1-IF2}}$) measured by the JVLA observations (see Section \ref{subsub:StokesI}) for the inner $\sim$100 au region.
To give a qualitative sense about the resolved spectral indices, we note that assuming a constant temperature in a line-of-sight and Rayleigh-Jeans limit, when the dust opacity spectral index\footnote{Defined as $\beta=d\log\kappa^{\mbox{\scriptsize abs}}_{\nu}/d\log\nu$, where $\nu$ is frequency.} $\beta$ is close to the interstellar value 1.75, given an optical depth $\tau_{\nu}=$1.0 (i.e., marginally optically thick/thin), the expected spectral index $\alpha$ at the frequency $\nu$ is 3.0.
The value of $\alpha$ becomes smaller than 3.0 when $\tau_{\nu}>$1.0.
Under the same assumptions, the value of $\alpha$ is smaller when $\beta$ is smaller.

From outer to inner radii, the resolved $\alpha_{\mbox{\tiny IF2-ALMA}}$ decreases from $\sim$3.0 to $\sim$2.0 gradually.
The resolved $\alpha_{\mbox{\tiny IF1-IF2}}$ is consistent with 3.75$\pm$0.35, where the uncertainty at the presented area estimated assuming Gaussian noise and standard error propagation is $\sim$0.3.
Given that the area presented with $\alpha_{\mbox{\tiny IF1-IF2}}$ is very small, the observed gradient of $\alpha_{\mbox{\tiny IF1-IF2}}$ may not be explained merely by noise.
Likely, we have resolved a small spatial variation of  $\alpha_{\mbox{\tiny IF1-IF2}}$ although the absolute values of  $\alpha_{\mbox{\tiny IF1-IF2}}$ can be biased by noise.

\begin{figure}
    \hspace{-2cm}
    \begin{tabular}{ p{9cm} }
       \includegraphics[width=10cm]{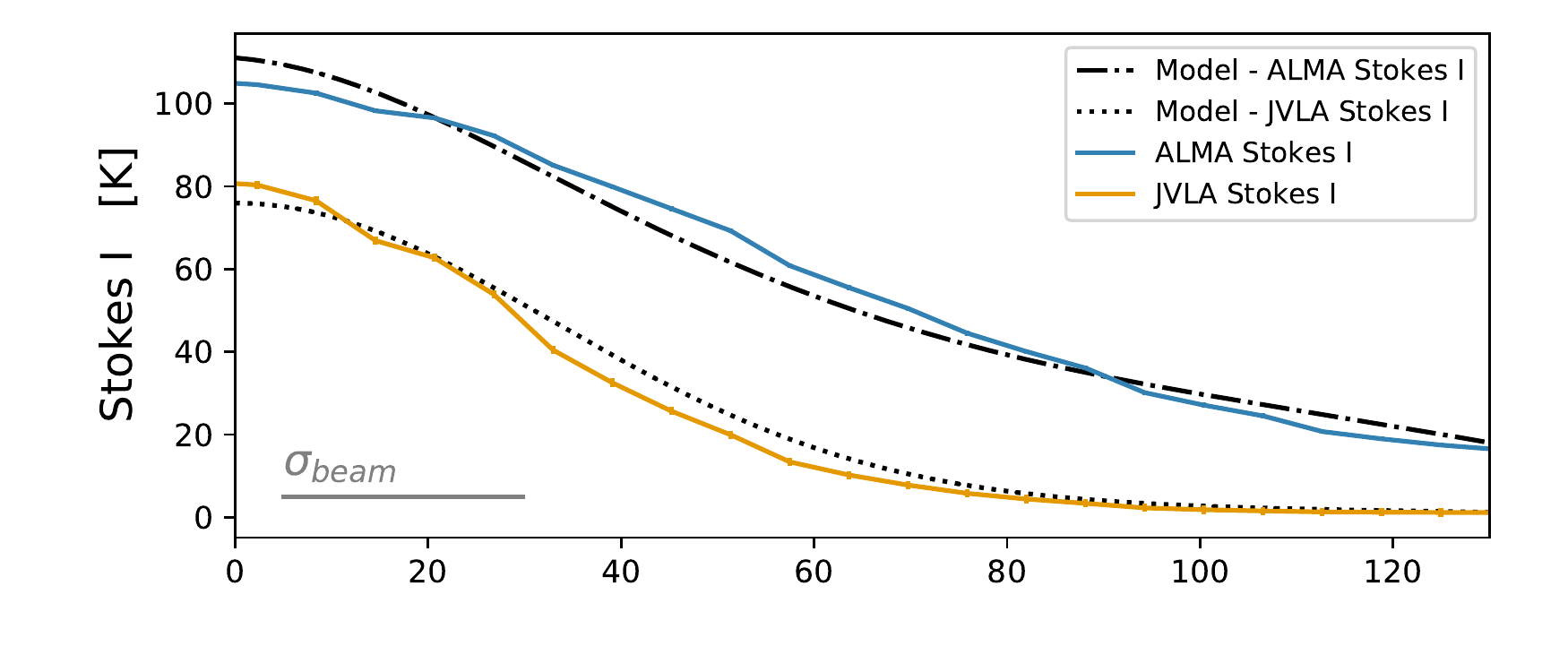} \\
       \vspace{-0.85cm}\includegraphics[width=10cm]{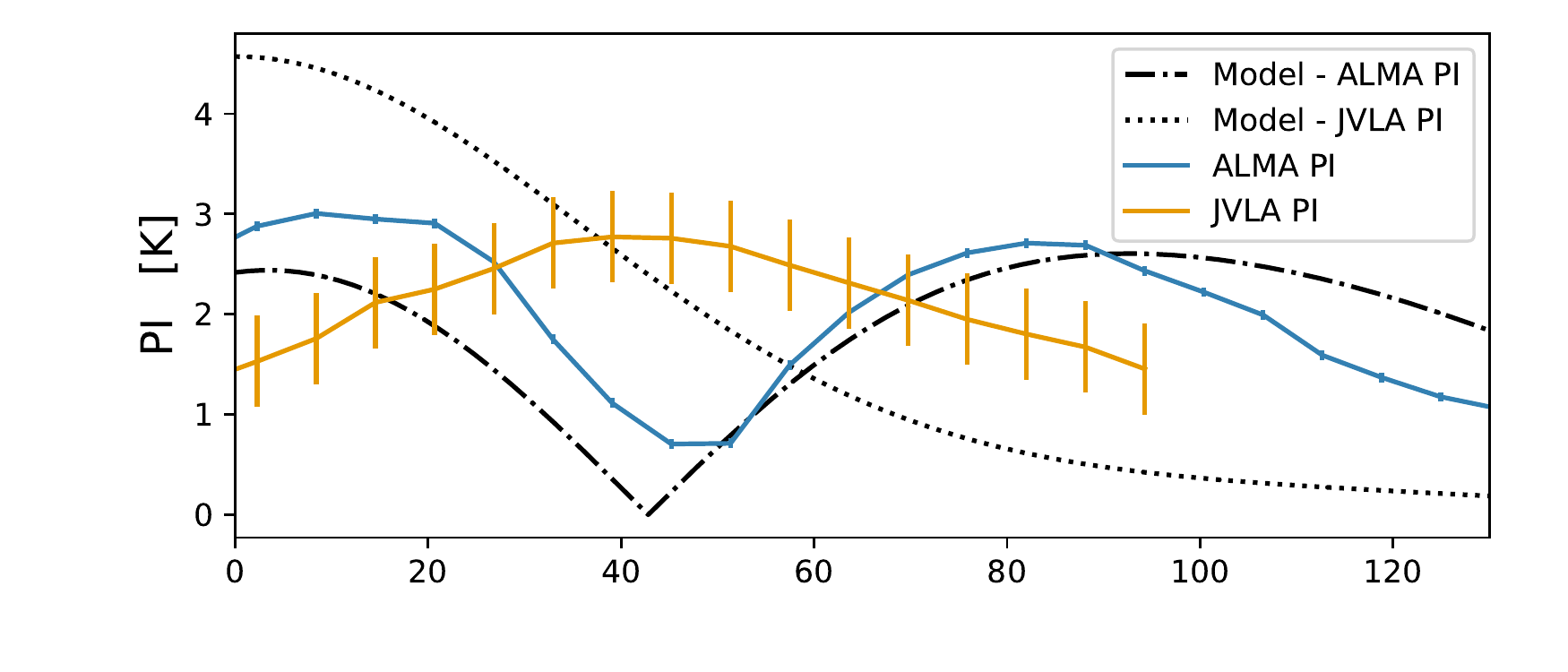} \\
       \vspace{-0.85cm}\includegraphics[width=10cm]{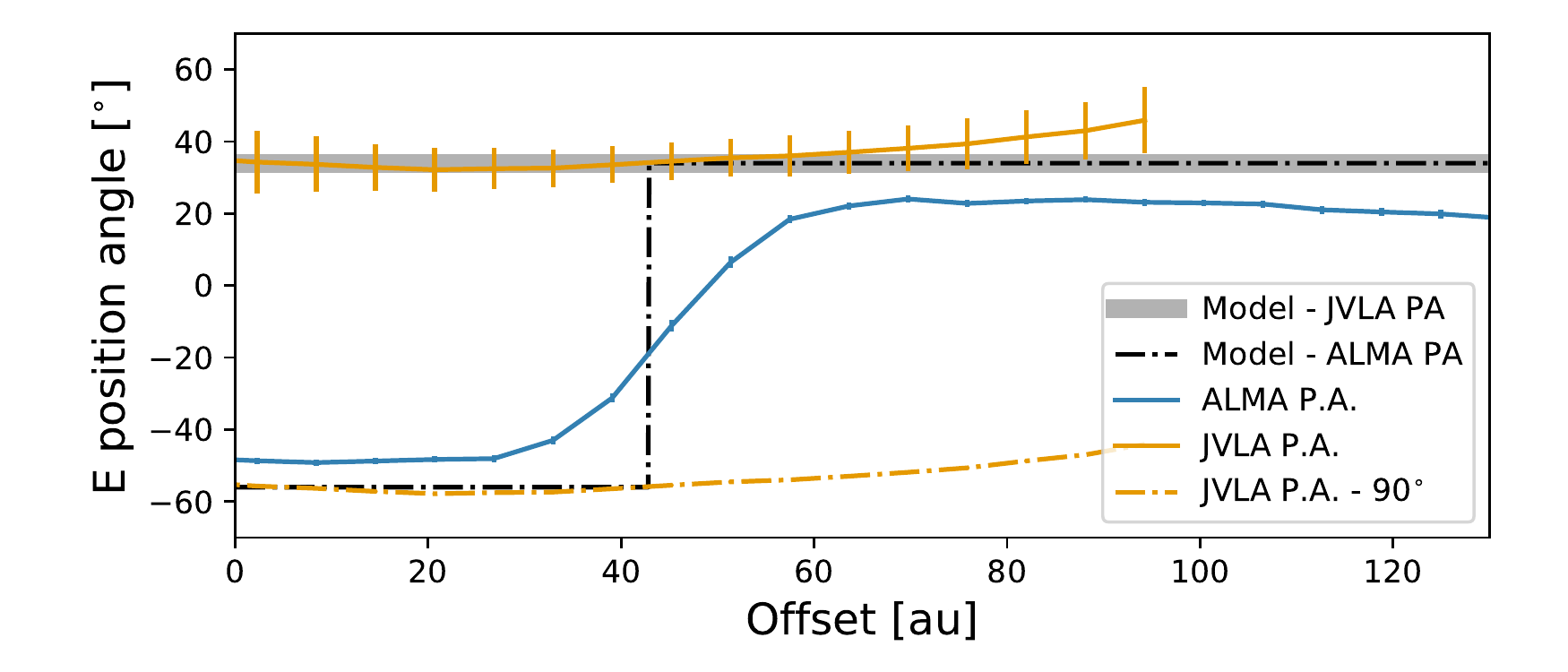} \\
    \end{tabular}
    \caption{The Stokes~I continuum brightness temperature ({\it top}), the linearly polarized intensities ({\it middle}), and the E-field polarization position angles measured along a slice of the ALMA (265 GHz) and JVLA (33 GHz) images at a position angle of 124$^{\circ}$ (see Figure \ref{fig:Bfield}). Horizontal axis is the projected separation from the location of the host protostar (R.A. (J2000) $=$ 05\rah35\ram23\ras.421, Decl. (J2000) $=$ Decl.=-05\decd01\decm30\farcs55). They are compared with the results of our radiative transfer toy model (see Figure \ref{fig:model}) which has taken the effect of synthesized beam smearing into consideration (Appendix Section \ref{sec:radiativetransfer}).}
    \label{fig:slice}
\end{figure}

\section{Discussion}\label{sec:discussion}
Observations tend to show that B-field are rather ordered instead of being random in star-forming regions.
This implies that the role of B-field pressure is at least comparable with turbulent energy.
In addition, it is also commonly considered that turbulent energy and gravitational potential energy are close to equipartition, which is the underlying assumption to permit estimating B-field strength based on the dispersion of B-field position angles (e.g., \citealt{Chandrasekhar1953,Houde2009}).
These observations have led to the so called standard model for low-mass star-formation, which describes the self-similar collapse of circumstellar core/envelope, regulated by well ordered and modestly strong B-field (for a review see \citealt{Shu1987}).
The expectations from this scenario are: (1) self-gravitational collapse is more rapid along the B-field lines, leading to the formation of a flattened (pseudo-)disk, (2) B-field on $\gtrsim$100 au scales may resembles an hourglass shape because of the drags of collapsing motion in the horizontal direction. 

Our interpretation and toy model for the OMC-3/MMS~6 system is tightly related to this scenario.
Specifically, we hypothesize that there is a dense (pseudo-)disk which is embedded in the parent gas envelope.
Both of them are pinched by the rather uniform B-field; the innermost part of the (pseudo-)disk may be similar with what was resolved from HH212 \citep{Lee2017} although OMC-3/MMS~6 appears to be not as edge-on.

Building a toy model to self-consistently interpret all the resolved features is involving.
We break down this problem into three parts.
In Section \ref{sub:dust}, we discuss the dust properties and dust optical depth in OMC-3/MMS~6, which will be related to our underlying assumptions about the polarization mechanisms.
We also provide our estimates of dust mass and compare with the estimates in the literature.
In section \ref{sub:radiativetransfer} we deal with the local radiative transfer problem and proposed the overall geometric configuration of gas structures.
In addition, we compare the radiative transfer results with the observational data.
This discussion intends to provide the rationale for our approaches of converting the resolved E-field PA distribution to B-field PA distribution.
We also demonstrate that our interpretation can be realized without assuming very extreme temperature, temperature contrasts, or other physical parameters (e.g., specific and unusual dust grain properties).
The global picture of the source is discussed in Section \ref{sub:Bfield}.
% We note that there are essential differences between our interpretation and what were given in  \citet{Takahashi2019}.

\subsection{Mass estimates and the maximum grain size}\label{sub:dust}

\subsubsection{Dust maximum grain size}\label{subsub:dustsize}

We base our discussion in this section on the DSHARP opacity table \citep{Birnstiel2018}, which is convenient for demonstrating the effects of changing the dust maximum grain size (\amax).
We quote the dust absorption and scattering opacities at the JVLA and ALMA observing frequencies in Figure \ref{fig:opacity}.

The intra-band spectral indices measured by JVLA ($\alpha_{\mbox{\tiny IF1-IF2}}$) is consistent with the interstellar value 3.75 (Figure \ref{fig:ALMAhiresSpid}).
OMC-3/MMS~6 is likely optically thin at $\sim$33 GHz except for an innermost, spatially unresolved region.
Since the 33 GHz observations towards YSOs is certainly well in the Rayleigh-Jeans limit, we can infer the dust opacity spectral index $\beta$ as $\alpha-$2.0.
From the comparison with Figure \ref{fig:opacity}, it is seen that the \amax in the 33 GHz emission source (e.g., the embedded (pseudo-)disk) can either be around $\sim$1 cm or smaller than $\sim$1 mm.

At the region where $\alpha_{\mbox{\tiny IF1-IF2}}$ is measured, the spectral indices measured between the ALMA and JVLA observing frequencies ($\alpha_{\mbox{\tiny IF2-ALMA}}$ ) is consistent with 2.0 (Figure \ref{fig:ALMAhiresSpid}).
If we assume that $\alpha$ is a monotonic function of frequency which is fair for our present case study, this implies that the value of $\alpha$ at 265 GHz ($\alpha_{\mbox{\tiny 265 GHz}}$) is well below 2.0 given that the value of $\alpha_{\mbox{\tiny IF2-ALMA}}$ is in between those of $\alpha_{\mbox{\tiny IF1-IF2}}$ and  $\alpha_{\mbox{\tiny 265 GHz}}$.
We refer to Figure 4 of \citet{Li2017} for the spectral energy distributions of another five Class 0/I YSOs which show such anomalously low $\alpha$ values, among which the source L1527 has recently been spatially resolved \citep{Nakatani2020}.
For the cases where the confusion by free-free or synchrotron emission is negligible, with the dust temperature of OMC-3/MMS~6 (more in Section \ref{sub:radiativetransfer} and \ref{sub:disk}), such an anomalously low $\alpha_{\mbox{\tiny 265 GHz}}$ can only be explained by (i) the effect of dust self-obscuration in the environments with high optical depths, high radial temperature gradient, and a high $\beta$ value at 265 GHz ($\sim$1.75; see \citealt{Li2017}, \citealt{Galvan2018}), or (ii) the anomalous effect of dust self-scattering in the optically thick environments with \amax$\sim$30-300 $\mu$m (\citealt{Liu2019}; \citealt{Zhu2019}).
For OMC-3/MMS~6, the confusion of free-free or synchrotron emission can be safely ruled out by the observed $\alpha_{\mbox{\tiny IF1-IF2}}$ values.
In general, these non-dust emission mechanisms are also not bright enough to significantly contribute to dust emission at 265 GHz (e.g. \citealt{Dzib2013}, \citealt{Liu2014}, \citealt{Coutens2019}, and references therein).
% We note that when interpreting with (i), using a smaller $\beta$ value may require a too steep to be physical radial temperature profile to reproduce the observed, anomalously low spectral indices.
Given the above discussion, it is safe to argue that \amax in the 33 GHz emission source is well below 1 mm.
%In this case, at the JVLA observing frequency ($\sim$33 GHz), the dust absorption opacity is higher than the dust scattering opacity and is not sensitive to \amax and ice coating (Figure \ref{fig:opacity}).
%Our assumptions of dust opacity for the dust mass estimates in Section \ref{subsub:dustmass} will be based on this point.

Based on the observations at only two frequency bands, we can not yet strictly rule out that in the 33 GHz emission source, \amax is as large as 100-1000 $\mu$m.
% This probability needs to be tested by the follow-up, high angular resolution observations at 40-48 GHz and 90-150 GHz bands, since we need at least three independent measurements to unambiguously constrain the three free parameters: dust temperature, dust column density, and \amax.
Nevertheless, we disfavor this probability since the large and maybe strongly frequency dependent dust albedo $\omega$ will make dust temperature a fine-tuning problem which may be energetically dis-favored.

In the following discussion, we will assume that the \amax in the 33 GHz emission source is well below 100 $\mu$m to avoid the fine-tuning problem with dust temperature and to avoid explicitly dealing with polarization due to dust self-scattering which required three-dimensional radiative transfer simulations (more in Section \ref{sub:Bfield}).
We assume that the values of \amax is smaller outside of the 33 GHz emission source, since grain growth is less efficient at lower density regions.
These assumption may be reasonable since over the spatial scales that can be resolved by the presented JVLA and ALMA observations, \amax may be only as large as $\sim$100 $\mu$m even in the relatively massive Class~II disks (e.g., \citealt{Kataoka2016a,Kataoka2016b,Stephens2017,Hull2018,Ohashi2018,Liu2019,Lin2020spid,Ohashi2019,Ueda2020,Ohashi2020}). 
It may be counter-intuitive if future observations do indicate that \amax in Class~0 objects can be in general considerably larger than 100 $\mu$m.

\subsubsection{Dust mass}\label{subsub:dustmass}

We based on the IF1 (29.45 GHz) observations of the JVLA, which has the lowest optical depth, to estimate the overall dust mass in the embedded (pseudo-)disk.
It is well in Rayleigh-Jeans limit and therefore the mass estimate is not sensitive to small errors in the assumed dust temperature.
The flux density $F_{\mbox{\scriptsize 29.45 GHz}}$ can be approximated by
\begin{equation}\label{eq:flux}
F_{\mbox{\scriptsize 29.45 GHz}} = d\Omega B_{\mbox{\scriptsize 29.45 GHz}}(\bar{T}_{\mbox{\scriptsize dust}}) \left(1 - e^{-\bar{\tau}_{\mbox{\tiny 29.45 GHz}}}\right),
\end{equation}
where $d\Omega$ is the solid angle of the disk, $B_{\mbox{\scriptsize 29.45 GHz}}(\bar{T}_{\mbox{\scriptsize dust}})$ is the Planck function at the mean dust temperature $\bar{T}_{\mbox{\scriptsize dust}}$, and $\bar{\tau}_{\mbox{\tiny 29.45 GHz}}$ is the mean dust optical depth at 29.45 GHz.
$\bar{\tau}_{\mbox{\tiny 29.45 GHz}}$ can be related to the dust mass surface density $\Sigma_{\mbox{\scriptsize dust}}$ by 
\[  
\bar{\tau}_{\mbox{\tiny 29.45 GHz}}=\kappa^{\mbox{\tiny abs}}_{\mbox{\tiny 29.45 GHz}}\Sigma_{\mbox{\scriptsize dust}}.
\]
Assuming \amax$<$100 $\mu$m, we self-consistently neglected the dust scattering opacity.
We adopted $\kappa^{\mbox{\tiny abs}}_{\mbox{\tiny 29.45 GHz}}=$0.0096 cm$^{2}$g$^{-1}$ which was quoted from the DSHARP project \citet{Birnstiel2018}.

In most of the applications of Equation \ref{eq:flux}, one has to assume $\bar{\tau}_{\nu}\ll$1 and has to adopt an arbitrarily assumed value for $\bar{T}_{\mbox{\scriptsize dust}}$.
In our present study, we approximated them by evaluating the intensity-weighted mean\footnote{In this estimate, we evaluated the intensity-weighted mean in a two dimensional model image, assuming that the target source is axi-symmetric.} values in the toy model introduced in Appendix Section \ref{sub:disk} ($\bar{T}_{\mbox{\scriptsize dust}}=145$ K, $\bar{\tau}_{\mbox{\tiny 29.45 GHz}}=0.6$; see Figure \ref{fig:model}).
Substituting them into Equation \ref{eq:flux}, we can then derive $d\Omega$ based on the observed $F_{\mbox{\scriptsize 29.45 GHz}}$.
% Since the disk is not optically thick at 29.45 GHz, the derivation of $d\Omega$ is not sensitive to a factor of unit errors in the adopted $\bar{\tau}_{\mbox{\tiny 29.45 GHz}}$ value.
Since the disk is not optically thick at 29.45 GHz, a $\sim$1 error in the adopted $\bar{\tau}_{\mbox{\tiny 29.45 GHz}}$ will not significant bias the derivation of $d\Omega$.
Assuming that the source is approximately axi-symmetric, our derived $d\Omega$ corresponds to a 43 au projected radius, which is identical to the measurement of radius reported in \citet{Tobin2020}.
This good comparison services as a consistency check for our toy disk model (Appendix Section \ref{sub:disk};  Figure \ref{fig:model}).
Finally, the overall dust mass $M_{\mbox{\scriptsize dust}}$ can be estimated by
\begin{equation}
   M_{\mbox{\scriptsize dust}} = d\Omega \Sigma_{\mbox{\scriptsize dust}} D^{2} = d\Omega \frac{\bar{\tau}_{\mbox{\tiny 29.45 GHz}}}{\kappa^{\mbox{\tiny abs}}_{\mbox{\tiny 29.45 GHz}}} D^{2},
\end{equation}
where $D$ is the distance to the target source.

Our derived $M_{\mbox{\scriptsize dust}}$ is $\sim$14000 $M_{\oplus}$, which is one order of magnitude higher than the mass estimate reported by \citet{Tobin2020} ($\sim$1500 $M_{\oplus}$).
We note that the mass estimates of \citet{Tobin2020} were based on the same JVLA data as what we are using.
Such a large discrepancy in our mass estimates is mainly due to that \citet{Tobin2020} adopted a much higher dust absorption opacity (0.13 cm$^{2}$\,g$^{-1}$) than us.
In the following discussion, we argue that \citet{Tobin2020} likely has underestimated dust masses by a factor of $\sim$10.
%It is noticed that this opacity assumption was specific in the sense that it is close to the highest possible value in the DSHARP opacity table.

\citet{Tobin2020} obtained the 9 mm dust absorption opacity by assuming the dust opacity spectral index $\beta=$1 and then scaling the dust absorption opacity from $\sim$1 mm wavelength ($\sim$1 cm$^{2}$\,g$^{-1}$) to 9 mm, ignoring the dust scattering opacity.
However, their assumption of $\beta=$1 can only be achieved when \amax is a lot greater than 1 mm (e.g., Figure \ref{fig:opacity}).
This statement is true also for the other commonly adopted dust opacity tables for protoplanetary disks (e.g., \citealt{Kataoka2015,Woitke2016}), unless one assumes a very significant inclusion of amorphous carbon (e.g., \citealt{Woitke2016, Yang2020}).
However, significant inclusion of amorphous carbon will make the values of $\alpha$ close to 3 also at 29-36 GHz frequency.
For OMC-3/MMS~6, this is only plausible on an unresolved spatial scale given the averaged spectral index at the JVLA band ($\alpha_{\mbox{\tiny IF1-IF2}}$) has been constrained to be 3.95$\pm$0.20 (Section \ref{sub:spidfeature}).

When \amax$>$1 mm, the dust albedo $\omega$ is close to 1.0 at the observing frequencies of the JVLA and ALMA data (e.g., Figure \ref{fig:opacity}).
In this case, if $\bar{\tau}_{\mbox{\tiny 29.45 GHz}}\lesssim1$, then $F_{\mbox{\scriptsize 29.45 GHz}}$ will be attenuated to become one order of magnitude fainter as compared with the case with $\omega\sim$0 (see Figure 9 of \citealt{Birnstiel2018}; c.f. \citealt{Liu2019,Zhu2019}) due to a very high dust total extinction.
In other words, even in the case that the assumptions of dust $\kappa^{\mbox{\scriptsize abs}}$ and $\beta$ in \citet{Tobin2020} are correct, they still needs to increase the overall dust masses by at least one order of magnitude to compensate the attenuation due to dust scattering.
Otherwise, their predicted flux densities will be considerably lower than what have been measured by JVLA.
The attenuation due to dust scattering is slightly less significant when $\bar{\tau}_{\mbox{\tiny 29.45 GHz}}\gtrsim10$. 
But for cases with such high optical depths, the masses estimates of \citet{Tobin2020} based on the optically thin assumption still needs to be corrected by a factor of $\sim$10.
Finally, the effect of dust scattering can be neglected if the optical depth evaluated based on total extinction (i.e., $\Sigma_{\mbox{\scriptsize dust}}\cdot(\kappa^{\mbox{\tiny abs}} + \kappa^{\mbox{\tiny sca, eff}})$) is well below 0.1 (\citealt{Birnstiel2018}), where $\kappa^{\mbox{\tiny sca, eff}}$ is the scattering opacity derived using the Henyey-Greenstein scattering approximation.
With the assumption of \amax$>$1 mm, this implies that  $\Sigma_{\mbox{\scriptsize dust}}\cdot\kappa^{\mbox{\tiny abs}}$ has to be 1-2 orders of magnitude lower than 0.1 (Figure \ref{fig:opacity}).
This is unlikely the case for the sources which have been constrained to have comparable angular scales with OMC-3/MMS~6 or are smaller.
Otherwise, the detection of these sources will imply too high dust temperatures, which can easily exceed the dust sublimation temperature ($\sim$1500 K).
The scaling of dust temperature versus $\Sigma_{\mbox{\scriptsize dust}}\cdot\kappa^{\mbox{\tiny abs}}$ can be roughly seen by comparing the top and middle panels of Figure \ref{fig:model}.
Since luminosity is expected to have a power equal to 4 dependence on temperature, the required high temperature also may not be realistically achieved around low-mass (e.g., $M_{*}<$ 8 $M_{\odot}$) Class~0 YSOs.

For our target source OMC-3/MMS~6, when the albedo is close to unity, we would also expect to detect linearly polarized dust scattered light.
The E-field position angles of the linearly polarized dust scattered light is expected to align in the northwest-southeast direction, which are not consistent with what was resolved by the JVLA observations (Figure \ref{fig:ALMAhires}).
We also note that presently there is no validated report of detecting linearly polarization dust scatter light at $>$7 mm wavelengths.
One possibility of reducing $\beta$ (and $\alpha$) but without enhancing the albedo is to consider dust grains of very high porosity, which is also disfavored by observations for this moment \citep{Tazaki2019}.
\citet{Li2017} analyzed the $\sim$40-350 GHz spectral energy distributions of nine Class 0/I YSOs and concluded that the directly measured $\beta$ values of four of them are consistent with the interstellar value; the $\beta$ values of the other five sources may be required to be consistent with the interstellar value, otherwise it is hard to reproduce the observed, anomalously low $\alpha$ at 230-345 GHz.
The assumption of $\beta=$1 may not be valid for a significant samples of Class 0/I YSOs.
As mentioned in Section \ref{sub:dust}, this assumption also appears counter-intuitive when compared to the observations towards spatially resolved Class II disks.

It is not unlikely that the rest of the samples presented in \citet{Tobin2020} and the earlier 9 mm surveys (e.g., \citealt{Tychoniec2018, Tychoniec2020}) are all subjective to such a systematic dust-mass underestimation by one order of magnitude given that they were based on similar assumptions of dust opacity which did not take dust scattering into consideration.
% The statistics of dust masses in Class 0/I objects derived merely based on observations at shorter wavelengths (e.g., \citealt{Sheehan2017,Williams2019}) are subject to the more serious concern.

Our assumption of dust opacity and that of \citet{Tobin2020} can be tested in the near future by improving the constraints on the spectral indice at 0.7-10 cm bands (e.g., with the JVLA Q, K, Ku, X, and C bands observations\footnote{Multi-wavelength observations at long wavelength bands are required to robustly constrain the confusion due to free-free emission}).
Nevertheless, this would be more of a matter about the fundamental dust properties.
It is expect to have little impact on our present discussion about dust mass, polarization mechanisms (Section \ref{sub:radiativetransfer}) and B-field configuration (Section \ref{sub:Bfield}).

Assuming a gas-to-dust mass ratio $f\sim$100, the corresponding disk gas mass according to our estimate is $\sim$4 $M_{\odot}$.
We note that on 500-1000 au scales, OMC-3/MMS~6 presents multiple clumps and spiral-arm like structures (\citealt{Takahashi2012}), which was hypothesized to be the signatures of disk self-gravitational instabilities.
This can be expected if the disk mass is indeed as high as our estimate, although the assumption of $f$ is not yet constrained by any observation.

We remark that for very optically thick and massive disks, such as OMC-3/MMS~6, the analysis of gas kinematics on $\lesssim$100 au scale can no more base on the approximation of a point-like gravitational source.
The previously resolved changes of velocity gradients in some Class~0/I YSOS at certain radii (e.g., \citealt{Ohashi2014}) may not be uniquely explained by a transition from rotating, infalling envelopes to Keplerian-rotating disks.
However, it may be partly attributed to that within those radii the enclosed gas mass is not negligible as compared with the protostellar mass (see also \citealt{Nakatani2020}).
Therefore, the analyses and interpretation of the rotation curves may need to be carried out in the fashion which is similar to what has been performed by the community of studying OB cluster-forming regions (e.g., \citealt{Liu2017}).
When dust is optically thick, the interpretation of the spectral line observations also needs to be careful that the spectra taken at a certain projected radius $r$ trace the gas kinematics and chemical composition at the radius $\rho>r$.
The analyses may be misleading if how to convert $r$ to $\rho$ is not correctly addressed.

%The analyses of gas kinematics and chemistry at optically thick bands are also required to address how the projected radius is inverted to radius otherwise can be misleading.

% kappaJVLA at 29.45 GHz= 0.0096
% diameter: 86.42657232017851
% dust mass: 13751 (round to 14000 Earth mass)
% Tobin mass: 1510 Earth mass   Tobin opacity: 0.13 cm^2 /g (not self-consistent due to the assumption of beta=1, need an order of magnitude correction for the scattering attenuation)

\begin{figure}
    \centering
    \includegraphics[width=8.5cm]{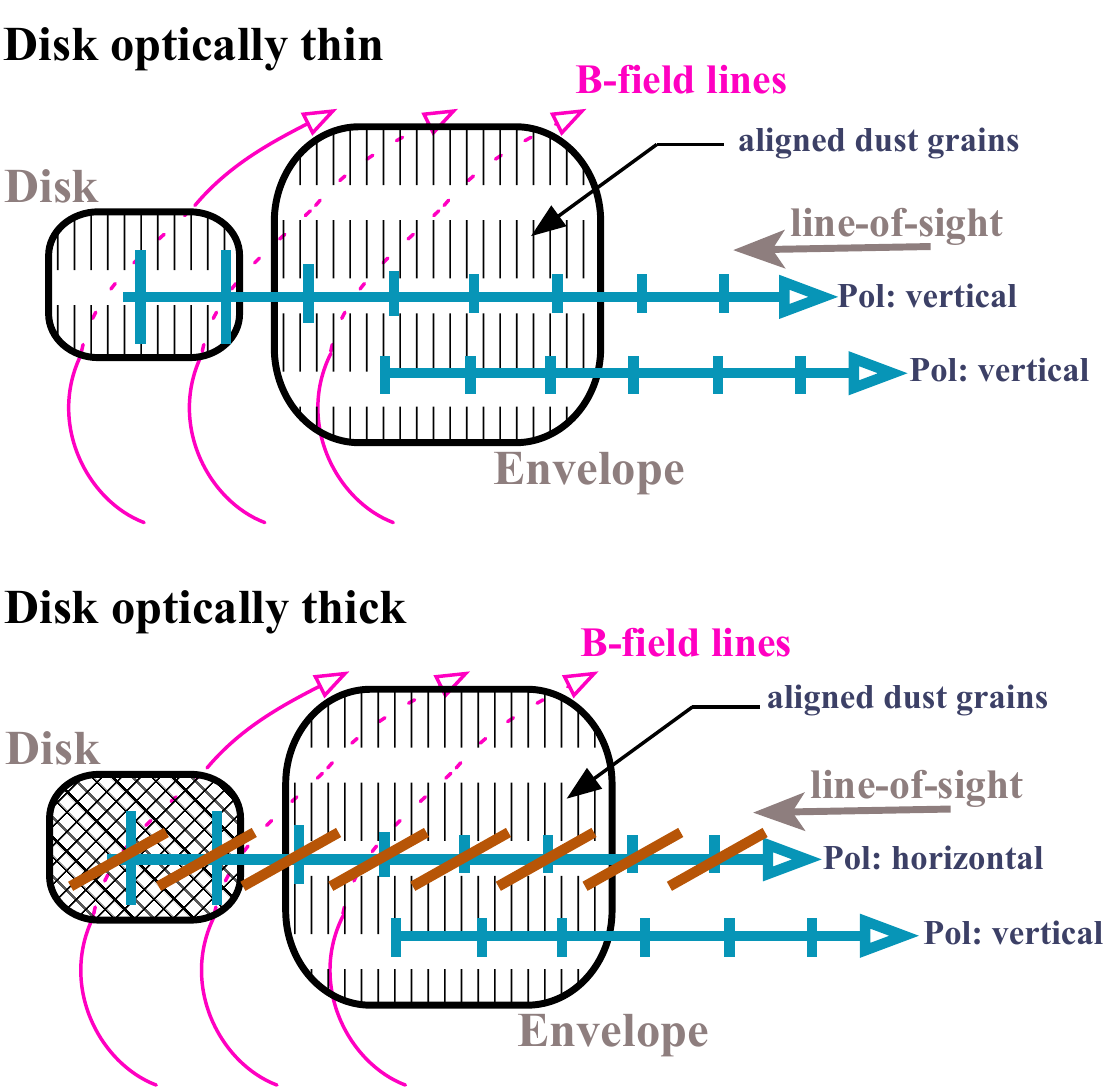}
    \caption{Schematic diagrams to explain the radiative transfer with aligned dust. The emission from the disk is attenuated by the envelope which is located in between the disk and the observer. Blue lines represent the radiation from the disk and from the envelope, which are overplotted with short segments to indicate the linear polarization states. When the disk is optically thick ({\it bottom panel}), the radiation originated from the disk is saturated to the essentially unpolarized black body emission.}
    \label{fig:schematic}
\end{figure}

\subsection{Polarized radiative transfer}\label{sub:radiativetransfer}

Our simplified, one dimensional (1D) radiative transfer model to explain the observed features was constructed based on the following assumptions:
\begin{enumerate}
    \item The density profiles of the system is described by a (pseudo-)disk component and an envelope component. Throughout this manuscript we refer to the former simply as disk for clarity. The disk is inclined, where the near-side is in the southeast.
    \item For the disk and the envelope, the temperature variation over a line-of-sight (los) is negligible.
    \item The B-field PA is aligned with the rotational axis of the disk. The projected long axes of dust grains are aligned perpendicular to B-field. The projected B-field position angle (34$^{\circ}$ by assumption) has no variation over each line-of-sight. To be consistent with the resolved polarization percentages by ALMA and JVLA, the dust polarization efficiency ($\epsilon$) are chosen to be 10\% and 15\% in the disk and in the envelope, respectively. We note that the values of $\epsilon$ are not difficult to pick if we are not required to consider too many significant digits, since the observed polarization percentages from the rather optically thin regions would be close to the value of $\epsilon$ (see \citealt{Liu2018}).
    \item We assumed \amax$<$100 $\mu$m and neglected $\kappa^{\mbox{\tiny sca, eff}}$. The absorption opacities $\kappa^{\mbox{\tiny abs}}_{\mbox{\scriptsize 33 GHz}}=$0.012 cm$^{2}$\,g$^{-1}$ and $\kappa^{\mbox{\tiny abs}}_{\mbox{\scriptsize 265 GHz}}=$0.49 cm$^{2}$\,g$^{-1}$ were quoted from the default DSHARP opacity table \citep{Birnstiel2018}.  Based on the opacity tables presented in Figure \ref{fig:opacity}, we neglect the differences made by ice coating.
\end{enumerate}
How we evaluated the radiative transfer is outlined in Appendix Section \ref{sec:radiativetransfer}.
Our disk and envelope models by construction are shown in Figure \ref{fig:model}.
The explicit functional forms of them and the rationales of how those functional forms were chosen are provided in Appendix Section \ref{sec:toy}.
These functional forms are not crucially important for our present science purposes.
We note that in Figure \ref{fig:model} we present the dust opacity depth of disk at 33 GHz and the envelope optical depth at 265 GHz instead of their dust column densities, since these optical depths are the more directly measurable quantities.
With this model, the envelope by itself will reproduce $\alpha_{\mbox{\tiny IF2-ALMA}}\sim$3.
Closer to the projected area of the disk, the effect of synthesized beam smearing will make the observed $\alpha_{\mbox{\tiny IF2-ALMA}}$ gradually decrease from 3 to 2.

We compare the Stokes I intensity, PI, and E-field PA predicted from this model with the measurements along a slice of the ALMA and JVLA images at a position  angle of 124$^{\circ}$ (Figure \ref{fig:slice}).
The zero-point (R.A.  (J2000)$=$05\rah35\ram23s\ras421,  Decl.   (J2000)$=$Decl.=-05$^{\circ}$01$'$30\farcs55) of the slice, which we regard as the stellar position, was chosen to be the Stokes I intensity peak of the JVLA IF2 image which was made with Briggs Robust$=0$ weighting.
The spatial coordinates are expressed as the relative offset from this zero-point.

We focus on the near-side where the JVLA PI image achieved better S/N.
We emphasize that this comparison is qualitative, mainly for demonstrating the radiative transfer effects.
We do not try to fine-tune the free-parameters and perform fittings due to the degeneracy as a consequence of the large degree of freedom and yet very limited observational constraints.
The quantitative interpretation of this comparison would need to be cautioned with the following major uncertainties: (i) the spatial resolution of the JVLA observations is limited such that the stellar position can not be determined precisely; for Class 0/I objects, it is in fact in general not even easy to know whether they are isolated stars or multiple systems, (ii) the disk position angle and inclination cannot be directly determined from observations at this moment, (iii) on small spatial scales the B-field PA likely do vary along the los, which cannot be taken into consideration with our present modeling strategy, (iv) the actual column density structures are likely a lot more complicated than what can be described with simplified functional forms (e.g., for the envelope this has been seen from the 265 GHz Stokes I intensity distribution; Figure \ref{fig:polmap}), (v) the dust polarization efficiency may also have stronger spatial dependence.
Nevertheless, qualitatively the radiative transfer results in Figure \ref{fig:slice} appears elucidating to us.

\begin{figure}
    \centering
    \includegraphics[width=8.5cm]{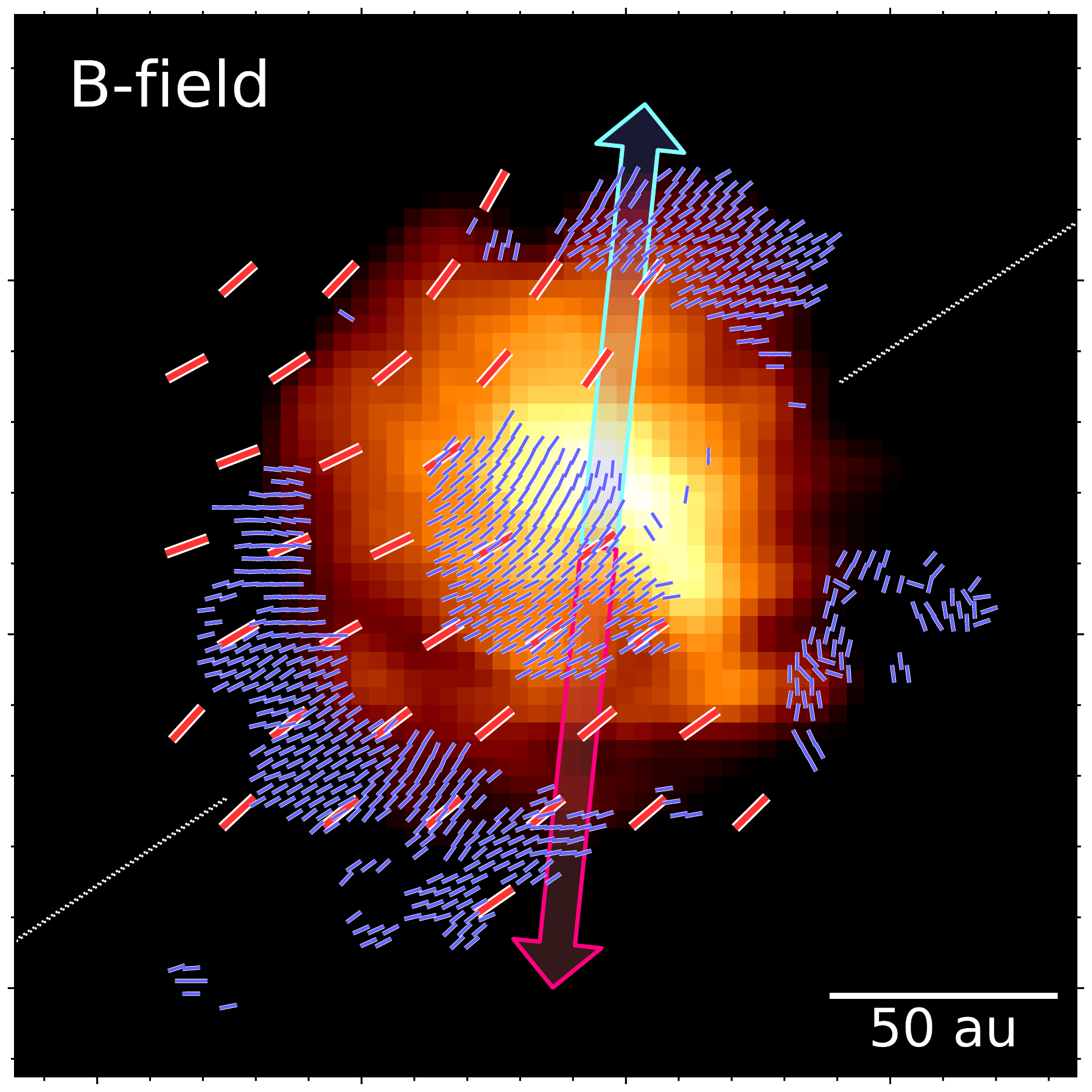}
    \caption{The suggested B-field configuration in the inner $\sim$100 au (radius) region around OMC-3/MMS~6.
    Color image is the same with the top left panel of Figure \ref{fig:ALMAhires}. Red line segments are those presented in the top left panel of Figure \ref{fig:ALMAhires} after a 90$^{\circ }$ rotation; blue line segments are those presented in that panel, but were 90$^{\circ}$ rotated outside of the 35-$\sigma$ contour of the JVLA image (i.e., as presented by the green contours) and were masked for the region between the 35-$\sigma$ and 100-$\sigma$ contours of the JVLA image. 
    Arrows show the position angles ($-6^{\circ}$) of the blue and redshifted CO outflows which were quoted from \citet{Takahashi2012}. White dotted line shows the proposed disk rotational axis at $\sim$50 au scale (P.A.$=$124$^{\circ}$) which passes through the proposed location of the host protostar.
    }
    \label{fig:Bfield}
\end{figure}

From Figure \ref{fig:slice}, it is seen that the 33 GHz and 265 GHz Stokes I intensities decrease monotonically with the projected offset from the central zero-point position.
While the 33 GHz PI was resolved with a small range ($\sim$1.5-3 K) as compared with its uncertainties, it might have a peak at the $\sim$50 au projected offset.
The resolved 33 GHz PA is consistent with a constant 34$^{\circ}$ angle.
Outward and inward of the $\sim$50 au projected offset, the 265 GHz PA is close to 34$^{\circ}$ and -56$^{\circ}$ (i.e., 34$^{\circ}$-90$^{\circ}$), respectively.
Finally, the 265 GHz PI has a minimum around the $\sim$50 au projected offset.

By comparing with the model presented in Figure \ref{fig:model}, the resolved features in Figure \ref{fig:slice} may be understood by the following radiative transfer phenomenon (see the schematic picture in Figure \ref{fig:schematic}).
When the embedded disk is optically thin (e.g., $\tau<$1) either because of low dust column density or because it was observed at long wavelengths, one expect to resolve the E-field PA which is perpendicular to the B-field PA.
However, when the embedded (pseudo-)disk is very optically thick (e.g., $\tau\gg$1), then the dust thermal emission originated from it is essentially unpolarized but then becomes linearly polarized after propagating through the foreground gas envelope, due to dichroic extinction.
When the disk is large or hot (\citealt{Lin2020}) such that the flux density of the (pseudo-)disk outweighs that of the envelope in a synthesized beam, then one expect to see the E-field PA which is parallel to the B-field PA.
This can explain why inward and outward of the disk radius the PA resolved by ALMA at 265 GHz presents a $\sim$90$^{\circ}$ relative offset, and also why the PA resolved by JVLA at 33 GHz is close to the PA resolved by ALMA outward of the disk radius.
We point out that, it is not unlikely that many previous dust polarization observations at $\sim$0.85-1.3 mm wavelengths towards star-forming regions (e.g., \citealt{Hull2014,Zhang2014,Cox2018,Galametz2018,Sadavoy2019}) were not interpreted correctly due to overlooking this effect.

Finally, we see that inward of the $\sim$30 au projected offset, the 33 GHz PI derived from our model shows a large discrepancy from what was detected (Figure \ref{fig:slice}). 
Given that our synthesized beam has a $\sim$58 au FWHM, when the spatial scales of interest is comparable with the synthesized beam size, it may be natural to expect that the B-field PA has variations within a synthesized beam, leading to canceling of PI and an effectively low $\epsilon$.
The high-resolution ALMA image indeed shows that the B-field PA has some variation in the innermost region of OMC-3/MMS~6 (Figure \ref{fig:ALMAhires}), although ALMA observations may not probe the B-field close to disk due to the high optical depth.
The discrepancy between the observed and the modeled 33 GHz PI could also be partly because that dust is not aligning well with B-field on small scales.

Moreover, the disk is inclined, and the disk temperature assumed in our model (Figure \ref{fig:model}) changes by one order of magnitude inward of the $\sim$30 au projected separation.
Therefore, it is likely that our simplifying assumption of negligible temperature variation over a line-of-sight breaks down in this region.
For example, the disk may harbor a very optically thick hot inner disk, or other sorts of sub-structures with high brightness (rings, spiral-arms, clumps, etc).
In such cases, at the projected area of the substructure(s), the temperature variation along the line-of-sight may still lead to a 90$^{\circ}$ PA flipping at 33 GHz, which in turn yields the canceling of 33 GHz PI.
Since presently the disk properties are constrained by observations at only one frequency band, and given the aforementioned uncertainties, it is not meaningful to fine-tune the model or add more components to the model within the inner 30 au radius (i.e., 60 au diameter) to fit the observations.
The constraints on the thermal profile of the disk can be dramatically improved by the follow-up observations at 40-48 GHz (e.g., JVLA Q band) and 90-150 GHz (e.g., ALMA Bands 3 and 4).

\subsection{B-field configuration and outflow alignment}\label{sub:Bfield}

Based on the radiative transfer model presented in Section \ref{sub:radiativetransfer}, we suggest that the 90$^{\circ}$ rotated 33 GHz PA trace the projected B-field orientation throughout the observed area.
The projected B-field PA is traced by the 265 GHz PA around the projected area of the disk, in particular, at the self-obscured dark lane (Figure \ref{fig:ALMAhires}); the projected B-field PA is traced by the 90$^{\circ}$ rotated 265 GHz PA at the spatially more extended, low brightness regions.
At the projected far-side of the disk, the 265 GHz PI is low simply due to the very high optical depth; at the projected area of the 265 GHz dark lane, the 265 GHz PI is low due to a transition from the regime of linearly polarized dust emission to the regime of linearly polarized dichroic extinction, and synthesized beam smearing.
We do not require a complicated B-field topology (e.g., due to strong turbulence) to explain the low PI "holes" resolved in the inner $\sim$100 au region of OMC-3/MMS~6.
Our hypothesis is that B-field is relatively well ordered on this spatial scale.
The "disk" part of our model may be analogous to the radiative transfer simulation presented in Figure 7g, 7h of \citet{Lin2020}.

In addition, according to the model presented in Section \ref{sub:radiativetransfer}, the 33 GHz Stokes I intensity contours may provide a good indication of the projected area of the embedded disk.
To be conservative, we use the 35-$\sigma$ contour to indicate the projected area at the (pseudo-)disk, which is smaller than the assumed disk radius in our model (Figure \ref{fig:model}; Appendix Section \ref{sub:disk}) and the 43 au radius derived in Section \ref{sub:dust}.
We avoid interpreting the 265 GHz PA at the area which is bounded by the 35-$\sigma$ and 100-$\sigma$ contours of the 33 GHz Stokes I intensity image, since we are less certain about the radiative transfer and beam spearing in that area.

Figure \ref{fig:Bfield} shows the overall B-field topology we suggest based on the above discussion.
From this figure, we see that the B-field topology can be described with an hourglass shape.
The B-field appears predominantly poloidal and does not present large pinch-angles, which may be expected from the non-ideal magnetohydrodynamics simulations (c.f. \citealt{Li2011}, \citealt{Zhao2016}, and references therein).
It is intriguing that this "hourglass" appears inclined with respect to the axis of the molecular outflow (\citealt{Takahashi2012}), although the ALMA high-resolution observations show that the B-field PA has a better consistency with the outflow axis at the innermost region.
The B-field may be mis-aligned with the disk (c.f., \citealt{Machida2020,Hirano2020}, and references therein).
Otherwise, the disk may be wrapped on small spatial scales (e.g., \citealt{Sakai2019}, \citealt{Bi2020}).

So, does dust scattering really matter?
With our assumption of disk position angle and inclination, the E-field PA of the scattered light is indeed consistent with what was resolved by the high-resolution ALMA image around the central PI peak (Figure \ref{fig:ALMAhires}).
Nevertheless, to reproduce the observed features in OMC-3/MMS~6, it does not seem to require dust self-scattering to dominate the linearly polarized intensity at 265 GHz.
In addition, merely reproducing the observed PA is not yet sufficient to justify that polarized scattered light is prominent.
Given the high optical depth at 265 GHz, to reproduce the observed polarization percentages by dust self-scattering, it may require fine-tuning the density structures to make the local 265 GHz radiation field sufficiently anisotropic over an extended range of disk radius  (\citealt{Takahashi2019}; c.f. \citealt{Kataoka2015}).
According to the discussion in Section \ref{sub:dust} (see also Figure \ref{fig:opacity}), the scattering opacity is certainly not high enough to produce linearly polarized scattered light at the observing frequency of the presented JVLA data.
The 33 GHz PI still needs to be reproduced with aligned dust.
Since \amax in the disk is not large, and since it is not easy to produce highly anisotropic radiation field in the very optically thick environment to align dust, it is more probable that dust in the disk is aligned with B-field.
Therefore, the B-field PA in the inner $\sim$100 au region of OMC-3/MMS~6 is still determined by the PA measured at 33 GHz.
This also explains why the PA observed at 33 GHz is consistent with the PA observed at 265 GHz outward of the $\sim$50 au projected offset.
Therefore, we conclude that our proposed overall B-field topology will not be affected by the consideration of dust scattering.

\section{Conclusion}\label{sec:conclusion}

We have re-analyzed the archival, JVLA ($\sim$9 mm) and ALMA ($\sim$1.2 mm) full polarization observations towards the Class~0 young stellar object (YSO) OMC-3/MMS~6 (also known as HOPS-87).
We found that inward of the $\sim$50 au projected radius, the observed polarization (E-field) position angles (PA) by JVLA and ALMA present a nearly 90$^{\circ}$ relative offset, while they are very well consistent with each other on larger spatial scales.
This region is likely very optically thick (e.g., $\tau\gg$1) at $\sim$1 mm wavelength such that the dominant polarization mechanism is dichroic extinction.
The system becomes a lot optically thinner (e.g., $\tau\lesssim$1) at $\sim$9 mm wavelength thus the observations can directly probe the polarized emission of non-spherical dust which is predominantly aligned with B-field.
The very different optical depths at 1 mm and 9 mm wavelengths disfavor a small dust opacity spectral index ($\beta$) value.

We make a general caution that many previous (sub)millimeter bands interferometric linear polarization observations towards Class~0 YSOs and high-mass star-forming cores may require re-interpretation if dichroic extinction has not been considered as one of the plausible polarization mechanisms.
Our development have revealed how difficult it is to correctly infer the B-field direction in the inner $\sim$100 au region around Class~0 YSOs based on ALMA observations.
It is unlikely one can make it based on the observations at only one single frequency band.
It is expected to be more challenging in high-mass star-forming regions given their higher optical depths and the likely more complicated thermal structures.

We interpret the inner $\sim$100 au region of OMC-3/MMS~6 as an inclined (pseudo-)disk which is pinched by the predominantly poloidal magnetic field (B-field) that has an hourglass shape.
Our proposed B-field configuration may be consistent with what is theoretically expected for a Class~0 YSO which is regulated by modestly strong B-field and ambipolar diffusion.
It is intriguing that the resolved B-field PA in the inner $\sim$50-100 au region are $\sim$40$^{\circ}$ offset from the previously reported outflow axis. 
The disk may be warped on a not yet resolved spatial scale.
It could also be possible that the B-field lines are not perfectly well aligned with the rotational axis of the disk.

Finally, the measurements of Stokes I spectral indices and the observed 90$^{\circ}$ flips of PA in OMC-3/MMS~6 both favor a maximum grain size of $a_{\max}\le$100 $\mu$m.
Assuming $a_{\max}\le$100 $\mu$m, the flux density detected by JVLA corresponds to an overall dust mass of $\sim$14000 $M_{\oplus}$ within a $\sim$43 au radius.
Based on these measurements, we argue that it is not unlikely that the recent 9 mm surveys towards Class~0/I YSOs have systematically underestimated dust mass by one order of magnitude, owing to that they assumed too high dust absorption opacity for $\sim$9 mm wavelengths but without self-consistently considering the dust scattering opacity.
Assuming the gas-to-dust mass ratio $f\sim$100, the inferred overall disk mass is $\sim$4 $M_{\odot}$ in OMC-3/MMS~6.
Given such a high disk mass, the enclosed gas mass in the disk cannot be neglected when studying the gas kinematics, and the previously resolved clumpy structures on $\sim$500-1000 au scales may be explained by self-gravitational instability concordantly.

\acknowledgments
%We thank the anonymous referee for the careful review and useful suggestions.
The National Radio Astronomy Observatory is a facility of the National Science Foundation operated under cooperative agreement by Associated Universities, Inc.
This article makes use of the following ALMA data: ADS/JAO.ALMA No. 2015.1.00341.S.
ALMA is a partnership of ESO (representing its member states), NSF (USA) and NINS (Japan), together with NRC (Canada), MOST and ASIAA (Taiwan), and KASI (Republic of Korea), in cooperation with the Republic of Chile.
The Joint ALMA Observatory is operated by ESO, AUI NRAO and NAOJ. 
% This work has made use of data from the European Space Agency (ESA) mission {\it Gaia} (\url{https://www.cosmos.esa.int/gaia}), processed by the {\it Gaia} Data Processing and Analysis Consortium (DPAC, \url{https://www.cosmos.esa.int/web/gaia/dpac/consortium}). Funding for the DPAC has been provided by national institutions, in particular the institutions participating in the {\it Gaia} Multilateral Agreement.
% This work is based [in part] on observations made with the Spitzer Space Telescope, which is operated by the Jet Propulsion Laboratory, California Institute of Technology under a contract with NASA.
H.B.L. is supported by the Ministry of Science and
Technology (MoST) of Taiwan (Grant Nos. 108-2112-M-001-002-MY3).

%% To help institutions obtain information on the effectiveness of their 
%% telescopes the AAS Journals has created a group of keywords for telescope 
%% facilities.
%
%% Following the acknowledgments section, use the following syntax and the
%% \facility{} or \facilities{} macros to list the keywords of facilities used 
%% in the research for the paper.  Each keyword is check against the master 
%% list during copy editing.  Individual instruments can be provided in 
%% parentheses, after the keyword, but they are not verified.

%\vspace{2cm}
\facilities{JVLA, ALMA}

%% Similar to \facility{}, there is the optional \software command to allow 
%% authors a place to specify which programs were used during the creation of 
%% the manuscript. Authors should list each code and include either a
%% citation or url to the code inside ()s when available.

\software{
          astropy \citep{2013A&A...558A..33A},  
          Numpy \citep{VanDerWalt2011}, 
          CASA \citep[v5.6.0; ][]{McMullin2007},
          }

%% Appendix material should be preceded with a single \appendix command.
%% There should be a \section command for each appendix. Mark appendix
%% subsections with the same markup you use in the main body of the paper.

%% Each Appendix (indicated with \section) will be lettered A, B, C, etc.
%% The equation counter will reset when it encounters the \appendix
%% command and will number appendix equations (A1), (A2), etc. The
%% Figure and Table counter will not reset.

% \clearpage

\vspace{5cm}
\bibliography{omc3mm6}{}
\bibliographystyle{aasjournal}

\appendix
\section{Full Stokes images}\label{sec:images}
Figure \ref{fig:fullstokes} shows the Stokes I, Q, U images and the PI images of OMC-3/MMS~6 which were produced from the data introduced in Section \ref{sec:obs}.
There were similarities and differences between the ALMA image and the JVLA A+C configuration images.
For example, they resolved positive Stokes Q intensities east of the Stokes I peak; they resolved positive Stokes U intensities southeast of the Stokes I peak.
The major difference is that the ALMA observations resolved very significant negative Stokes U intensities around the Stokes I peak while this is not seen in the JVLA A+C configuration image.
% We do not think this difference can be attributed to artifacts.
Our interpretation for these similarities and differences is given in Section \ref{sec:discussion}.

\section{Radiative transfer}\label{sec:radiativetransfer}
In this section we outline how we perform our 1D radiative transfer modeling. % in a rather abstracted sense.
Our notation mostly follows that of the Section 4.1 of \citet{Liu2018}, except that we use $\epsilon$ instead of $\alpha$ to denote polarization efficiency in order to avoid confusing with the spectral indices $\alpha_{\mbox{\tiny IF1-IF2}}$ and $\alpha_{\mbox{\tiny IF1-ALMA2}}$ introduced within the present paper.

We assumed that the innermost part of our target source is a flatted (pseudo-)disk which has the (projected) radial dust column density and temperature profiles $\Sigma^{\mbox{\scriptsize disk} }(r)$ and $T^{ \mbox{\scriptsize disk} }(r)$, where $r$ is the projected offset from the location of the host protostar.
For observers, the near side of this (pseudo-)disk is obscured by the inner circumstellar envelope which has the (projected) radial dust column density and temperature profiles $\Sigma^{ \mbox{\scriptsize env} }(r)$ and $T^{ \mbox{\scriptsize env} }(r)$.
The explicit forms of $\Sigma^{\mbox{\scriptsize disk} }(r)$, $T^{ \mbox{\scriptsize disk} }(r)$, $\Sigma^{ \mbox{\scriptsize env} }(r)$, $T^{ \mbox{\scriptsize env} }(r)$ are introduced in Appendix Section \ref{sec:toy}.

The dust optical depth ($\tau^{ \mbox{\scriptsize disk} }_{\nu}(r)$, $\tau^{ \mbox{\scriptsize env} }_{\nu}(r)$) at the mean ALMA (265 GHz) and JVLA (33 GHz) observing frequencies ($\nu$) can be estimated by $\Sigma^{ \mbox{\scriptsize disk} }(r)\times\kappa^{\mbox{\scriptsize abs}}_{\nu}$ and $\Sigma^{ \mbox{\scriptsize disk} }(r)\times\kappa^{\mbox{\scriptsize abs}}_{\nu}$.
Following our discussion in Section \ref{sub:dust}, we assume that the maximum grain size ($a_{\mbox{\scriptsize max}}$) is well below $\sim$100 $\mu$m both in the disk and in the envelope.
In this case, the scattering opacity $\kappa^{\mbox{\scriptsize sca, eff}}_{\nu}$ can be safely neglected.
In addition, $\kappa^{\mbox{\scriptsize abs}}_{\nu}$ is not sensitive to $a_{\mbox{\scriptsize max}}$ and has a negligibly weak dependence on the ice coating (see Figure \ref{fig:opacity}).
For the sake of clarity, We drop the explicit $r$ dependence in the following discussion.

We assumed that the projected B-field position angle is 34$^{\circ}$ in both the envelope and the disk, and assumed that the long axis of the dust is aligned perpendicular to B-field. 
We express the dust optical depth at the orientation of B-field ($\tau_{\nu}^{\mbox{\scriptsize B-disk,env}}$) and at the orientation perpendicular to B-field ($\tau_{\nu}^{\mbox{\scriptsize E-disk,env}}$) by the assumed effective polarization efficiency ($\epsilon^{\mbox{\scriptsize disk}}$, $\epsilon^{\mbox{\scriptsize env}}$):

\begin{equation}
    \tau_{\nu}^{\mbox{\scriptsize E-disk,env}} = \tau_{\nu}^{\mbox{\scriptsize disk,env}} \times (1 + \epsilon^{\mbox{\scriptsize disk,env}}), \,\,\,\,\,\,\,\,\,\,
    \tau_{\nu}^{\mbox{\scriptsize B-disk,env}} = \tau_{\nu}^{\mbox{\scriptsize disk,env}} \times (1 - \epsilon^{\mbox{\scriptsize disk,env}}).
\end{equation}
In order to match observations, we assumed $\epsilon^{\mbox{\scriptsize disk}}$ and $\epsilon^{\mbox{\scriptsize env}}$ to be 10\% and 15\%, respectively.
Our assumed polarization efficiency may be achieved by the Radiative Grain Alignment (RAT) mechanism (e.g., \citealt{Hoang2016} and references therein).

The intensity in the E and B orientations ($I_{\nu}^{\mbox{\scriptsize E,B}}$) can be estimated by

\begin{equation}
    I_{\nu}^{\mbox{\scriptsize E,B}} = \frac{1}{2}\left(
    (1-e^{-\tau_{\nu}^{  \mbox{\tiny E,B-disk}  }})B_{\nu}(T^{\mbox{\scriptsize disk}})e^{-\tau_{\nu}^{\mbox{\tiny E,B-env}}} +
    (1-e^{-\tau_{\nu}^{  \mbox{\tiny E,B-env}  }})B_{\nu}(T^{\mbox{\scriptsize env}})
    \right),
\end{equation}
where $B_{\nu}(T)$ is the Planck function for a certain temperature $T$.

The synthesized beam standard deviation ($\sigma_{\mbox{\scriptsize beam}}$) is 25 au.
Given that the inner $r < 0.5 \sigma_{\mbox{\scriptsize beam}} $ region is very seriously beam smeared, we additionally adopt the following regulation
\begin{eqnarray}
I_{\nu}^{\mbox{\scriptsize E,B}}(r<0.5\sigma_{\mbox{\scriptsize beam}}) = I_{\nu}^{\mbox{\scriptsize E,B}}(r=0.5\sigma_{\mbox{\scriptsize beam}}) \nonumber \\ 
%PI_{\nu}(r<0.5\sigma_{\mbox{\scriptsize beam}}) = PI_{\nu}(r=0.5\sigma_{\mbox{\scriptsize beam}}) \nonumber \\ 
\end{eqnarray}
to avoid overestimating the flux density.
We argue that such an approximation is acceptable since the details on the unresolved spatial scales are uncertain in any case.
To mimic observations of finite angular resolution, we evaluated the smoothed intensity $\hat{I}_{\nu}^{\mbox{\scriptsize E,B}}$ by convolving $I_{\nu}^{\mbox{\scriptsize E,B}}$ with a 1D normal distribution of which the standard deviation is $\sigma_{\mbox{\scriptsize beam}}$.

Finally, the models for the observed Stokes~I intensity ($I_{\nu}$) and polarized intensity ($PI_{\nu}$) are constructed as
\begin{eqnarray}
\hat{I}_{\nu} = \hat{I}_{\nu}^{\mbox{\scriptsize E}} + \hat{I}_{\nu}^{\mbox{\scriptsize B}} \nonumber \\ 
\hat{PI}_{\nu} = | \hat{I}_{\nu}^{\mbox{\scriptsize E}} - \hat{I}_{\nu}^{\mbox{\scriptsize B}}  | 
\end{eqnarray}
We can see that the observed polarization position angle is 34$^{\circ}$ at where $\hat{I}_{\nu}^{\mbox{\scriptsize E}} > \hat{I}_{\nu}^{\mbox{\scriptsize B}}$ and is 124$^{\circ}$ otherwise.

\section{Target source toy model}\label{sec:toy}
\begin{figure}
    \centering
    \includegraphics[width=6cm]{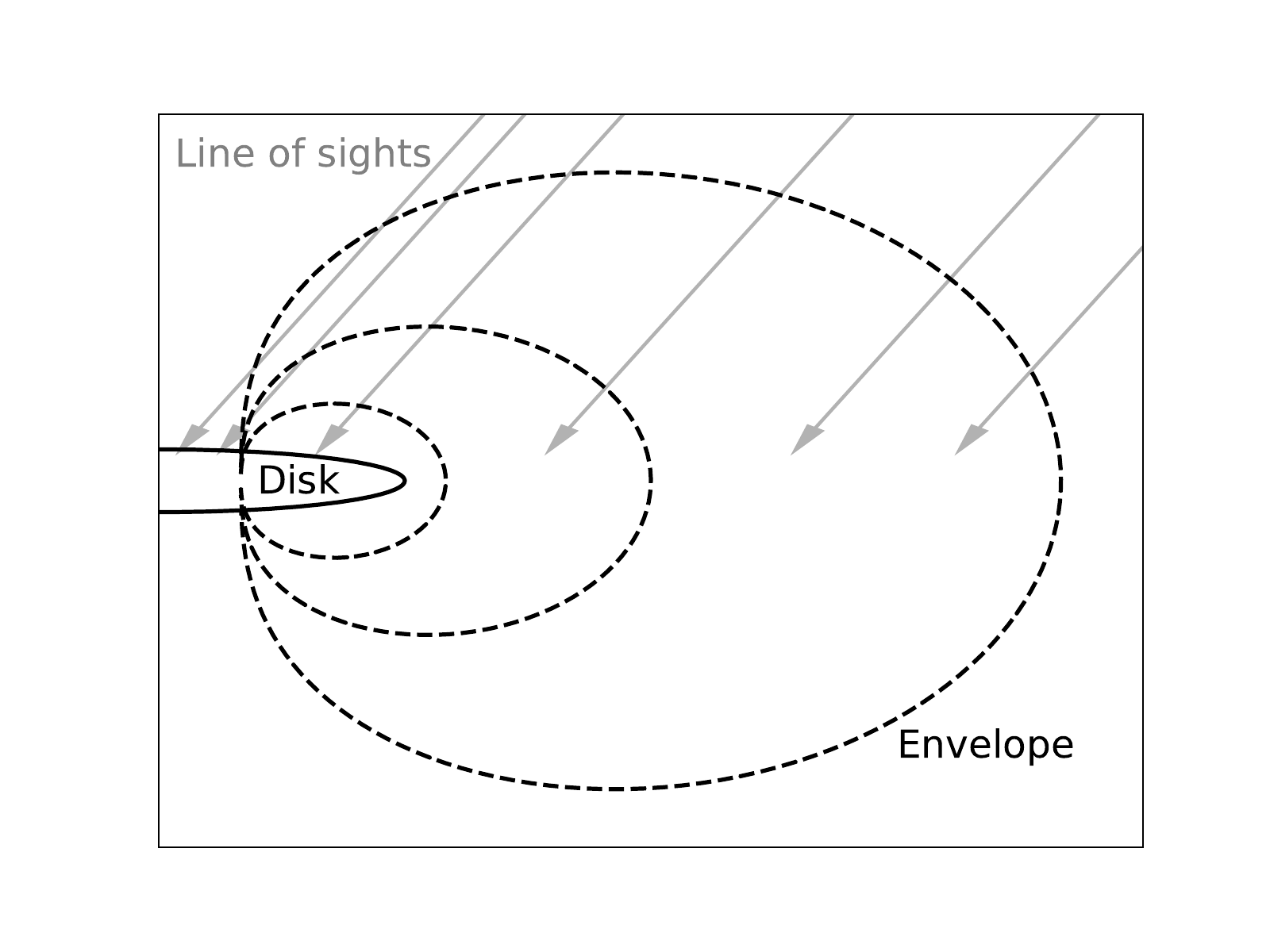}
    \vspace{-0.6cm}
    \caption{Schematic picture for our toy disk+envelop model. Solid and dashed lines represent the iso-density contours of the disk and envelope, respectively.}
    \label{fig:schematic_model}
\end{figure}

\begin{figure}
    \centering
    \begin{tabular}{ p{4.2cm} p{4.2cm} p{4.2cm} p{4.2cm}}
         \includegraphics[width=4.3cm]{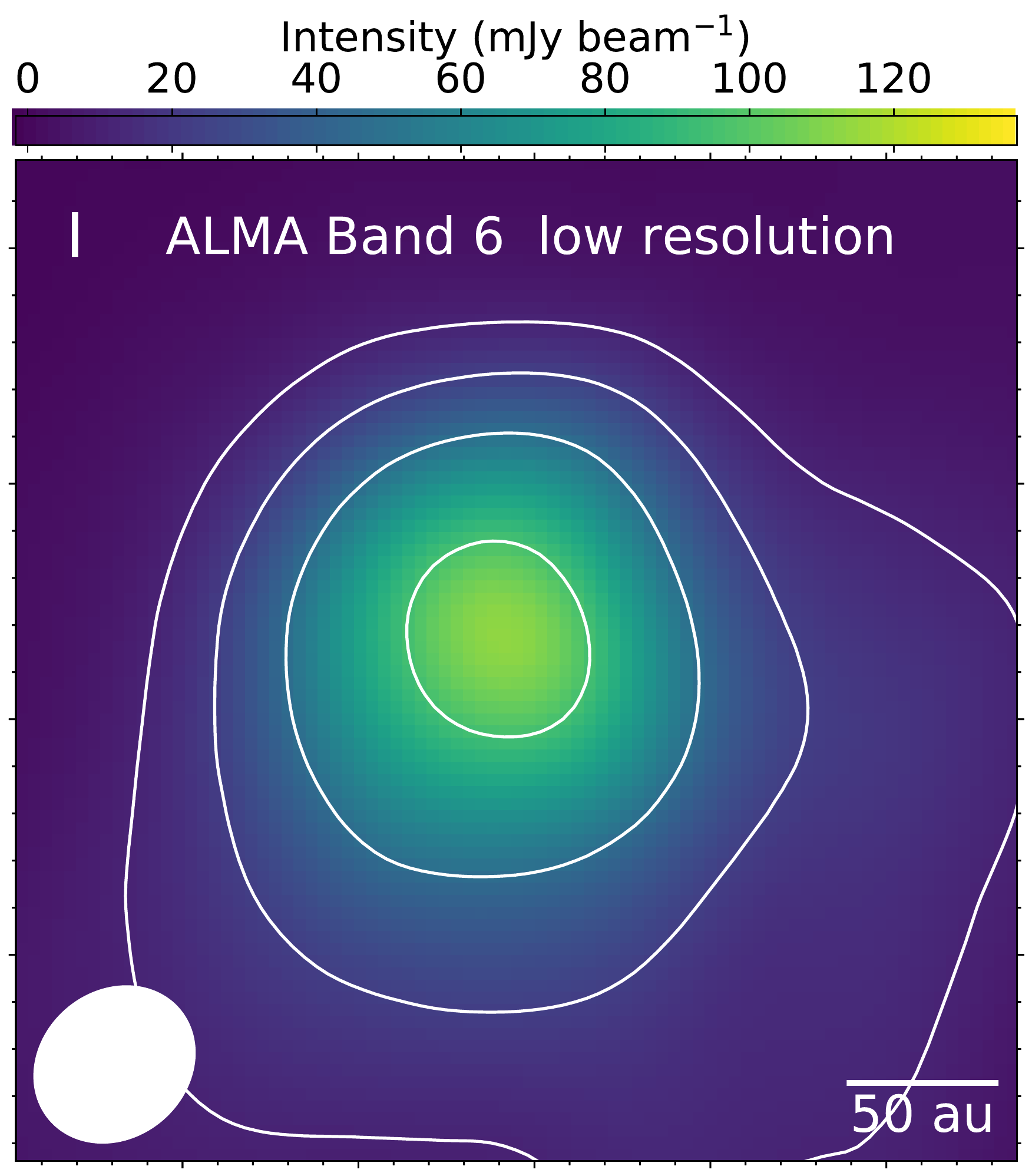} &
         \includegraphics[width=4.3cm]{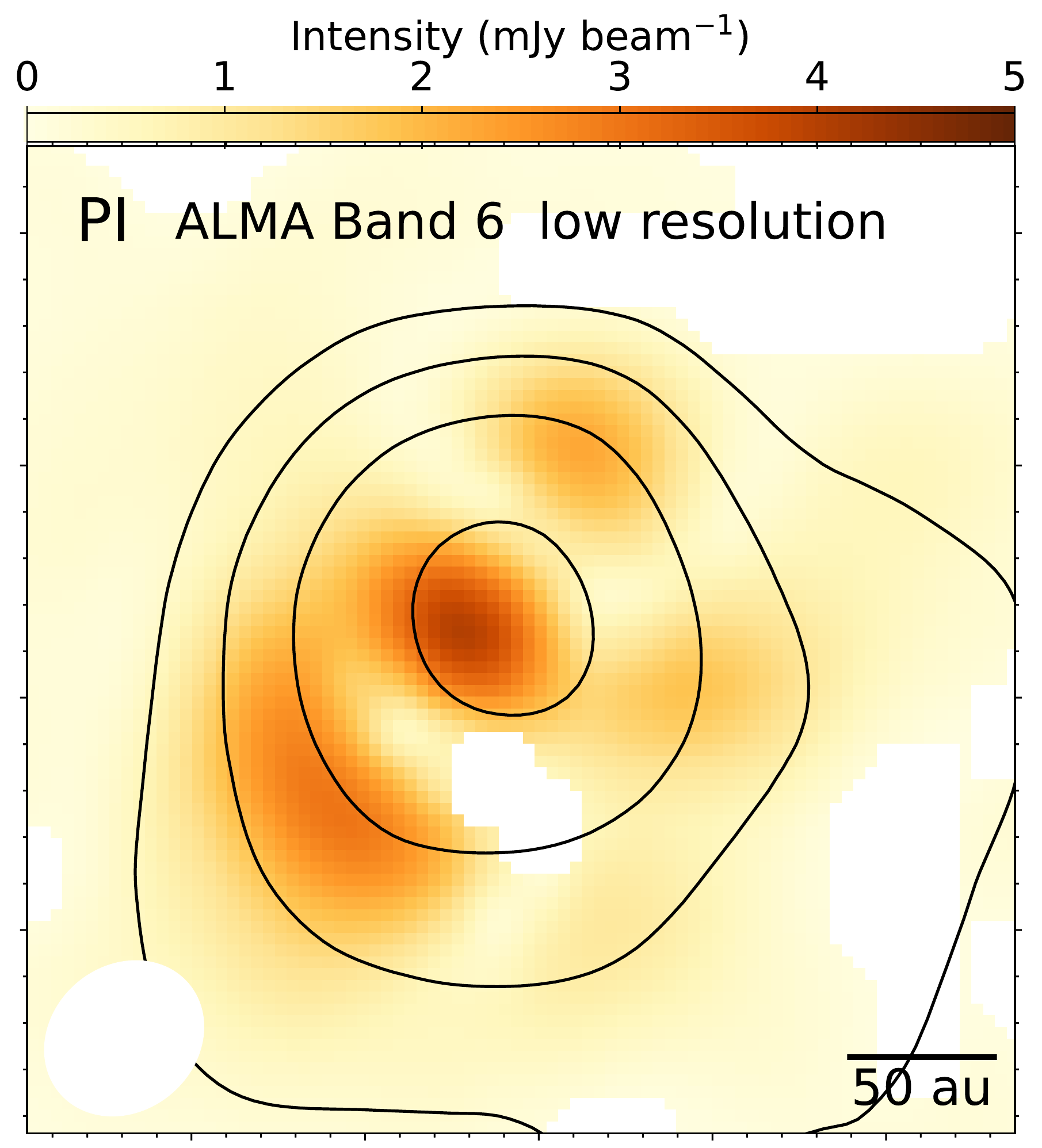} &
         \includegraphics[width=4.3cm]{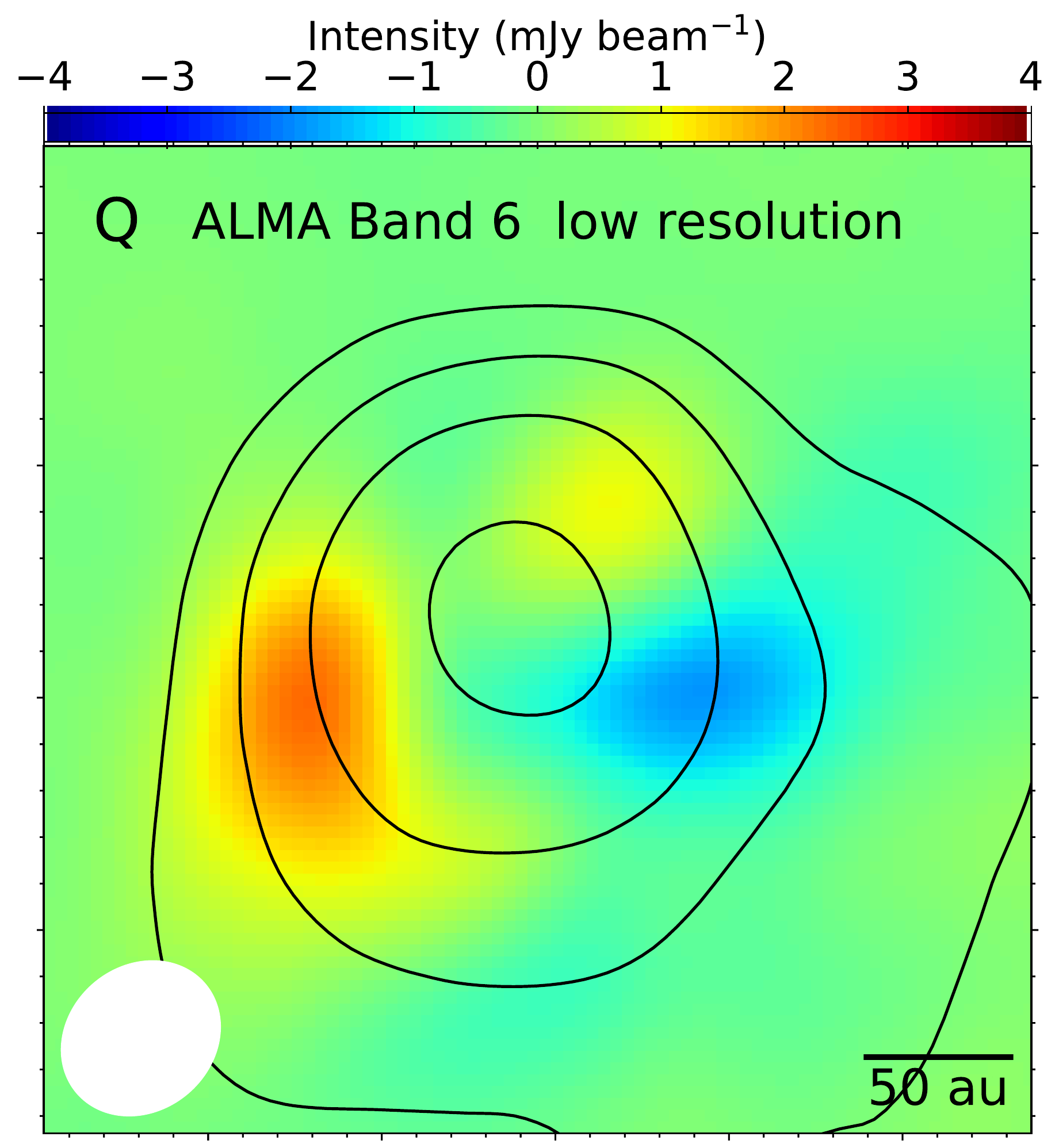} &
         \includegraphics[width=4.3cm]{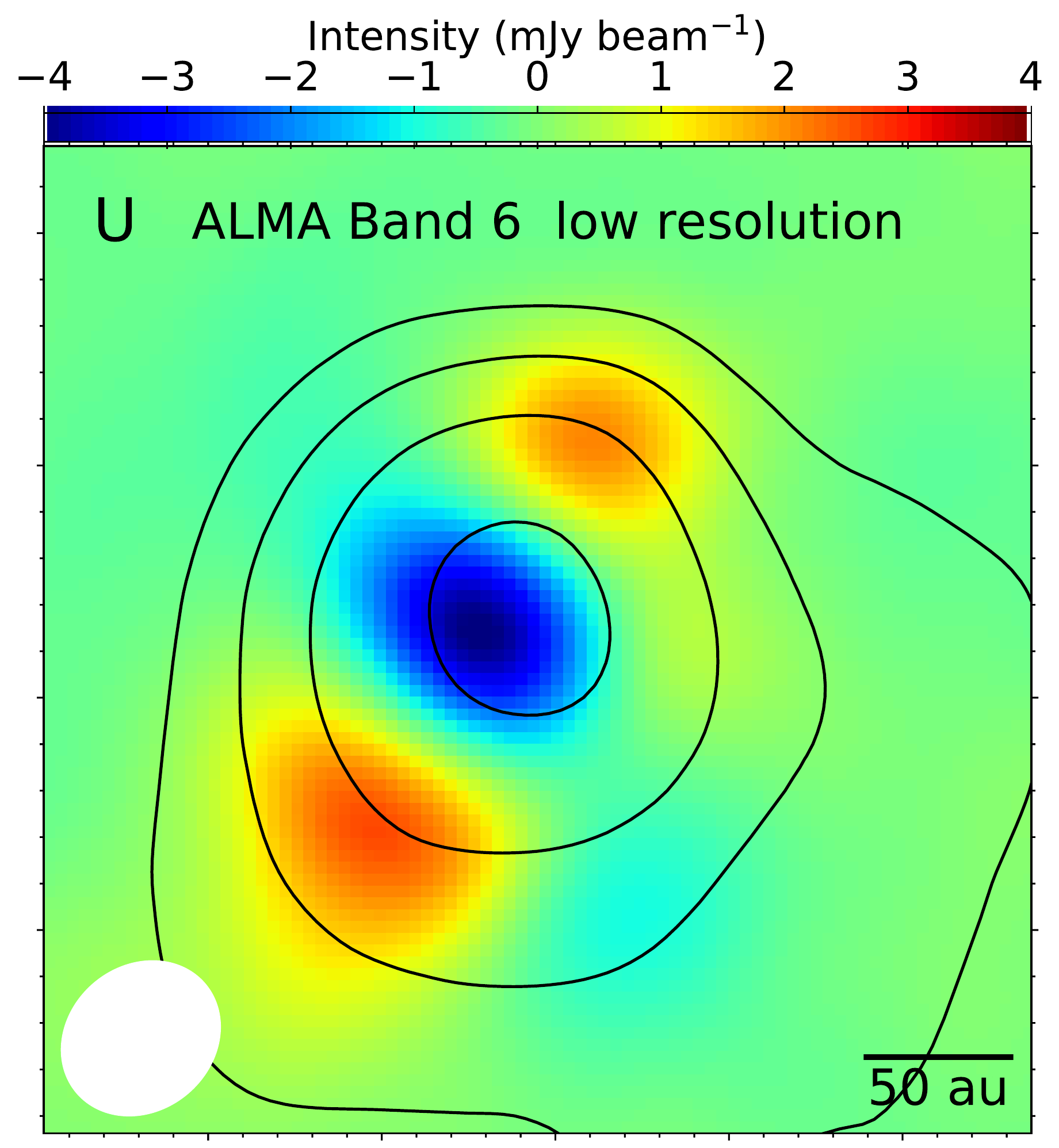} \\
         \includegraphics[width=4.3cm]{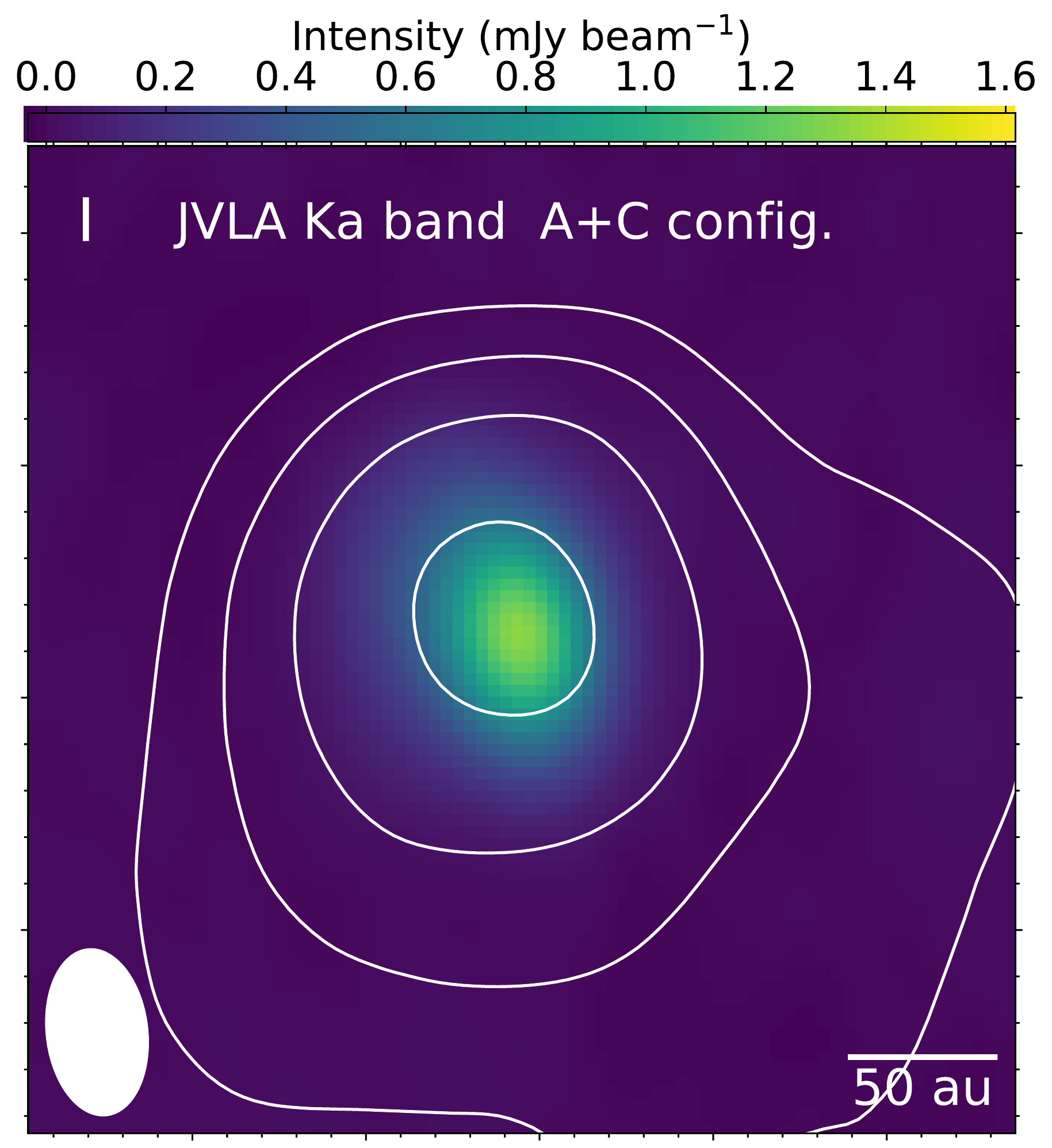} &
         \includegraphics[width=4.3cm]{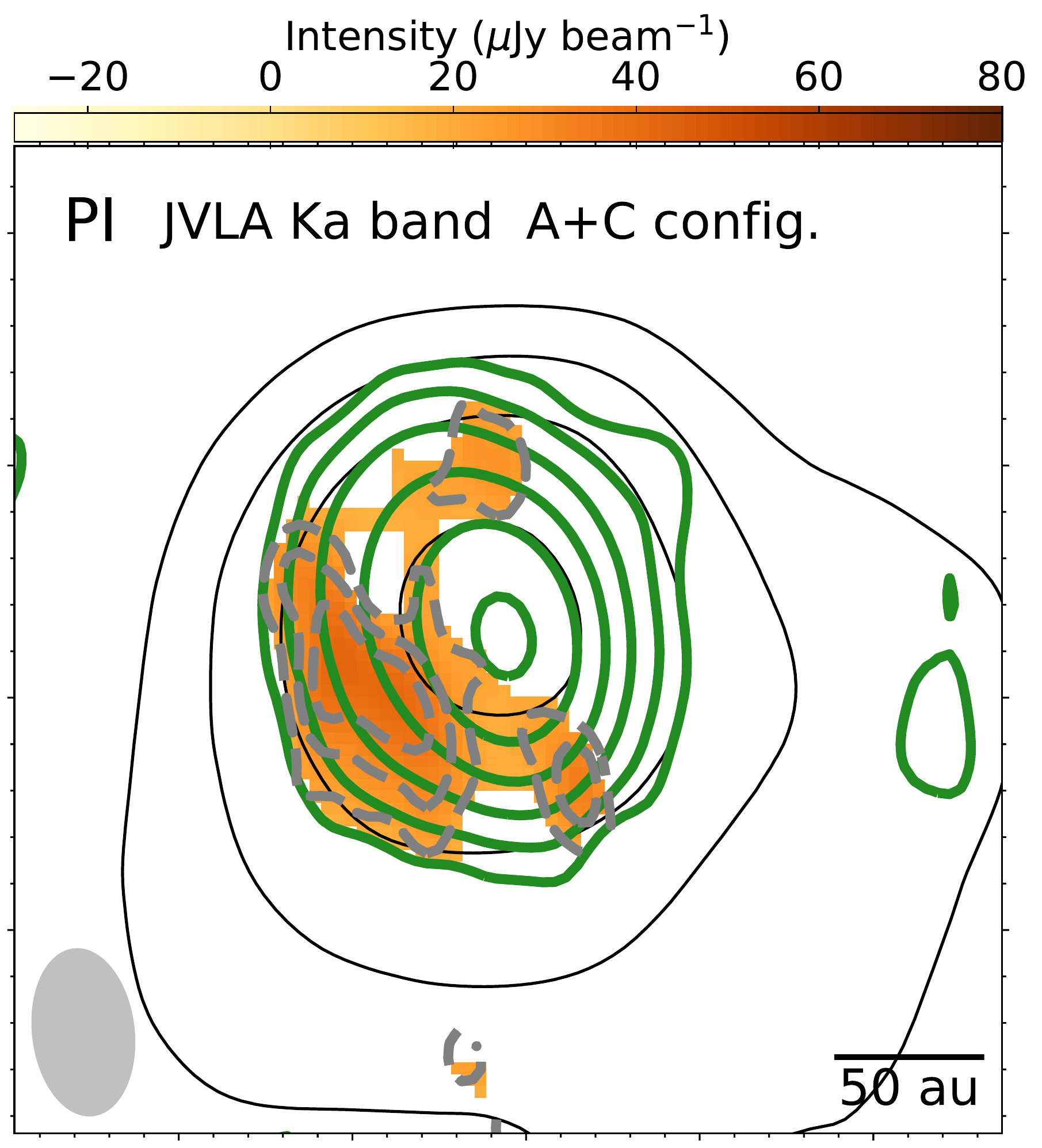} &
         \includegraphics[width=4.3cm]{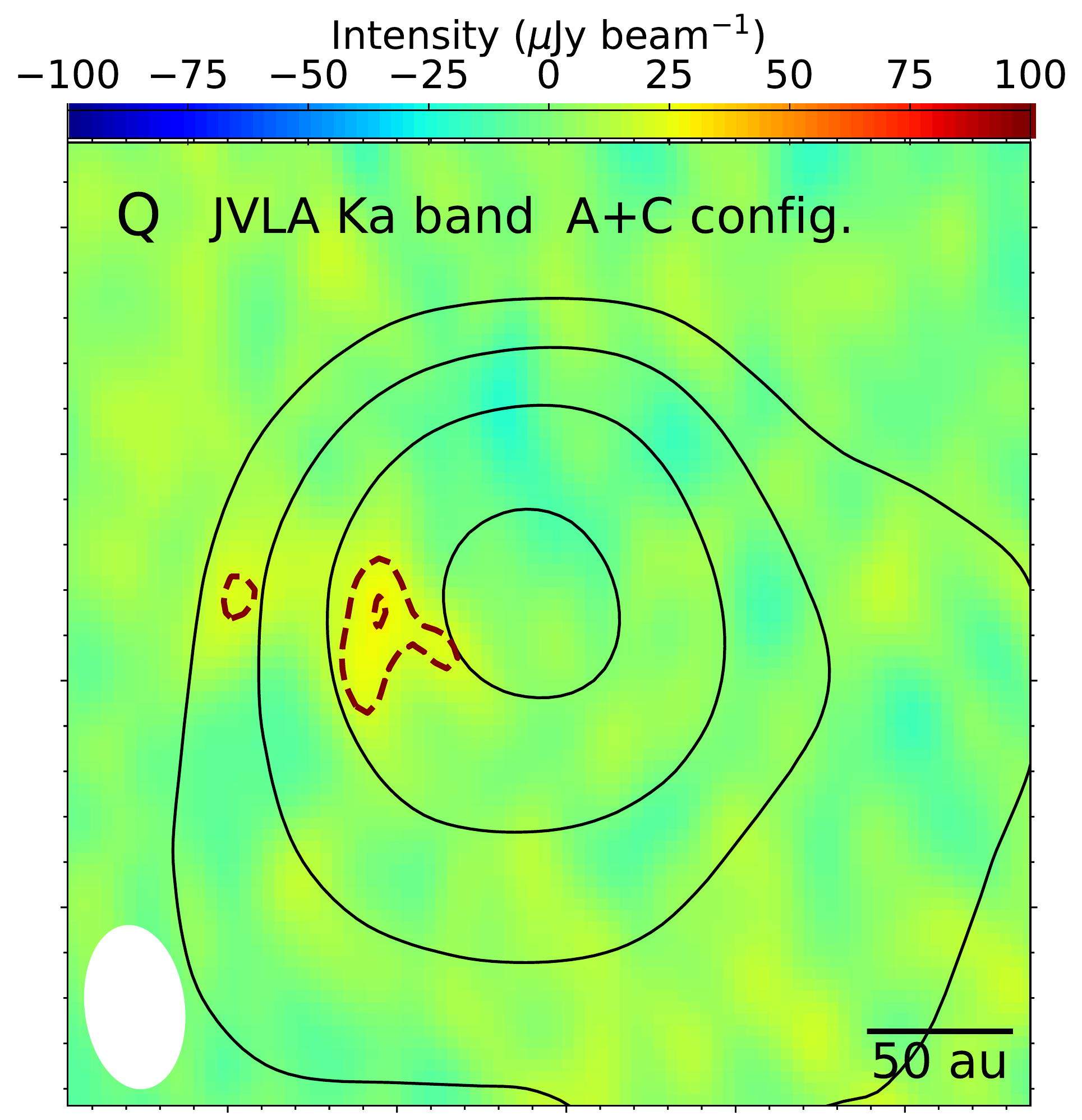} &
         \includegraphics[width=4.3cm]{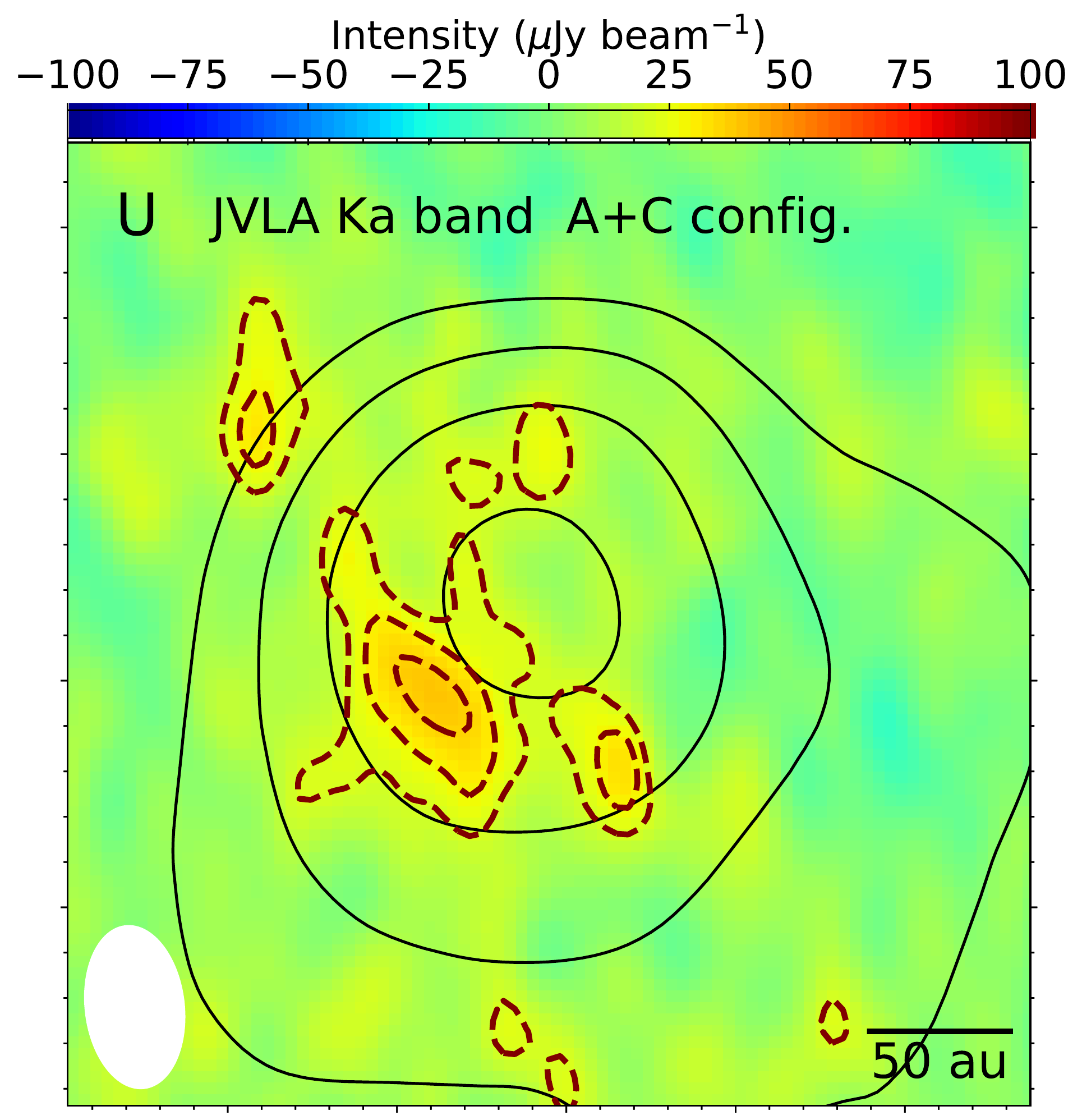} \\
         \includegraphics[width=4.3cm]{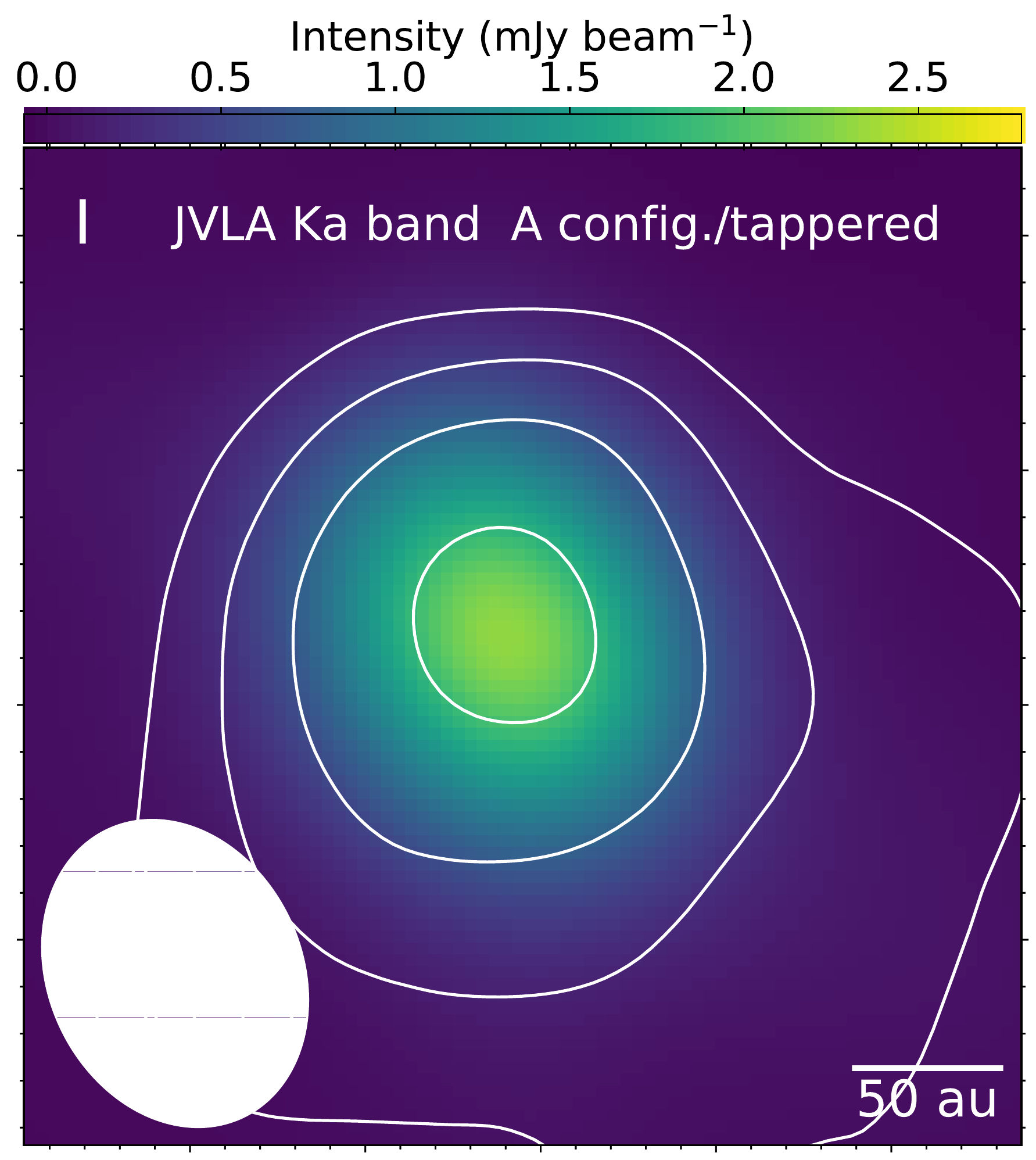} &
         \includegraphics[width=4.3cm]{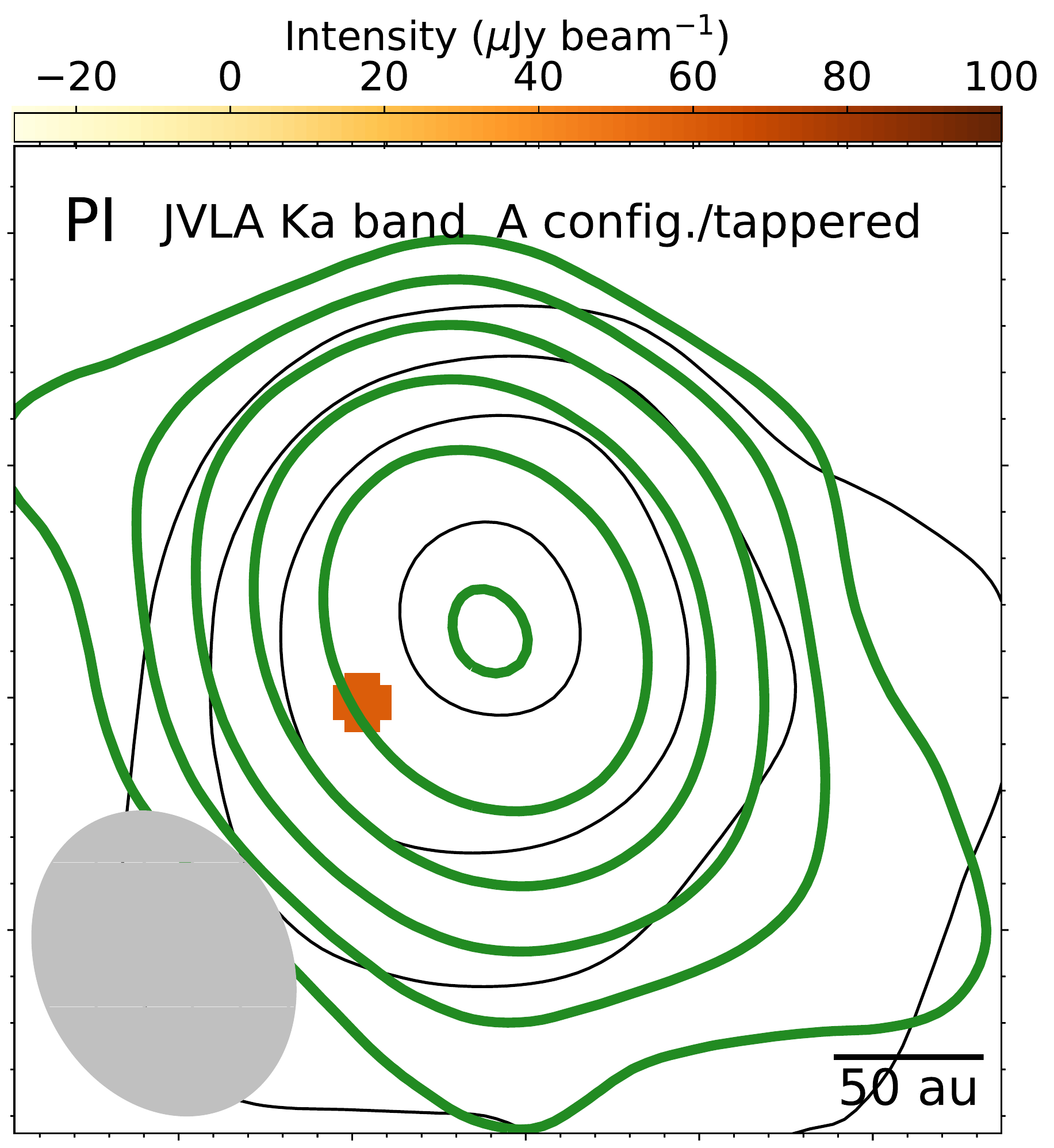} &
         \includegraphics[width=4.3cm]{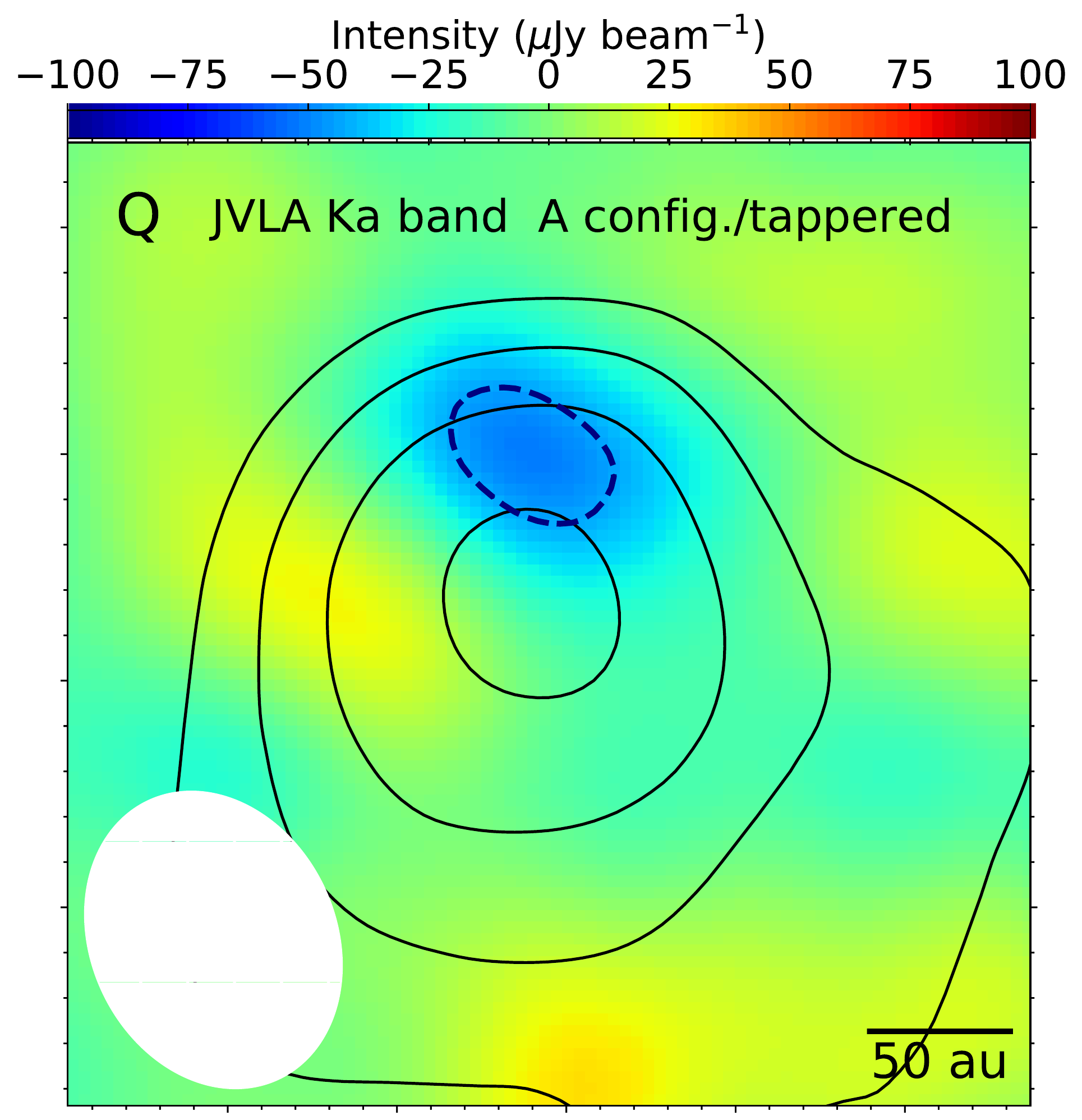} &
         \includegraphics[width=4.3cm]{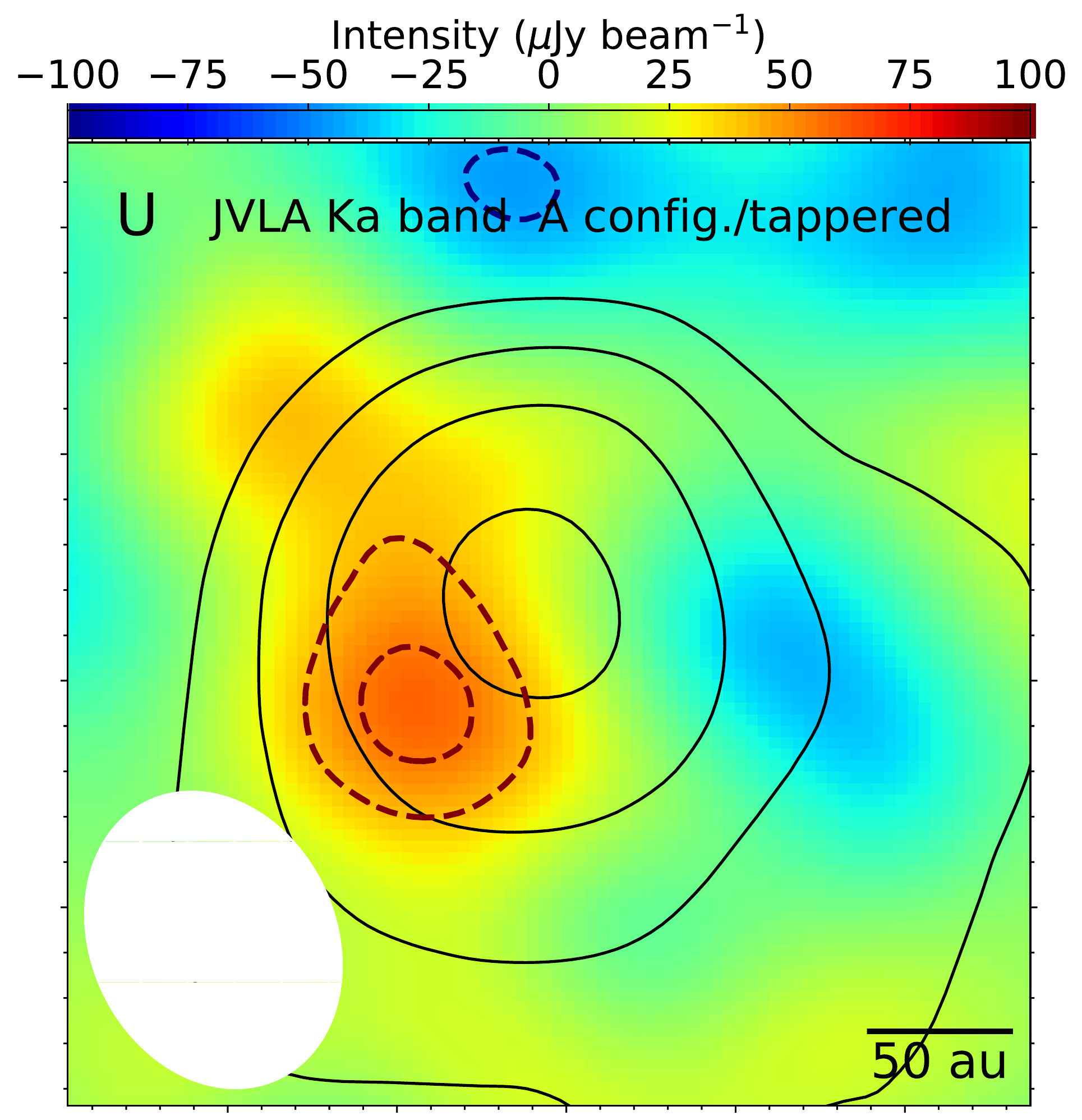} \\
         \includegraphics[width=4.3cm]{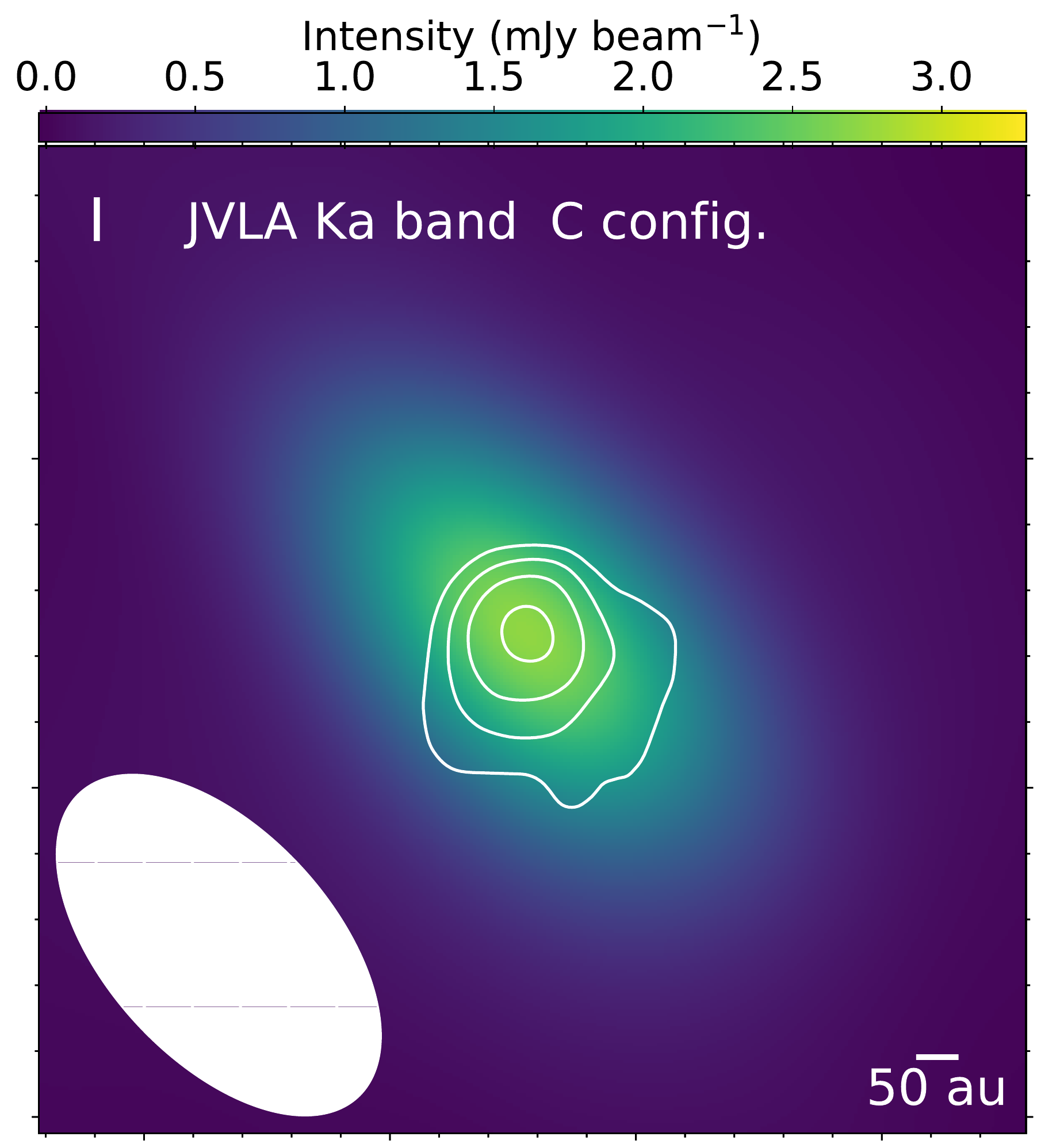} &
        \includegraphics[width=4.3cm]{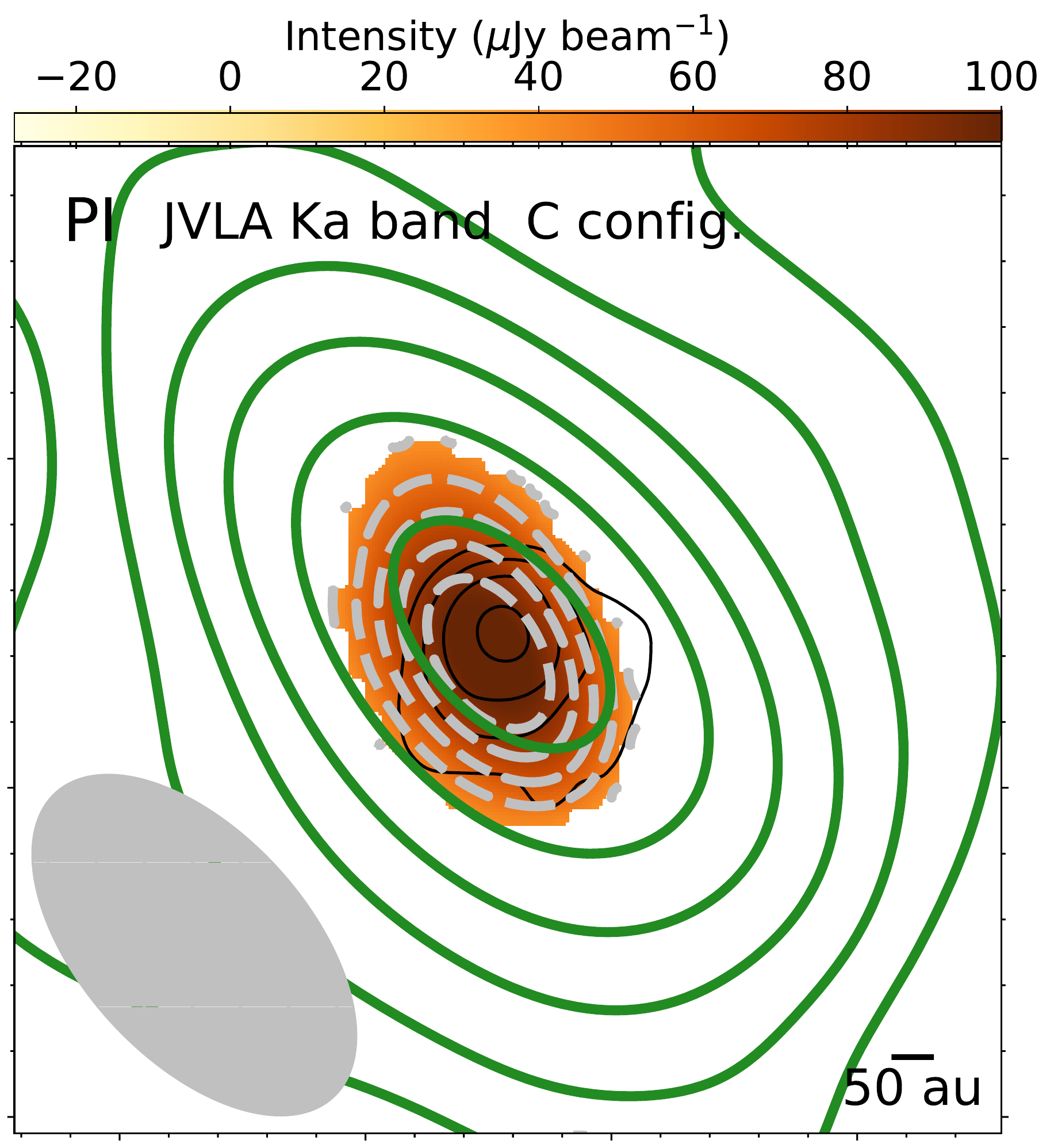} &
         \includegraphics[width=4.3cm]{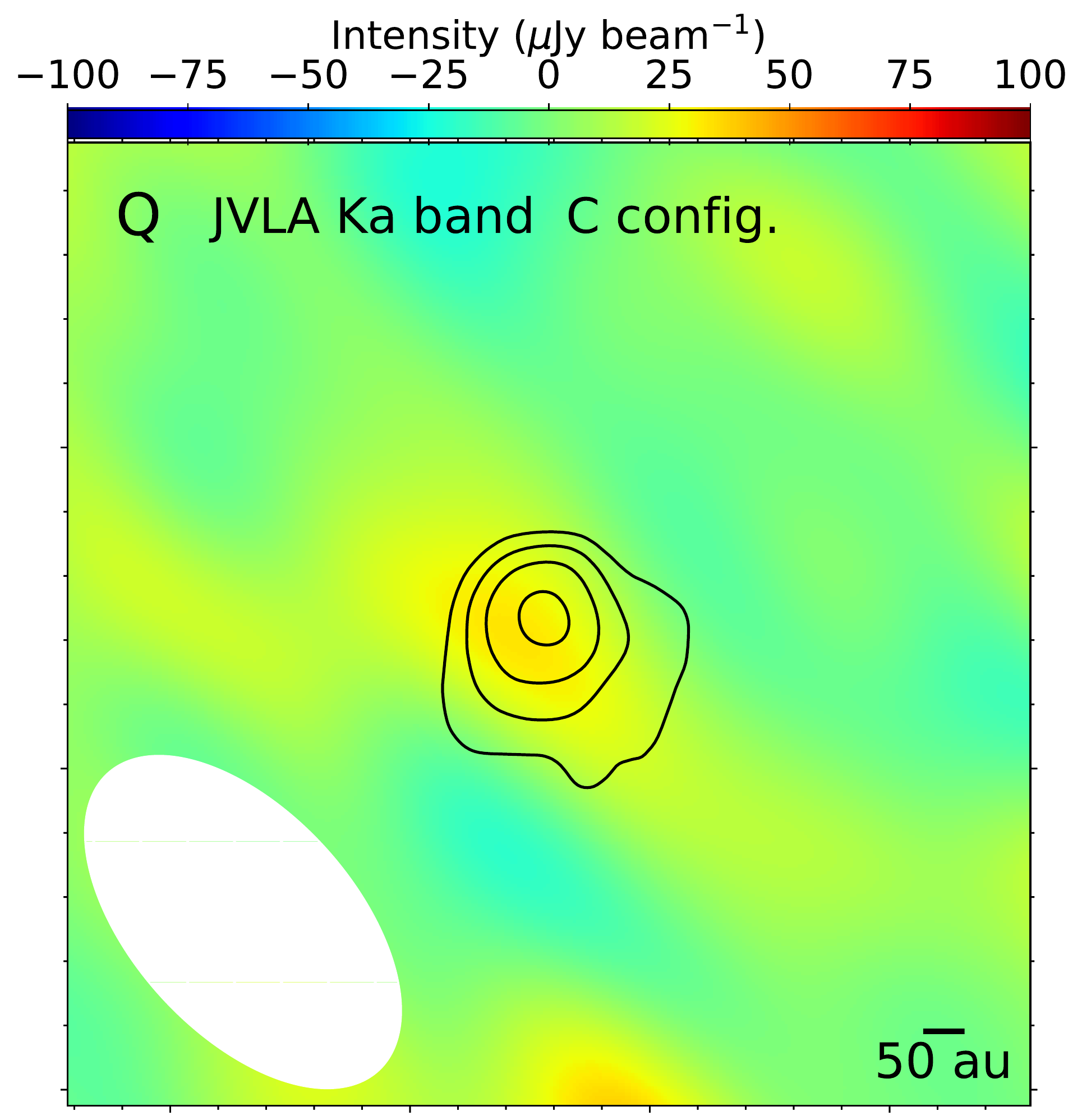} &
         \includegraphics[width=4.3cm]{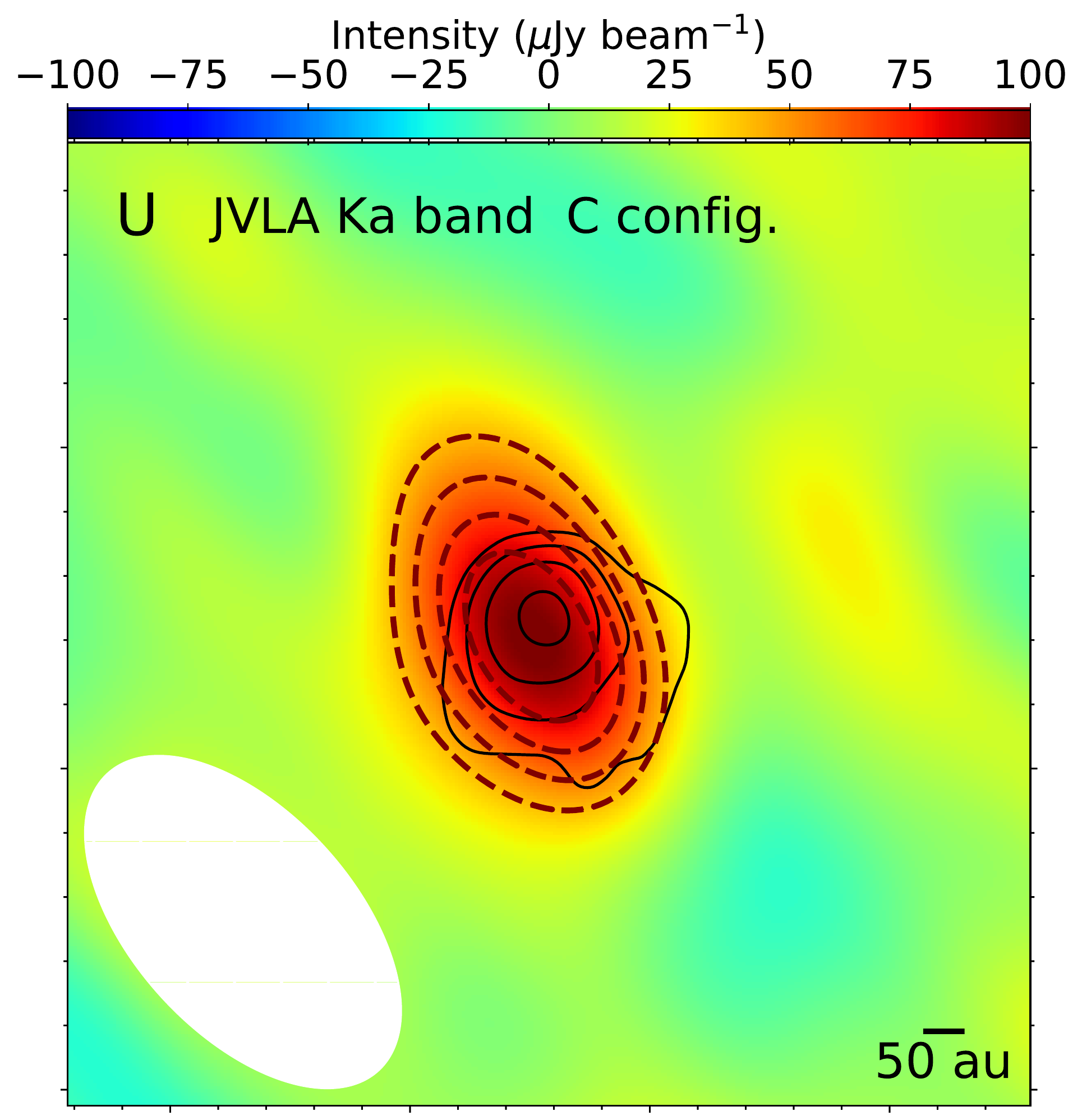} \\
    \end{tabular}
    \caption{Stokes I, Q, U images and the polarization intensity (PI) images (color maps) produced from the ALMA (top row; \beam$=$0\farcs14$\times$0\farcs12, P.A.=-50$^{\circ}$; RMS noise: 32 \ujyperbeam), JVLA A+C array configuration (2nd row; \beam$=$0\farcs14$\times$0\farcs087, P.A.=6.6$^{\circ}$; RMS noise: 7.5 \ujyperbeam), JVLA A array configuration (3rd row; tapperd; \beam$=$0\farcs27$\times$0\farcs22, P.A.=25$^{\circ}$; RMS noise: 14 \ujyperbeam), and JVLA C array configuration observations (bottom row; \beam$=$1\farcs3$\times$0\farcs59, P.A.=42$^{\circ}$; RMS noise: 14 \ujyperbeam). 
    The plotting areas of the top three rows are the same with that of Figure \ref{fig:polmap}.
    The JVLA Stokes I images presented in the first column are also overplotted in the second column in green contours ([1, 2, 4, 8, $\cdots$]$\times$5-$\sigma$).
    Gray dashed contours in the second column show the JVLA PI images, starting from 3-$\sigma$ level and with 1-$\sigma$ intervals.
    Red and blue dashed contours in the third and fourth columns show the positive and negative Stokes Q and U intensities, starting from $\pm$3-$\sigma$ level and with 1-$\sigma$ intervals.
    In all panels we overplot the ALMA Stokes I image in black or white solid contours ([375-$\sigma$, 750-$\sigma$, 1500-$\sigma$, 3000-$\sigma$,]) for referencing positions.
    The synthesized beams are shown in the bottom left.
    The details of these observations are provided in Section \ref{sec:obs}.
    }
    \label{fig:fullstokes} 
\end{figure}

In this section, we introduce how we constructed the explicit forms of the radial column density and temperature profiles of the disk and the envelope (c.f., Appendix Section \ref{sec:radiativetransfer}).
A schematic picture of our model is presented in Figure \ref{fig:schematic_model} for illustration.
In spite that we provide some physical rationale about how this model was constructed, we note that the role of the rationale was to guide us towards some formulae that work reasonably well for the specific target source OMC-3/MMS~6 empirically.
Without the guide, the degree of freedom would be too large to manage.
However, it would not be particularly meaningful if one base on the comparison of observations with the radiative transfer model introduced in this section and Appendix Section \ref{sec:radiativetransfer} to argue that the physical rationale we are based on has been verified or falsified.

\subsection{Envelope}\label{sub:envelope}
We assumed that the envelope is flattened yet is geometrically thick.
Inward of a innermost radius $r^{\mbox{\scriptsize env}}_{\mbox{\scriptsize in}}$, the volume density of the envelope drops to 0; for the rest of the region, the radial surface density profile (e.g., when viewed face-on) is described by a power law.
Physically, the truncation of the envelope at small radii can be made by the outflow/jet.
More specifically, we fix the power law index ($\gamma^{\mbox{\scriptsize env}}$) to be -0.5, which might be understood since the volume density profile in the innermost part of a rapidly collapsing envelope is expected to have a $r^{-1.5}$ dependence \citep{Tereby1984}.
When the source is inclined, at a certain radius $\rho$ at the near side, dust at larger scale height is seen at smaller projected radius $r$.
If the envelope is supported by thermal pressure or microturbulence (c.f., \citealt{deJager1954}), then one may expect that the observed, projected dust column density profile is smeared by approximated Gaussian distributions.
For simplicity, we consider an identical Gaussian smoothing kernel $G(r-r')$ for all radii.
Given that the scale height of the envelope was marginally spatially resolved, we fixed the standard deviation of $G(r-r')$ to $\sigma_{\mbox{\scriptsize beam}}$.
The explicit form of the projected dust column density profile ($\Sigma^{ \mbox{\scriptsize env} }(r)$) we adopted was constructed as
\begin{equation}\label{eq:envSigma1}
\begin{split}
\tilde{\Sigma}^{ \mbox{\scriptsize env} }(r) \mbox{ [g\,cm$^{-1}$] }  & =  0 \,\,\,\,\, (r \le r^{\mbox{\scriptsize env}}_{\mbox{\scriptsize in}} )  \\
& =  2.1\cdot\left( \frac{r \mbox{[au]}}{50} \right)^{\gamma^{\mbox{\scriptsize env}}}  \,\,\,\,\, (r > r^{\mbox{\scriptsize env}}_{\mbox{\scriptsize in}} ),
\end{split}
\end{equation}

\begin{equation}\label{eq:envSigma2}
\Sigma^{ \mbox{\scriptsize env} }(r) = \tilde{\Sigma}^{ \mbox{\scriptsize env} }(r) \otimes G,
\end{equation}
where $\otimes$ denotes convolution, and $r^{\mbox{\scriptsize env}}_{\mbox{\scriptsize in}}$ was chosen to be 10.8 au. 
The realization of $\Sigma^{ \mbox{\scriptsize env} }(r)$ is shown in Figure \ref{fig:model}.
The value of $r^{\mbox{\scriptsize env}}_{\mbox{\scriptsize in}}$ and the coefficients Equation \ref{eq:envSigma1} were chosen to make the predicted intensity comparable to what was observed by ALMA and JVLA.
We note that Equation \ref{eq:envSigma1} and \ref{eq:envSigma2} are just a way we constructed the column density profile of the envelope.
Whether or not the envelope is indeed supported by any physical mechanisms, or whether or not microturbulence is physical, are not real issues for us.

We found that varying $\gamma^{\mbox{\scriptsize env}}$ in the range of $-$0.1-$-$1.0 without changing the other parameters do not lead to qualitatively different radiative transfer (Appendix Section \ref{sec:radiativetransfer}) results.
This is because that the envelope is relatively optically thick at 265 GHz while it does not emit much at 33 GHz, which is consistent with the resolved spectral index distribution $\alpha_{\mbox{\tiny IF1-ALMA2}}$ (Section \ref{sec:result}; Figure \ref{fig:ALMAhires}, left right) and the 33 GHz brightness temperature profile of OMC-3/MMS~6.
Therefore, to compare with the JVLA and ALMA data presented in this paper, it is more critical to fine-tune the radial dust temperature profile in the envelope rather than the dust column density profile.
Nevertheless, if dust in the envelope is primarily heated by protostellar irradiation, then based on Stefan–Boltzmann law, one would not expect to have too much of freedom in fine-tuning the details of the radial temperature profile since temperature has a power equal to $1/4$ dependence on the radiative flux.
In this sense,  we are not actually dealing with a particularly bad fine-tuning model in spite that there is a large number of free parameters introduced in this section.
The column density profile of the envelope can be more tightly constrained if provided with $\lesssim$10 au resolution observations at 100-150 GHz bands.
 
In this paper, we assumed that the temperature profile of the envelope has a $r^{-1/2}$ dependence, which can be understood if the radiative flux has a $r^{-2}$ dependence. 
This assumption of $r$ dependence has a concern of confusing radius with projected radius, which cannot be resolved without self-consistently solving the envelope scale height and the thermal structure iteratively in  three dimensional space. 
Nevertheless, we found that based on such an assumption, it is not difficult to yield an intensity profile model which is comparable with what was observed from OMC-3/MMS~6.
Explicitly, the temperature profile we adopted was: 
\begin{equation}\label{eq:envT}
\begin{split}
T^{ \mbox{\scriptsize env} }(r) \mbox{ [K] }  & =  67.5 \,\,\,\,\, (r \le r^{\mbox{\scriptsize env}}_{\mbox{\scriptsize in}} ),  \\
& =  70\cdot\left( \frac{r \mbox{[au]}}{50} \right)^{-0.5}  \,\,\,\,\, (r > r^{\mbox{\scriptsize env}}_{\mbox{\scriptsize in}} ).
\end{split}
\end{equation} 
For $r > r^{\mbox{\scriptsize env}}_{\mbox{\scriptsize in}}$, the coefficient was chosen such that the predicted $\hat{I}_{\mbox{\scriptsize 265 GHz}}$ approximately matches what was resolved from ALMA observations.
This projected power-law temperature profile is expected for an $\sim$10-100 $L_{\odot}$ protostar (\citealt{Whitney2003}), while the previous observations have constrained the bolometric luminosity of OMC-3/MMS~6 to be 32 $L_{\odot}$ (\citealt{Tobin2020}; c.f. \citealt{Chini1997}).
Inward of $r^{\mbox{\scriptsize env}}_{\mbox{\scriptsize in}}$ we fix the dust temperature to that at $r^{\mbox{\scriptsize env}}_{\mbox{\scriptsize in}}$.
The value of $r^{\mbox{\scriptsize env}}_{\mbox{\scriptsize in}}$ was chosen to be identical to that of the outermost radius of the embedded disk (see Appendix Section \ref{sub:disk}), which is comparable with the assumed scale height (i.e., the full Gaussian width above the mid-plane).
For projected radius $r<r^{\mbox{\scriptsize env}}_{\mbox{\scriptsize in}}$, it is no more self-consistent to consider that the envelope is geometrically flat.
The assumed temperature profile in Equation \ref{eq:envT} is to take into account that the lower temperature dust at larger radii can significantly contribute to the emission at small $r$ when the envelope is optically thick (see Figure \ref{fig:schematic_model}).

The dust temperature described by this profile has a small range.
We found that as long as the the dust temperature in the envelope is within such a small range, the predicted $\hat{PI}_{\nu}$ is in fact not very sensitive to small variations of the coefficient or power law index in Equation \ref{eq:envT}.
Therefore, we consider our present approach of constructing $T^{ \mbox{\scriptsize env} }(r)$ to be adequate for our present purposes.

\subsection{Disk}\label{sub:disk}
The dust thermal emission from the embedded disk dominates the continuum intensity at 33 GHz.
The JVLA observations with a synthesized beam $\sigma_{\mbox{\scriptsize beam}}$ (see Section \ref{sec:result}) have marginally spatially resolved the  33 GHz emission source.
Therefore, we constructed the column density profile of disk ($\Sigma^{ \mbox{\scriptsize disk} }(r)$) as a Gaussian distribution of which the standard deviation ($\sigma^{\mbox{\scriptsize disk}}$) is close to the value of $\sigma_{\mbox{\scriptsize beam}}$.
The detailed functional form of $\Sigma^{ \mbox{\scriptsize disk} }(r)$ is not particularly important given that its emission is seriously smeared by the synthesized beam.
However, we found that it is necessary to truncate $\Sigma^{ \mbox{\scriptsize disk} }(r)$ at a certain radius $r^{\mbox{\scriptsize disk}}$ to avoid yielding a spatially too extended $\hat{I}_{\mbox{\scriptsize 33 GHz}}$.
The explicit form of $\Sigma^{ \mbox{\scriptsize disk} }(r)$ is:

\begin{equation}\label{eq:diskSigma}
\begin{split}
\Sigma^{ \mbox{\scriptsize disk} }(r) \mbox{ [g\,cm$^{-1}$] }  & =  0 \,\,\,\,\, (r > r^{\mbox{\scriptsize disk}} )  \\
& =  115\cdot e^{-\frac{1}{2}(\frac{r}{ \sigma^{\mbox{\tiny disk}} })^{2} }  \,\,\,\,\, (r \le r^{\mbox{\scriptsize disk}} ),
\end{split}
\end{equation}
where $\sigma^{\mbox{\scriptsize disk}} $ and  $r^{\mbox{\scriptsize disk}}$ are chosen to be 27 au and 54 au, respectively.
With our assumed $\kappa^{\mbox{\scriptsize abs}}_{\nu}$ (Section \ref{sub:radiativetransfer}), the corresponding peak $\tau^{ \mbox{\scriptsize disk} }_{\mbox{\scriptsize 33 GHz}}$ value is $\sim$1
This will yield a minimum $\alpha_{\mbox{\tiny IF1-IF2}}$ value of $\sim$3.0, which is consistent with the JVLA observations (Section \ref{sec:result}; Figure \ref{fig:ALMAhires}, bottom right).

To reproduce the observed PA and $PI$ profiles, the dust temperature in the disk is required to be considerably higher than that of the envelope in a line-of-sight (c.f. \citealt{Liu2018}, \citealt{Ko2020}).
Given the typical high accretion rate of the Class~0 protostar, one natural way to led to a high temperature contrast is to consider that dust in the disk is dominated by viscous heating.
In this case, the radial temperature profile has a $r^{-0.75}$ dependence \citep{Chiang1997}.

We assumed that the radial temperature profile  $T^{ \mbox{\scriptsize disk} }(r)$ has a $r^{-0.75}$ dependence.
A caveat is that viscous dissipation is a heating mechanism for gas instead for dust.
How efficient the dust in the disk can be heated through the collision with H$_{2}$ required more careful calculations.
When gas density is high, and when the grain sizes are small thus dust and gas remains dynamically well coupled, it may not be unrealistic to assume that they are also thermally coupled.
Some recent observations of dust continuum emission towards the inner disks of certain actively accreting YSOs also indicated that viscous heating may be required to explained the observed dust continuum emission (e.g., \citealt{Liu2019b}, \citealt{Takami2019}; c.f. \citealt{Lin2020}).
The explicit form of  $T^{ \mbox{\scriptsize disk} }(r)$ in our model is:
\begin{equation}\label{eq:Tdisk}
   T^{ \mbox{\scriptsize disk} }(r) = 65 \mbox{[K]} \cdot \left( \frac{r \mbox{[au]}}{50} \right)^{-0.75},
\end{equation}
which can yield a 265 GHz brightness temperature distribution that is consistent with the peak value resolved from the high-resolution ALMA image (Figure \ref{fig:model}).
Such a radial temperature profile is consistent with the radiation temperature profile of a optically thick viscous disk (see \citealt{Chiang1997}), if assuming a $\sim$4 $M_{\odot}$ host protostar (i.e., comparable with the disk gas mass estimated in Section \ref{sub:dust}) and a constant 1$\times$10$^{-4}$ M$_{\odot}$\,yr$^{-1}$ over radius.

The mass of the host protostar in OMC-3/MMS~6 is presently uncertain.
For a Class~0 YSO, the mass accretion rate of as high as a few times 10$^{-4}$ M$_{\odot}$\,yr$^{-1}$ is indeed plausible  (e.g., \citealt{Jayawardhana2001}).
In addition, we note that reducing the accretion rate by one order of magnitude would not lead to a qualitatively different radiative transfer result since temperature has a power equal to $1/4$ dependence on the mechanical work.
Moreover, the disk cannot be vertically isothermal if it is dominated by viscous heating: dust closer to the disk mid-plane is expected to have higher temperature due to the vertical heat diffusion.
Given that $T^{ \mbox{\scriptsize disk} }(r)$ was mainly constrained by the 33 GHz observations which is not very optically thick, the dust temperature profile described by Equation \ref{eq:Tdisk} should be understood as the weighted-average of dust temperature over the projected scale-height, which is higher than the radiation temperature of the the $\dot{M}=$ 10$^{-4}$  M$_{\odot}$\,yr$^{-1}$ viscous disk  (\citealt{Chiang1997}).
Finally, the estimated mass-loss rate from OMC-3/MMS~6 based on the observations of the CO outflows is 4.4$\times$10$^{-6}$ M$_{\odot}$\,yr$^{-1}$, which needs to be supported by the protostellar mass accretion of comparably higher than 4$\times$10$^{-5}$ M$_{\odot}$\,yr$^{-1}$ if assuming a comparably lower than 10\% mass-loading to outflow.
Therefore, we regard the assumed $T^{ \mbox{\scriptsize disk} }(r)$ profile to be realistic.

We remark that our 1D radiative transfer modeling (Section \ref{sec:radiativetransfer}) did inevitably ignore the vertical thermal profile of the disk.
This is fair when comparing with the ALMA observations and the JVLA observations over spatially extended area.
However, for the central $\sim$20 au region where the 33 GHz optical depth reaches $\sim$1 or higher, ignoring the vertical thermal structure of a viscous-heating dominated disk would lead to significant overestimates of polarization percentages (e.g., on the spatial scales unresolved by the presented JVLA observations, the $PI_{\mbox{\scriptsize 33 GHz}}$ may trace linearly polarized dichroic extinction instead of polarized emission), which may partly explain the discrepancy between our radiative transfer model and the observations (Section \ref{sec:discussion}; Figure \ref{fig:slice}).

%\clearpage

\begin{deluxetable*}{ l l c c c c c c c }
\tablecaption{Polarization measurements taken from the JVLA A+C configuration image$^{1}$ \label{tab:jvla_segments}}
  \tablehead{
  \colhead{R.A.} & \colhead{Decl.}  & \colhead{Stokes I} & \colhead{Stokes Q} & \colhead{Stokes U} & \colhead{PI}   &   \colhead{PA}  &   \colhead{$P$} \\
  (J2000)   & (J2000)   &   (\mjyperbeam)   &   (\mjyperbeam)   &   (\mjyperbeam)   &   (\mjyperbeam) &   ($^{\circ}$)    &   (\%)
  } 
  \startdata
5\rah35\ram23.435\ras  & -5\decd1\decm30.701\decs   & 0.047 & 0.0019 & 0.032 & 0.03 & 43 & 66 \\
5\rah35\ram23.437\ras  & -5\decd1\decm30.651\decs   & 0.048 & -0.0027 & 0.029 & 0.026 & 48 & 58 \\
5\rah35\ram23.435\ras  & -5\decd1\decm30.6\decs   & 0.12 & 0.019 & 0.034 & 0.036 & 31 & 31 \\
5\rah35\ram23.437\ras  & -5\decd1\decm30.55\decs   & 0.081 & 0.029 & 0.024 & 0.035 & 20 & 45 \\
5\rah35\ram23.435\ras  & -5\decd1\decm30.5\decs   & 0.16 & 0.031 & 0.028 & 0.039 & 21 & 26 \\
5\rah35\ram23.437\ras  & -5\decd1\decm30.45\decs   & 0.061 & 0.016 & 0.024 & 0.026 & 28 & 45 \\
5\rah35\ram23.435\ras  & -5\decd1\decm30.399\decs   & 0.068 & 0.0039 & 0.033 & 0.03 & 42 & 47 \\
5\rah35\ram23.431\ras  & -5\decd1\decm30.701\decs   & 0.095 & 0.009 & 0.038 & 0.037 & 38 & 40 \\
5\rah35\ram23.433\ras  & -5\decd1\decm30.651\decs   & 0.13 & 0.01 & 0.042 & 0.04 & 38 & 32 \\
5\rah35\ram23.431\ras  & -5\decd1\decm30.6\decs   & 0.35 & 0.029 & 0.05 & 0.055 & 30 & 16 \\
5\rah35\ram23.433\ras  & -5\decd1\decm30.55\decs   & 0.28 & 0.034 & 0.037 & 0.047 & 23 & 18 \\
5\rah35\ram23.431\ras  & -5\decd1\decm30.5\decs   & 0.46 & 0.024 & 0.033 & 0.038 & 27 & 8.7 \\
5\rah35\ram23.433\ras  & -5\decd1\decm30.45\decs   & 0.22 & 0.013 & 0.031 & 0.031 & 33 & 15 \\
5\rah35\ram23.431\ras  & -5\decd1\decm30.399\decs   & 0.2 & -0.0025 & 0.029 & 0.026 & 47 & 14 \\
5\rah35\ram23.427\ras  & -5\decd1\decm30.701\decs   & 0.15 & 0.0081 & 0.041 & 0.039 & 39 & 27 \\
5\rah35\ram23.429\ras  & -5\decd1\decm30.651\decs   & 0.27 & 0.014 & 0.05 & 0.049 & 37 & 19 \\
5\rah35\ram23.427\ras  & -5\decd1\decm30.6\decs   & 0.72 & 0.026 & 0.054 & 0.056 & 32 & 8.1 \\
5\rah35\ram23.429\ras  & -5\decd1\decm30.55\decs   & 0.64 & 0.033 & 0.042 & 0.05 & 26 & 8.2 \\
5\rah35\ram23.427\ras  & -5\decd1\decm30.5\decs   & 0.88 & 0.011 & 0.031 & 0.03 & 35 & 3.6 \\
5\rah35\ram23.429\ras  & -5\decd1\decm30.45\decs   & 0.45 & 0.0065 & 0.034 & 0.032 & 40 & 7.4 \\
5\rah35\ram23.427\ras  & -5\decd1\decm30.399\decs   & 0.32 & -0.0093 & 0.032 & 0.031 & 53 & 10 \\
5\rah35\ram23.425\ras  & -5\decd1\decm30.751\decs   & 0.058 & 0.01 & 0.027 & 0.027 & 35 & 48 \\
5\rah35\ram23.423\ras  & -5\decd1\decm30.701\decs   & 0.21 & 0.0069 & 0.038 & 0.036 & 40 & 18 \\
5\rah35\ram23.425\ras  & -5\decd1\decm30.651\decs   & 0.5 & 0.0098 & 0.05 & 0.048 & 39 & 10 \\
5\rah35\ram23.423\ras  & -5\decd1\decm30.6\decs   & 1.2 & 0.015 & 0.043 & 0.043 & 35 & 3.7 \\
5\rah35\ram23.425\ras  & -5\decd1\decm30.55\decs   & 1.3 & 0.02 & 0.037 & 0.039 & 31 & 3.1 \\
5\rah35\ram23.425\ras  & -5\decd1\decm30.45\decs   & 0.73 & -0.0046 & 0.034 & 0.032 & 49 & 4.5 \\
5\rah35\ram23.423\ras  & -5\decd1\decm30.399\decs   & 0.38 & -0.013 & 0.036 & 0.036 & 55 & 9.9 \\
5\rah35\ram23.425\ras  & -5\decd1\decm30.349\decs   & 0.14 & -0.014 & 0.024 & 0.025 & 60 & 18 \\
5\rah35\ram23.419\ras  & -5\decd1\decm30.701\decs   & 0.19 & 0.0057 & 0.038 & 0.035 & 41 & 19 \\
5\rah35\ram23.421\ras  & -5\decd1\decm30.651\decs   & 0.61 & 0.0079 & 0.043 & 0.041 & 40 & 6.9 \\
5\rah35\ram23.419\ras  & -5\decd1\decm30.6\decs   & 1.1 & 0.0091 & 0.03 & 0.028 & 37 & 2.7 \\
5\rah35\ram23.421\ras  & -5\decd1\decm30.55\decs   & 1.6 & 0.0096 & 0.028 & 0.026 & 35 & 1.7 \\
5\rah35\ram23.421\ras  & -5\decd1\decm30.45\decs   & 0.77 & -0.01 & 0.027 & 0.026 & 55 & 3.6 \\
5\rah35\ram23.419\ras  & -5\decd1\decm30.399\decs   & 0.29 & -0.0094 & 0.03 & 0.028 & 54 & 10 \\
5\rah35\ram23.415\ras  & -5\decd1\decm30.701\decs   & 0.1 & 0.00032 & 0.034 & 0.031 & 45 & 32 \\
5\rah35\ram23.417\ras  & -5\decd1\decm30.651\decs   & 0.42 & 0.011 & 0.033 & 0.033 & 36 & 8.1 \\
  \enddata
  \tablecomments{(1) Robust$=$2 weighted and smoothed to a 0\farcs15 circular synthesized beam. The RMS noise of the Stokes I, Q, U, and PI images after smoothing is 8.7 \ujyperbeam. The errors of PA and $P$ can be evaluated following standard error propagation and based on the assumption of Gaussian noise statistics. More details are given in Section \ref{sub:JVLA}. Samplings were made at the centroids of the polarization line segments plotted in Figure \ref{fig:polmap}.}
\end{deluxetable*}

\end{document}